\documentclass{comnet-mod}

\usepackage{amssymb}
\usepackage{amsmath}
\usepackage{amsfonts}
\usepackage{amstext}

\usepackage{bbm}
\usepackage{bm}
\usepackage{bbold}

\usepackage{array}
\usepackage{booktabs}
\usepackage{multirow} 

\usepackage{mathtools} 
\usepackage{makecell} 
\usepackage{todonotes} 

\newcommand{\eq}[1]{Eq.\,(\ref{eq:#1})}
\newcommand{\Eq}[1]{Equation (\ref{eq:#1})}

\newcommand{\eqrange}[2]{Eqs.\,(\ref{eq:#1}) to (\ref{eq:#2})}

\newcommand\trace{\textrm{tr}}

\newcommand\Frag{S} 
\newcommand\effi{\mathcal{E}}

\newcommand\smallestpad{} 

\newcommand\citecell[2]{\parbox[t][][t]{#1\columnwidth}{\begin{flushleft}\vspace{-0.6cm}\cite{#2}\end{flushleft}}}
\newcommand\relvspace{\vspace{-0.25cm}}
\newcommand\bigrelvspace{\vspace{-0.3cm}}
\title{Structural robustness and vulnerability of networks}
\shorttitle{Robustness and vulnerability of networks} 
\shortauthorlist{Schwarze, Jiang, Wray, Porter} 

\author{
\name{Alice C.\,Schwarze$^*$}
\address{Dartmouth College, Hanover, NH, USA\email{$^*$Corresponding author: alice.c.schwarze\@ dartmouth.edu}}
\name{Jessica Jiang}
\address{Dartmouth College, Hanover, NH, USA}
\name{Jonny Wray}
\address{Nucleome Therapeutics plc, Oxford, UK}
\and
\name{Mason A.\,Porter}
\address{University of California Los Angeles, Los Angeles, CA, USA}
\address{Santa Fe Institute, Santa Fe, NM, USA}
}

\begin{document}

\maketitle

\begin{abstract}
{
Networks are useful descriptions of the structure of many complex systems. Unsurprisingly, it is thus important to analyze the robustness of networks in many scientific disciplines. In applications in communication, logistics, finance, ecology, biomedicine, and many other fields, researchers have studied the robustness of networks to the removal of nodes, edges, or other subnetworks to identify and characterize robust network structures. A major challenge in the study of network robustness is that researchers have reported that
different and seemingly contradictory network properties are correlated with a network's robustness. 
Using a framework by Alderson and Doyle~\cite{Alderson2010},  we categorize several notions of network robustness and we examine these ostensible contradictions.
We survey studies of network robustness with a focus on (1)~identifying robustness specifications in common use,
(2)~understanding when these specifications are appropriate, and (3)~understanding the conditions under which one can expect different notions of robustness to yield similar results. With this review, we aim to give researchers an overview of the large, interdisciplinary body of work on network robustness and develop practical guidance for the design of computational experiments to study a network's robustness.
}
{System design, robustness, vulnerability, reliability, resilience, stability, performance, random failures, targeted attacks, paths, walks}
\end{abstract}

\tableofcontents

\begin{table}[!h]
\begin{center}
\begin{tabular}{l| l} 
\begin{tabular}{p{0.65cm} p{4.9cm}} 
\toprule
Symbol & Meaning \\
\midrule
$G$ or $(V,E)$ & a network (i.e., a graph) with node set $V$ and edge set $E$ \smallestpad\\
$v_i$ & the $i$th node of a network \smallestpad\\
$e_{i,j}$ & an edge from node $v_i$ to node $v_j$ \smallestpad\\
$k_i$ & the degree of node $v_i$ \smallestpad\\
$N$ & the number of nodes of a network \smallestpad\\
$m$ & the number of edges of a network \smallestpad\\
$(V_S,E_S)$ & a subnetwork with node set $V_S\subseteq V$ and edge set $E_S\subseteq E$ \smallestpad\\
$(V_P,E_P)$ & a path in a network \smallestpad\\
$(V_H,E_H)$ & a hypergraph with node set $V_H$ and hyperedge set $E_H$ \smallestpad\\
$w$ & a walk on a network \smallestpad\\
$\ell$ & the length of a path or walk \smallestpad\\
${\bf A}~=~(a_{i,j})$ & a network's adjacency matrix \smallestpad\\
${\bf L}~=~(l_{i,j})$ & a network's combinatorial Laplacian matrix \smallestpad\\
$X$ & a performance measure \smallestpad\\
$N_{\textrm{LCC}}$ & the size of a network's largest connected component \smallestpad\\
$n_c$ & the number of components of a network \smallestpad\\
$\overline{N}_c$ & the mean component size of a network \smallestpad\\
$\Frag$ & the relative size of a network's largest connected component \smallestpad\\
$r$ & a network's reachability \smallestpad\\
$p(k)$ & a network's degree distribution \smallestpad\\
$H$ & the graph entropy of a network
  \smallestpad\\
$C$ & a network's mean local clustering coefficient \smallestpad\\
$T$ & a network's global clustering coefficient \smallestpad\\
$L$ & a network's mean shortest-path length \smallestpad\\
$\effi$ & a network's efficiency \smallestpad\\
$\Omega$ & a network's resistance distance \smallestpad\\
$K$ & a network's natural connectivity \smallestpad\\
$\bf x$ & the state of a dynamical system\smallestpad\\
\bottomrule
\end{tabular} &
\begin{tabular}{p{0.4cm} p{4.9cm}} 
\toprule
Symbol & Meaning \\
\midrule
    $\lambda_i$ & the $i$th eigenvalue of a matrix\smallestpad\\
    $\boldsymbol\eta_i$ & the $i$th eigenvector of a matrix\smallestpad\\ 
    $\mathcal P$ & a set of perturbations\smallestpad\\
    $p$ & a perturbation \smallestpad\\
    $X_p$ & the performance of a system after a perturbation $p$\smallestpad\\
    $I_p$ & the impact of a perturbation $p$ on a system's performance\smallestpad\\
    $\textrm{BC}_i$ & the geodesic betweenness centrality of node $v_i$\smallestpad\\
    $\textrm{CC}_i$ & the closeness centrality of node $v_i$\smallestpad\\
    $\textrm{EC}_i$ & the eigenvector centrality of node $v_i$\smallestpad\\
    $\textrm{KC}_i$ & the Katz centrality of node $v_i$\smallestpad\\
    $\textrm{PRC}_i$ & the PageRank centrality of node $v_i$\smallestpad\\
    $\textrm{SC}_i$ & the subgraph centrality of node $v_i$\smallestpad\\
    $\textrm{DC}_i$ & the damage centrality of node $v_i$\smallestpad\\
    $\textrm{BC}_{i,j}$ & the edge betweenness centrality of $e_{i,j}$\smallestpad\\
    $C_{i,j}$ & the edge clustering coefficient of edge $e_{i,j}$\smallestpad\\
    $f_v$ & the critical fraction of nodes to be removed\smallestpad\\
    $f_e$ & the critical fraction of edges to be removed\smallestpad\\
    $R$ & Schneider's robustness index\smallestpad\\
    $V$ & Schneider's vulnerability index \smallestpad\\
    $\kappa_v$ & a network's node connectivity\smallestpad\\
    $\kappa_e$ & a network's edge connectivity\smallestpad\\
    $\partial V_S$ & the edge boundary of a node set $V_S$\smallestpad\\
    $\partial_{\textrm{out}} V_S$ & the outer node boundary of a node set $V_S$\smallestpad\\
    $h_v$ & the node expansion of a network\smallestpad\\
    $h_e$ & the edge expansion of a network\smallestpad\\
    $a$ & the algebraic connectivity (i.e., Fiedler value) of a network\smallestpad\\
    $\rho$ & the spectral radius of a matrix\smallestpad\\
    $P(I,p)$ & the joint probability distribution of perturbations and impacts\smallestpad\\
\bottomrule
\end{tabular}\\
\end{tabular}
\end{center}
\vspace{-0.3cm}
\caption{List of symbols.}
\label{tab:los}
\end{table}

\section{Introduction}

Across many scientific disciplines, researchers have modeled various real-world systems as networks and have examined the robustness of networks to perturbations \cite{Larhlimi2011, Cuadra2015, Gao2015, Catanese2016, Williams2016, Caccioli2018}. Indeed, robustness is a crucial concept in many applications of networks. The study of the robustness of an infrastructure network (e.g., a power grid or a road network) can help anticipate the performance of such systems in the event of component failures or adversarial attacks \cite{Cuadra2015, Gao2015}. Examining the robustness of a social network can help understand how communities of people, animals, or other individuals react to various perturbations (e.g., a natural disaster or the loss of one or several community members) \cite{Berkes2013, Mourier2017}. Investigating the robustness of biological systems is an important part of understanding the persistence of life on earth \cite{Whitacre2010, Noon2016}. Organisms are affected by environmental changes, and there are genetic changes from one generation of individuals to the next~\cite{Nachman2000}. The study of genetic robustness can help understand how a genome that is subjected to many changes can robustly code for a phenotype \cite{Masel2010, Whitacre2010}. Studying the robustness of an ecological system can inform forecasts of the effects of climate change or species extinction on it \cite{Dunne2002, Schleuning2016}. The robustness of biological systems is also relevant to applications in biomedicine \cite{Kitano2004a}. 

Depending on the application and context, one can view robustness as a desirable property and seek ways to increase it. For example, one can view the human body as a system whose function is to survive and stay healthy. By contrast, one can view a cancer, a viral infection, and other diseases as biological systems that one wants to disrupt by a therapeutic intervention. Understanding the robustness of these systems can help in the design of therapeutic interventions that can circumvent fail-safe mechanisms in these systems and can thereby disrupt them \cite{Kitano2007a}.

The aforementioned applications and many others have generated much interest in identifying `robust network structures' \cite{VanMieghem2005, Herrmann2011, Peixoto2012, Zeng2012}. A major problem in this endeavor is that many researchers have reported that various (and sometimes seemingly contradictory) network properties are correlated with a network's robustness. For example, Dunne et al.~\cite{Dunne2002} investigated the robustness of a food web (i.e., in which directed edges indicate that one species eats another species) to species loss and concluded that robustness increases with increasing network density. By contrast, Vieira and Almeida-Neto~\cite{Vieira2015} examined the robustness of a mutualistic network (i.e., in which undirected edges indicate that two species interact in a way that is beneficial to both of them) to species loss and concluded that robustness increases with decreasing network density. Newman \cite{Newman2002} reported that a network's robustness to node removal increases with degree--degree assortativity. However, Zhou et al.~\cite{Zhou2012} reported that the robustness of a multilayer network to node removal decreases with degree--degree assortativity. These examples demonstrate that it is difficult --- and perhaps impossible --- to identify network properties that necessarily correlate positively with robustness in all or even many systems. In the preface to the first edition of Piet Van Mieghem's book on `Graph Spectra of Complex Networks', which was originally published in 2010, the author even stated that `the rather simple but highly relevant question `What is a robust network?' seems beyond the realm of present understanding' \cite[p.\,xiii]{VanMieghem2023}. More than a decade later, many problems concerning network robustness remain open \cite{pirani2022}.

Alderson and Doyle~\cite{Alderson2010} argued that one can resolve the supposed contradictions between studies of robustness by distinguishing different types of robustness using a few specifications. They proposed to define `robustness' as a measure of invariance of a property of a system to a set of perturbations. In general, different choices of the measure of invariance, the system, the system property, and the set of perturbations lead to different notions of robustness that do not need to be correlated with each other. Their framework helps one to recognize that many studies of robustness consider different notions of robustness and thus need not lead to similar conclusions for the characteristics of a `robust network structure'. After recognizing these different notions of robustness, one can ask whether or not one can learn any general lessons from research on network robustness. Are there conditions under which one can expect different notions of robustness to lead to similar results? Given a system, which notion of robustness is most appropriate to use for it? To study these questions, we review existing research on network robustness. The focus of our review is to (1) identify robustness specifications that researchers use frequently, (2) develop an understanding of when these specifications are appropriate, and (3) develop an understanding of the conditions under which one can expect different studies of robustness to yield similar results.

To discuss these issues, it is important to distinguish between real-world systems and the models that researchers use to study those systems. Although the terms "system" and "model" may be used in other ways, we will generally use the word "system" to refer to a real-world object that a researcher studies. By contrast, we will use the term "model" to refer to an abstraction of a system.
We consider network models, which are abstract models that capture the interactions between the entities or elements (i.e., nodes) of a system. These models include a \textit{network} (see Section \ref{sec:rev:simple_networks}), which encodes the structure that underlies these interactions, and also can include additional features (see Section \ref{sec:networks_plus}).

In our review, we aim to give network scientists --- including aspiring network scientists --- guidance to filter and categorize the network-robustness results that are relevant to their field and advice on how to select network models for their own research. The robustness of many real-world systems is very different from the robustness of well-studied models. In discussing these discrepancies between real-world systems and models, we aim to highlight research areas that we expect to benefit from further theoretical work.

Our paper proceeds as follows. In Section \ref{sec:robustness_def}, we adapt Alderson and Doyle's definition of robustness \cite{Alderson2010} and discuss the necessary specifications for a robustness problem. In Section \ref{sec:concepts}, we examine several concepts that are related to robustness. In Section \ref{sec:review:graphs}, we give definitions of graph-theoretic concepts that are useful for the study of the structural robustness of networks. In Sections \ref{sec:models}--\ref{sec:measures}, we survey robustness-problem specifications that researchers have considered in the network-robustness investigations and discuss the relevance of these specifications to real-world systems. We conclude our review in Section \ref{sec:review:conclusion} by offering practical guidance to researchers who are interested in investigating the robustness of real networked systems.

\section{Defining robustness}\label{sec:robustness_def}

In a sequence of several publications, Carlson, Doyle, Alderson, and their collaborators developed a commonly referenced approach to study system robustness \cite{Carlson1999, Carlson2000, Doyle2000, Carlson2002, Doyle2005, Alderson2010}. Their approach characterizes most systems as `robust yet fragile' \cite{Doyle2005}. They proposed that, for most systems, one can find some aspects that are `robust' and other aspects that are `fragile' and that an assessment of `robustness' and `fragility' in a system requires several specifications \cite{Alderson2010}. More specifically, Alderson and Doyle~\cite{Alderson2010} proposed that a `[property] of a [system] is \textit{robust} if it is [invariant] with respect to a [set of perturbations]', where they used square brackets to indicate the concepts that require specification before one can attempt to study a system's robustness. For example, it is impossible to answer the question `Are bacteria robust?' without specifying the system (e.g., the bacterial species, the initial bacteria population, and the environment), the system property (i.e., some property of bacteria, such as movement, growth, or reproduction rate), the set of permissible perturbations (e.g., changes in temperature, pressure, or nutrient availability), and how one measures the `invariance' of a system property under a perturbation. In the present discussion, we use the term `invariant' to be faithful to the definition in \cite{Alderson2010}. However, in most practical settings, system properties almost always change under perturbations, so they are very rarely strictly invariant (i.e., unchanged). Therefore, we slightly modify the definition of robustness in \cite{Alderson2010} and instead use the following definition:

\begin{center}\textit{Robustness} is the [insensitivity] of a [property] of a [model of a system]\\ to a [set of perturbations].\end{center}

\begin{figure}[t]
\centering
\includegraphics[trim={0cm 0cm 0.1cm 0.5cm},clip,width=1\textwidth]{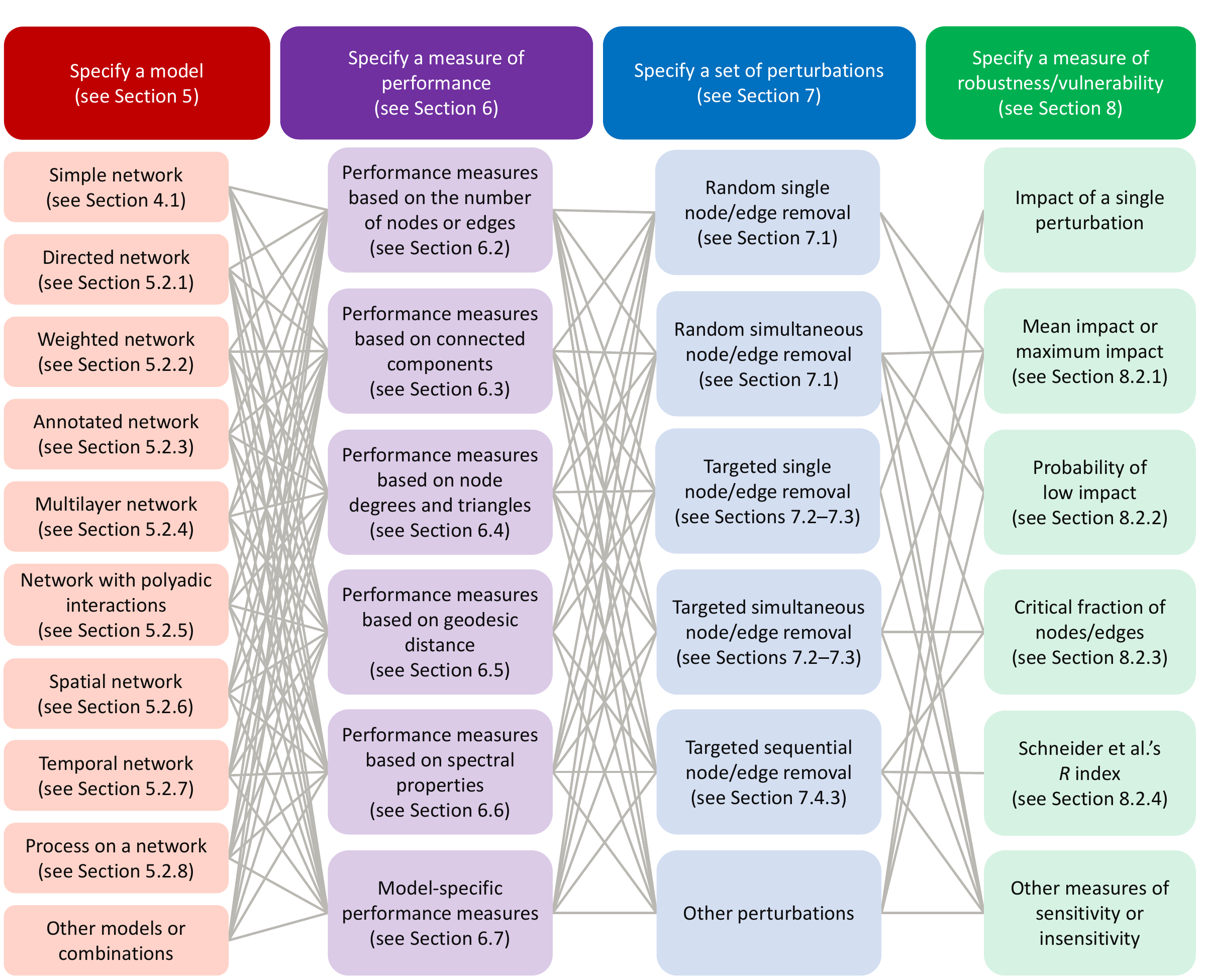}  
\caption{{\bf Specifications of a robustness problem for a networked system.} In each column, we categorize widely used choices for a specification of a network-robustness problem. Gray lines connect two categories for different specifications if, to the best of our knowledge, scholars have chosen specifications from this pair of categories in a study of network robustness. We omit edges between concepts in non-consecutive chapters in this visualization.
} 
\label{fig:flowchart}
\end{figure}

We refer to the problem of determining a system's robustness (e.g., the robustness of a bacterium) as a \textit{robustness problem}. The question `Are bacteria robust?' poses an \textit{underspecified} robustness problem. In other words, it is a problem that one cannot solve because one lacks the necessary information to solve it. To solve a robustness problem, the following specifications are necessary:
\begin{enumerate}
\item the specification of a {\bf system} and (if the study is not an observational study) the specification of an experimental, computational, or theoretical {\bf model} of the specified system;
\item the specification of a system property or model property as a {\bf measure of system performance};
\item the specification of a {\bf set of perturbations} to which one subjects the system;
\item the specification of a {\bf measure of insensitivity} (i.e., robustness) of system performance under the specified set of perturbations.
\end{enumerate}
Specifying these attributes can help resolve many supposed contradictions between the conclusions of different studies of robustness \cite{Alderson2010}. We dedicate a large portion of our review to the various ways that researchers have specified attributes of robustness problems.  In Fig.~\ref{fig:flowchart}, we indicate a variety of attributes and categorize common ways to specify them. The task of specifying a robustness problem is modular in the sense that one can combine most possible choices for one attribute (e.g., the network model) with most of the possible choices for another attribute (e.g., the performance measure or the set of perturbations). However, the small number of gray edges between the choices of perturbation sets and the choices of robustness measures (i.e., insensitivity measures) signifies that the specification of a perturbation set can impose constraints on the possible robustness measures. 
For example, using the mean impact, maximum impact, or the probability of a low impact as a robustness measure requires that the perturbation set includes more than one perturbation. Another example is Schneider et al.'s $R$ index, which is the mean loss of performance  as one sequentially removes all nodes or all edges of a network \cite{Schneider2011}. When computing Schneider's $R$ index, one thus necessarily considers perturbation sets  of sequentially increasing size (i.e., the number of nodes or edges that are removed).

Selecting the most fitting specifications for a robustness problem is important, but it can be a difficult task. In the remainder of our review, we aim to provide guidance on this selection process. Following an overview of related concepts (see Section \ref{sec:concepts}) and an introduction to graph-theoretic terminology (see Section \ref{sec:review:graphs}), we review common choices that researchers have made to specify robustness problems. Sections \ref{sec:models}, \ref{sec:performance_measures}, \ref{sec:perturbations}, and \ref{sec:measures} each concern one robustness-problem attribute. We end each of these sections with a discussion of questions that can arise when attempting to specify the associated attribute of a robustness problem.

\section{Related concepts}\label{sec:concepts}

\begin{figure}[p]
\centering
\includegraphics[trim={0cm 2.cm 2.75cm 3cm},clip,width=0.85\textwidth]{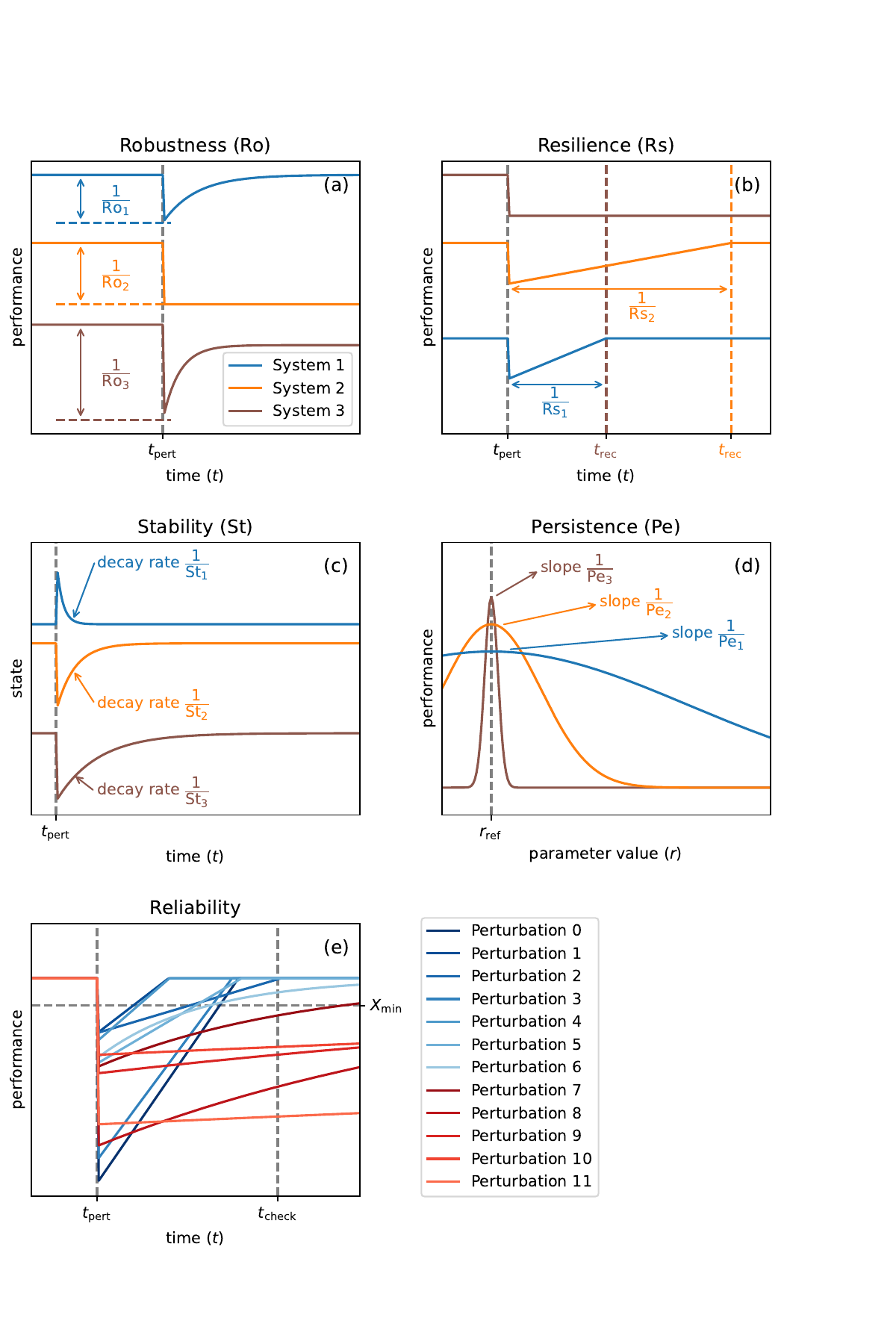}  
\caption{{\bf Examples of formalizations of robustness and related concepts.}
In panel (a), we show performance--time curves for three example systems for which we use the inverse performance drop at the time $t_{\textrm{pert}}$ of perturbation as a measure of robustness. In panel (b), we use the inverse length of the time interval between $t_{\textrm{pert}}$ and the time $t_{\textrm{rec}}$ of complete performance recovery as a measure of resilience. In panel (c), we show state--time curves for three example systems for which we use the inverse decay rate of a small displacement as a measure of stability. In panel (d), we show performance over a system parameter $r$ for three example systems for which we use the slope of the curve close to a reference value $r_{\textrm{ref}}$  of $r$ as a measure of persistence. In panel (e), we illustrate that a single system can respond differently to different perturbations, and we use the fraction of perturbations after which the system recovers a minimal performance given by $X_{\textrm{min}}$ by time $t_{\textrm{check}}$.
}
\label{fig:concepts}
\end{figure}

In studies of attack tolerance, failure tolerance, system design, and other questions, many researchers have worked with concepts with related, overlapping, or identical meanings with the notion of robustness of Alderson and Doyle \cite{Alderson2010}. In Section \ref{sec:synonyms}, we review some of these concepts. Other concepts have been presented in previous studies as synonymous with robustness, although these concepts differ from our focal robustness notion in the present review (and also differ from the robustness notion in \cite{Alderson2010}). In Section \ref{sec:homonyms}, we review some of these concepts. In Section \ref{sec:more_reviews}, we point readers to other reviews of robustness and related concepts. In Fig.~\ref{fig:concepts}, we provide a schematic of 
formalizations of some of the concepts in Sections \ref{sec:synonyms} and \ref{sec:homonyms}.

\subsection{Different name, but similar concept?}\label{sec:synonyms}

Researchers in different fields have used a variety of different terms to describe the responses of systems to perturbations. Examples of such terms include `stability' \cite{Dorato1987, Stelling2004, Lesne2008}, `reliability' \cite{Little2003, Rausand2004, Cuadra2015}, and `resilience' \cite{Bhamra2011}. Researchers have often used these terms as synonyms for robustness and for each other \cite{Dorato1987, Bhamra2011, Walsh2013, Cuadra2015}. Therefore, a clear disambiguation of these terms is not possible without creating conflicts with at least some of the literature on network robustness. In this subsection, we consider the language that researchers have used to refer to robustness, special cases of robustness, and closely related concepts.

\subsubsection{Resilience.}\label{sec:resilience}

At its heart, robustness is the insensitivity of some property of a system to some perturbation. Many scholars consider `resilience' to be a measure of the ability of a system to recover its previous state after a perturbation has changed a system property \cite{Cuadra2015, Hosseini2016, Krakovska2024}. A system's resilience can thus include processes that occur over some period of time. A system property can be very sensitive to a set of perturbations (thereby leading to low robustness of that property to that set of perturbations) but can quickly retain its original state through one or several fail-safe mechanisms (thereby leading to a high reliability of that property with respect to the set of perturbations). For example, Cuadra et al.~\cite{Cuadra2015} considered a power grid to be `robust' if blackouts are rare events. They considered a power grid to be `resilient' if blackouts are short-lived because the power grid is able to quickly recover its functioning state. Crespi et al.~\cite{Crespi2021} gave an example of the difference between robustness and resilience in plant biology. They considered a plant to be a robust system if it is able to grow under many different light conditions, and they considered it to be a resilient system if it is able to grow under changing light conditions. In their view, a plant that can grow under many light conditions but cannot tolerate switching light conditions is robust, but it is not resilient. By contrast, a plant that can compensate for periods of darkness with efficient growth during sunny periods but cannot grow in a permanently shaded environment is resilient, but it is not robust. Other researchers who have examined resilience include Bhamra et al.~\cite{Bhamra2011}, who surveyed the resilience of various systems, and Liu et al.~\cite{Liu2022}, who connected aspects of resilience in engineered systems to dynamical properties of network models of biological and social systems.

To illustrate the conceptual difference between robustness and resilience, we show system performance over time for three example systems in Fig.~\ref{fig:concepts}~(a) and for another three example systems in Fig.~\ref{fig:concepts}~(b). In Fig.~\ref{fig:concepts}~(a), System 1 is more robust than System 2, which in turn is more robust than System 3. In Fig.~\ref{fig:concepts}~(b), System 3 has a resilience of $\textrm{Rs}_1 = 0$, because its performance after the perturbation never reaches its initial value during the observed time period. The resilience of System 2 is nonzero, but it is smaller than the resilience of System 1. 

In practice, it can sometimes be difficult to distinguish between the concepts of robustness and resilience. We illustrate this difficulty using the example of power grids. Consider a setting in which we can only query the state of a power grid once per day. We may decide to check for a blackout every morning at 8:00 am. If we detect a blackout very rarely, we may perhaps conclude that our power grid is robust to various environmental conditions and/or to the random failures that are likely to occur at some point on most days. However, it is possible that our power grid experiences many blackouts every day that last only a few seconds. Such short-lived blackouts indicate that our power grid is not robust, but that instead it is resilient. Our limited measurements do not make it possible to distinguish between these two different scenarios. Therefore, despite the starkly different definitions of robustness and resilience, it can be hard to separate these two concepts in practice, and both (1) the temporal and other resolutions of measurements and (2) model complexity can affect whether one concludes that a system or a model of a system is robust, resilient, both robust and resilient, or neither robust nor resilient.

For further discussions of resilience and/or its connection to robustness, we point our readers to several other review articles on these topics. Bhamra et al.~\cite{Bhamra2011} surveyed the resilience of systems. Crespi et al.~\cite{Crespi2021} compared a variety of definitions of robustness and resilience, with a focus on biological systems. Liu et al.~\cite{Liu2022} connected aspects of resilience in engineered systems to dynamical properties of network models of biological and social systems.

\subsubsection{Stability.}

In the study of dynamical systems, it is very common to investigate the \textit{asymptotic stability} of a mathematical model with respect to small perturbations \cite[p.\,143]{Strogatz2018}. A model is \textit{asymptotically stable} to small perturbations if a small perturbation to a model's state dies out over time, such that the model eventually returns to the state that it had before the perturbation \cite[p.\,3]{Guckenheimer2013}. We can understand asymptotic stability to perturbations as a special case of resilience; asymptotic stability is the resilience of a dynamical system's state to a small displacement of the state \cite{Stelling2004, Lesne2008}.

To illustrate how stability relates to the time evolution of a system's state, we show a one-dimensional system state as a function of time in Fig.~\ref{fig:concepts}~(c). In this example, System 1 is more stable than System 2, which in turn is more stable than System 3.

Boolean networks are coupled dynamical systems of Boolean variables, which are variables that take values in the set $\{0,1\}$ \cite{Schwab2020}. The calculus-based approach that is commonly used to analyze the stability of of dynamical systems with continuous-valued variables does not lend itself to the stability analysis of Boolean systems because the smallest possible perturbation to a Boolean network is a bit-flip (i.e., a change of variable from 0 to 1 or vice versa). Several researchers have proposed measures of impact of bit-flips on Boolean network dynamics \cite{Gershenson2005, Pineda2019, Schwab2020}. For example, Gershenson et al.~used the Hamming distance between trajectories of perturbed and unperturbed systems to measure the impact of a perturbation \cite{Gershenson2005}. Pineda et al.~proposed a robustness measure, which they called `antifragility' \cite{Pineda2019}. A Boolean network's antifragility is given by negative of the change of the mean entropy of node states divided by the bit-flip frequency. Among the concepts discussed in this section, the insensitivity of Boolean dynamics to bit-flips is most closely related to the concept of stability because

\subsubsection{Reliability.}

In the engineering sciences, the reliability of a human-made system is the `ability of [a system] to perform a required function, under given environmental and operational conditions and for a stated period of time' \cite[p.\,5]{Rausand2004}. Rausand and H{\o}yland~\cite[p.\,5]{Rausand2004} stated that, until the 1960s, it was common to define reliability as the probability (rather than the ability) of a system to perform a required function. They also noted that many scholars still use a probabilistic definition of reliability. There are several examples of such probabilistic approaches in studies of communication networks \cite{Colbourn1987, Bhamra2011, Walsh2013, Gu2020}. Colbourn~\cite{Colbourn1987} defined reliability as the probability of successful communication between specified terminals. Gu et al.~\cite{Gu2020} and Zhou et al.~\cite{Zhou2020} defined reliability of a transportation network as `the probability that the transportation network remains satisfactory [...] under perturbation'. Cuadra et al.~\cite{Cuadra2015} characterized the reliability of power grids as `a beneficial property for a power grid that refers to its ability to supply electric loads with a high level of probability, during a given time interval'. These are three of many examples that suggest that a possible distinction between reliability and robustness is that measures of reliability are commonly probabilities, whereas measures of robustness are commonly measures of insensitivity of a system's performance (e.g., a quantity that is related to the mean or maximum performance drop under a perturbation). 

When one measures the reliability of a model or a system, it is common to specify a criterion for what one considers to be an acceptable system performance, such as a  minimum growth rate for a bacterial colony or a plant or a minimum number of households that receive power through a power grid. One then considers a set of conditions and associates the reliability of the system with the probability that the criterion is satisfied under a randomly\footnote{For example, perhaps all conditions occur with equal probability; one then samples uniformly at random from the set of conditions. Alternatively, perhaps some conditions are more likely than others; one then samples conditions in a biased way.} selected condition from the set of conditions. 
In other words, reliability is a measure of \textit{how likely} it is that system performance declines beyond a stated minimum performance, whereas robustness is a measure of the insensitivity of system performance and thus commonly includes some information about \textit{how much} a system's performance changes from a perturbation. 

In Fig.\,\ref{fig:concepts}~(e), we show a system's performance over time for 11 perturbations of an example system. The red-shaded curves indicate perturbations after which the system is not able to recover a minimum performance level $X_{\textrm{min}}$ before the temporal checkpoint $t_{\textrm{check}}$. The blue-shaded curves indicate perturbations after which the system is able to recover the minimum performance $X_{\textrm{min}}$ level before the temporal checkpoint $t_{\textrm{check}}$. The reliability of the system is a function of the cumulative probability of perturbations that lead to one of the red-shaded curves and/or the cumulative probability of perturbations that lead to one of the blue-shaded curves.

We also note that many researchers have defined reliability and/or robustness in ways that do not align with this distinction \cite{Rausand2004, Woods2015} or have even considered reliability and robustness as synonymous \cite{Perez2018}.

\subsection{Same name, but different concept?}\label{sec:homonyms}

The word `robustness' is used widely in many research areas. In this subsection, we list concepts that researchers have been called `robustness' but differ from the definition of robustness in Section \ref{sec:robustness_def}.

\subsubsection{Performance.}\label{sec:performance_concept}

Measures of performance indicate how well a system achieves a desired function \cite{Townley2003}. Performance is a system property that one can define without specifying a set of perturbations. By contrast, measures of robustness describe the change of a system's performance under perturbations. A definition of robustness thus requires that one specifies --- either explicitly or implicitly --- a set of perturbations. Some researchers have considered robustness to be equivalent to the performance after a given perturbation \cite{Gu2020}.

\subsubsection{Mechanisms of robustness, resilience, and stability.}

In many disciplines, it is of great interest to investigate the mechanisms by which the performance of a system can become robust to some type of perturbation \cite{Kitano2004, Stelling2004}. Researchers have proposed many mechanisms by which systems can achieve robustness or resilience \cite{Edelman2001, Kitano2004, Stelling2004, Masel2009, Araujo2018, Jeynes-Smith2024}. In some studies, the indicators and requirements for such mechanisms --- such as measures of network redundancy \cite{Krakauer2002, Little2003, Kitano2004, Stelling2004, Wagner2005} or the modular structure of networks \cite{Simon1991, Kitano2004, Stelling2004, Ma2006, Wang2012} --- have been used as synonyms of robustness. For example, in several disciplines, researchers have proposed that a system can become robust by having several structurally different components that fulfill the same function within a system \cite{Lawton1994, Tononi1999, Wagner2005, Macia2008, Whitacre2010, Whitacre2010a}. Different researchers have referred to this property of a system's or network's structure as `functional redundancy' \cite{Lawton1994, Rosenfeld2002, Loreau2004}, `degeneracy' \cite{Tononi1999, Whitacre2010}, or `distributed robustness' \cite{Wagner2005, Macia2008, Whitacre2010a}. 

Many researchers have examined the ability of dynamical systems return to a state after a perturbation \cite{Golubitsky2009, Ma2009, Ferrell2016, Reed2017, Antoneli2018, Araujo2018, Curto2019, golubitsky2023, Krakovska2024, Jeynes-Smith2024}. Holling proposed `adaptive cycles' of population growth, conservation, collapse, and reorganization as a mechanism of resilience in ecological systems \cite{Holling1985, Burkhard2011}. Several researchers have associated resilience of biochemical systems to  `homeostatic mechanisms` \cite{Reed2017, Antoneli2018} or `robust adaptation' \cite{Ma2009, Ferrell2016, Araujo2018, Jeynes-Smith2024}. Curto et al.~identified `domination' as an important concept for identifying stable states in coupled dynamical systems that model neural networks \cite{Curto2019}. In many of these studies, the authors identified short cycles (see Section \ref{sec:rev:walks}) and non-overlapping paths (see Section \ref{sec:paths}) that connect the same point of origin to the same destination as structural network properties that are either advantageous or necessary for the proposed mechanism to take effect in a coupled dynamical system\cite{Ma2009, Ferrell2016, Reed2017, Antoneli2018, Araujo2018, Curto2019, Jeynes-Smith2024}. Across many fields, researchers have proposed connections between these structural network properties and network robustness \cite{Wagner2005, Randles2011, Braunstein2016, Radicchi2017, Ozel2018, Xu2021, Yazdani2011}.

\subsubsection{Insensitivity of a system to parameter changes.}

Stelling et al.~\cite{Stelling2004} defined robustness as the `persistence of a system's characteristic behavior under perturbations or conditions of uncertainty'. This definition includes the robustness definition in Section \ref{sec:robustness_def}. However, it also includes the insensitivity of system properties with respect to parameter uncertainty. Robustness (i.e., insensitivity to perturbations) and persistence (i.e., insensitivity to parameter uncertainty) are different concepts. In studies of robustness, one compares the state of a system before and after a perturbation. This requires that one chooses an initial (i.e., unperturbed) system state. The study of persistence to parameter uncertainty does not necessarily require one to choose  an initial state. Instead, it is common to consider some distribution of parameter values. Measures of persistence to parameter uncertainty are then functions of the corresponding performance distribution \cite{Stelling2004, Walsh2013}.

In Fig.\,\ref{fig:concepts} (d), we illustrate the concept of persistence by showing the system performance for different values of a system parameter $r$ for three example systems. System 1 is more persistent than System 2, which in turn is more persistent than System 3.

\subsection{Further reviews of robustness and related concepts}\label{sec:more_reviews}

The study of robustness and resilience is very active in many disciplines, and we cover only a small portion of the relevant literature in our review, whose aim is to overview and present a framework to compare studies of the robustness of networks. Many other researchers have reviewed the robustness and resilience of both network models \cite{Gao2015, Catanese2016, Williams2016, Caccioli2018, Oehlers2021, Freitas2022, Lou2023, Artime2024} and other models \cite{Kitano2004, Stelling2004, Kitano2007, Walsh2013, Urruty2016, Keller2018, Krakovska2024} in a variety of research areas. In this subsection, we briefly overview some of these reviews and their respective focus areas.

Many researchers have studied the robustness of systems in biology and biomedicine \cite{Kitano2004, Kitano2004a, Stelling2004, Kitano2007, Kitano2007a, Wagner2007, Larhlimi2011}.
Kitano~\cite{Kitano2004} examined robustness in biological systems and discussed several mechanisms that can lead to robustness in them. He later used the expectation of the impact of a perturbation \cite{Kitano2007} as a measure to help classify systems as robust or vulnerable. He also considered cancers to be robust biological systems \cite{Kitano2004a} and argued that an improved understanding of the origins of the robustness of cancers can help in the design of cancer drugs \cite{Kitano2007a}. Stelling et al.~\cite{Stelling2004}  reviewed the robustness and persistence of cellular processes. Wagner~\cite{Wagner2007} reviewed the robustness of biological systems to genetic perturbations. Larhlimi et al.~\cite{Larhlimi2011} surveyed several definitions of robustness in the study of network models of metabolic systems.

Robustness and resilience are also important concepts in social systems \cite{Tobin1999, Cutter2008, Magis2010, Berkes2013, Plough2013, Agreste2016, Catanese2016}. In the social sciences, many researchers have investigated the resilience of human communities \cite{Tobin1999, Cutter2008, Magis2010, Berkes2013}. Plough et al.~\cite{Plough2013} reviewed approaches for improving the resilience of human communities to, for example, natural disasters. Catanese et al.~\cite{Catanese2016} reviewed studies of the resilience of crime networks. In the networks that they studied, nodes represent criminals and edges encode some interaction between two criminals (e.g., a phone call or a crime that they committed together). Many of the studies that they discussed focused on efficient strategies to disrupt crime networks as much as possible. Agreste et al.~\cite{Agreste2016}examined the robustness of social networks of Sicilian Mafia organizations. Their article also includes a brief review of the role of social ties in criminal organizations.

It is also relevant to study the robustness of systems of economic ties between people, businesses, banks, and other entities \cite{Bhamra2011, Tukamuhabwa2015, Hosseini2016, Kamalahmadi2016, Caccioli2018, Monostori2018}. For example, supply chains are systems of economic ties between producers, distributors, and consumers of goods \cite{Tang2016}. Many researchers have studied the robustness and resilience of supply systems to various perturbations \cite{Bhamra2011, Tukamuhabwa2015, Hosseini2016, Kamalahmadi2016, Monostori2018}. Financial markets provide another important family of examples of systems of economic ties. Caccioli et al.~\cite{Caccioli2018} reviewed studies of robustness of network models of financial markets to cascading failures that arise from debtor default or other perturbations. In studies of the effects of perturbations on financial systems, it is very common to consider concepts such as `systemic risk' and `contagion risk', which relate to the likelihood of failure propagation in a system \cite{Helbing2013}. Several researchers have reviewed studies of systemic risk \cite{deBandt2000, Eisenberg2001, Haldane2011, Bisias2012, Altar2015, Benoit2017} and contagion risk \cite{Hasman2013, Pei2017}. Neveu~ \cite{Neveu2018} surveyed network-based approaches to measuring systemic risk.

Robustness and resilience are also important in infrastructure systems \cite{Gay2013, Cuadra2015, Huang2015, Urruty2016, Williams2016}. Gay and Sinha~\cite{Gay2013} reviewed system resilience in civil infrastructure systems. Cuadra et al.~\cite{Cuadra2015} reviewed models of network robustness, resilience, and reliability in power grids. Their article includes a critical discussion of the potential of network analysis to provide insights into the robustness of real power grids. Williams and Musolesi~\cite{Williams2016} examined the robustness of spatiotemporal network models of urban transportation systems. Urruty et al.~\cite{Urruty2016} proposed a disambiguation of the concepts of stability, robustness, vulnerability, and resilience in  agricultural systems and reviewed various studies of these concepts in agronomy. Huang et al.~\cite{Huang2015}
reviewed studies of network robustness and resilience in wireless sensor networks. With a focus on studies of the internet, Oehlers and Fabian~\cite{Oehlers2021} reviewed structural properties of networks that researchers have used in studies of network robustness.

Other areas of research in which scholars have reviewed notions of robustness and resilience include the psychological sciences \cite{Walsh2013}, ergonomic design \cite{Hoffman2017}, software design \cite{Shahrokni2013}, and music perception \cite{Keller2018}.

Some scholars have reviewed robustness from a theoretical perspective and focused on a specific type of model. Cohen and Havlin explored the robustness of simple networks \cite{Cohen2010}.
Artime et al.~reviewed connections between network robustness and several percolation processes \cite{Artime2024}. Gao et al.~\cite{Gao2015} reviewed studies of robustness of `networks of networks'. Their review focused on the robustness of random-graph models, but they also discussed networks of networks (see Section \ref{sec:models}) that model infrastructure systems (e.g., streets in cities). Williams and Musolesi~\cite{Williams2016} reviewed spatiotemporal network models (see Section \ref{sec:models}) of urban traffic systems. Shekhtman et al.~\cite{Shekhtman2018} reviewed network robustness in spatially embedded infrastructure networks and networks of networks.

Several recent reviews of network robustness include comparisons of various robustness specifications and discuss aspects to consider as one specifies a robustness problem \cite{Schaeffer2021, Freitas2022}. In their mini-review of network robustness measures, Schaeffer et al.~propose to give consideration to the computational complexity of algorithms that contribute to the runtime of computational perturbation experiments because runtimes that increase greatly with network size can make the study of some robustness measures on very large networks infeasible \cite{Schaeffer2021}. In their survey of robustness specifications, Freitas et al.~emphasize that the selection of robustness measures and other specifications of a robustness problem should be specific to the one's research question \cite{Freitas2022}. They include hypothetical case studies to demonstrate their approach and conclude that one should consider more than one robustness measure when assessing whether a change to a network's structure is likely to increase or decrease its robustness.

\section{Simple networks, subnetworks, and matrix representations of networks}\label{sec:review:graphs}

\begin{figure}[t]
\centering
\includegraphics[trim={0.5cm 7cm 0.3cm 0cm},clip,width=1\textwidth]{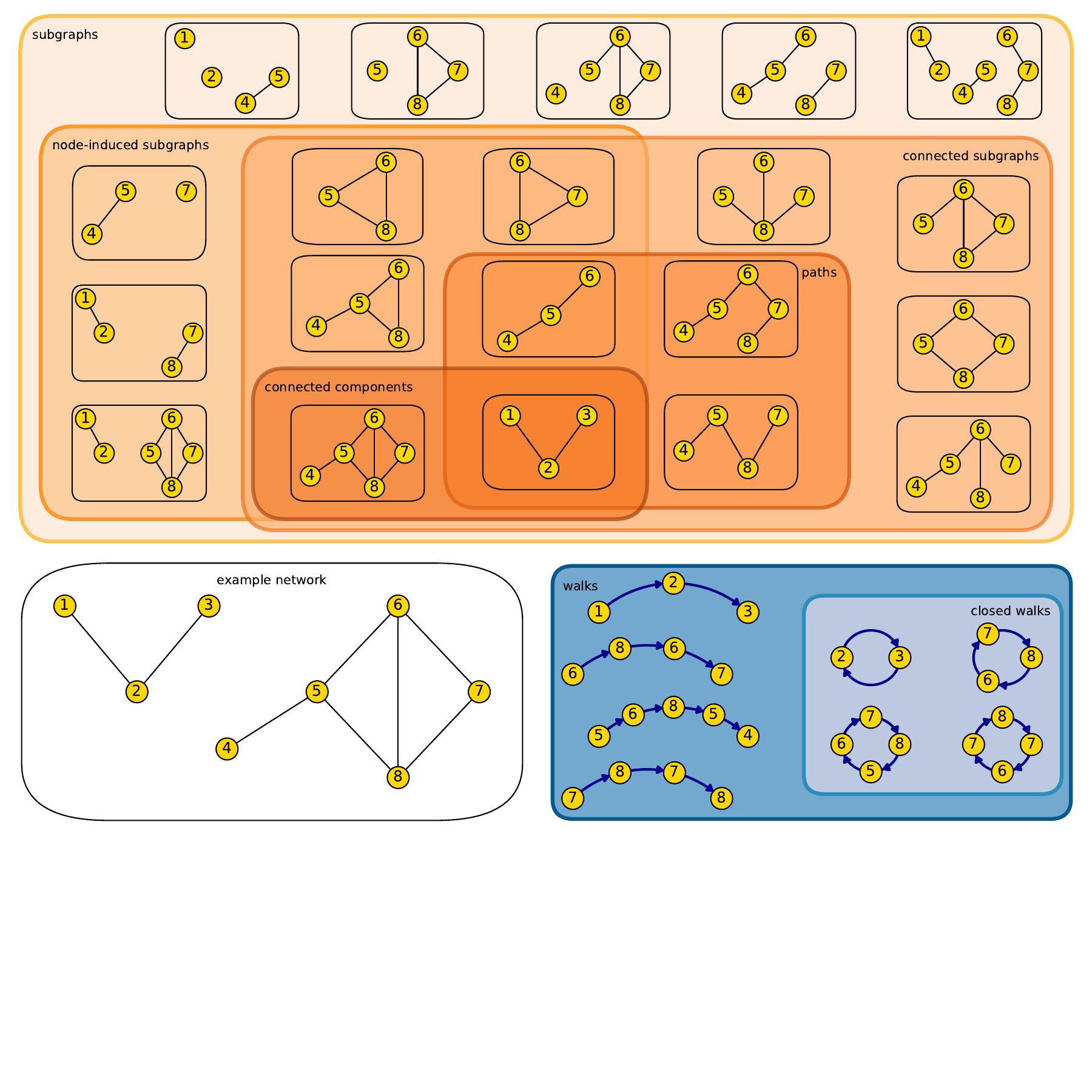}  
\caption{{\bf Subgraphs and walks on a network.} We show an example network with $N = 8$ nodes and $m = 8$ edges in a black box in the bottom-left corner. In the top half of the figure, we show several examples of subgraphs, connected subgraphs, node-induced subgraphs, paths, and connected components of the example graph. In the bottom-left corner, we show several examples of walks and closed walks on the example network. 
}
\label{fig:graphs}
\end{figure}

In this section, we define several graph-theoretic concepts that are useful for the study of networks and their robustness to structural perturbations. We illustrate several of these concepts in Fig.\,\ref{fig:graphs}; we will refer to this figure throughout the section. Helpful textbooks on network analysis that give more details about fundamental graph-theoretic concepts include \cite{Newman2018} and \cite{bullo2022}.

\subsection{Networks and simple networks}\label{sec:rev:simple_networks}

A \textit{network} $G=(V,E)$ is an object that is defined by a set $V = \{v_i\}_i$ of nodes (i.e., vertices) and a set $E = \{e_{i,j}\}_{i,j}$ of edges. In a \textit{directed} network, edges are directed, so the order of the indices $i$ and $j$ matters. In our paper, we use $e_{i,j}$ to denote an edge from node $v_i$ to node $v_j$ in a directed network. In an \textit{undirected} network, the order of the indices $i$ and $j$ does not matter, so $e_{i,j}$ and $e_{j,i}$ denote the same edge. The edge $e_{i,i}$ from a node to itself is a \textit{self-edge}. One can construct network models $G=(V,E,\ldots)$ that include additional information about a system by, for example, including a set of node labels, a set of edge weights, or other elaborations. In Section \ref{sec:models}, we briefly discuss examples of such models. We refer to an undirected network $G$ with distinct edges $e_{i,j}$ and no self-edges as a \textit{simple} network.

We denote the \textit{size} (i.e., the number of elements) of a set $S$ by $|S|$. For a network $G=(V,E)$, we write $N := |V|$ for the number of nodes of a network and $m := |E|$ for the number of edges of a network. We refer to $N$ as the \textit{size of a network}.

In the lower-left box of Fig.\,\ref{fig:graphs}, we show an example of an undirected network with node labels and no self-edges. This network has $N = 8$ nodes and $m = 8$ edges.

For a simple network $G(V,E)$, one can construct a corresponding \textit{line graph}, which is a network $\Lambda(G)$ in which each node corresponds to an edge in $G$ \cite[p.\,93]{Bollobas2013}. Two nodes in the line graph are connected by an edge if the two corresponding edges in $G$ are incident to the same node.

\subsection{Neighbors and neighborhoods.}\label{sec:rev:neighbors}

In an undirected network, a node $v_j$ is a \textrm{neighbor} of node $v_i$ in a network $G=(V,E)$ if the edge $e_{i,j}\in E$. The \textit{neighborhood} of $v_i$ is the set of neighbors of $v_i$. The \textit{degree} $k_i$ of a node $v_i$ is the number of edges that are attached $v_i$ \cite[p.\,9]{Newman2018}. In a simple network, the degree of a node is equal to the number of nodes in its neighborhood.

In a directed network, the node $v_j$ is an \textit{out-neighbor} of $v_i$ if the network's edge set includes a directed edge from node $v_i$ to node $v_j$ (i.e., $e_{i,j}\in E$) \cite[p.\,58]{VanSteen2010}. Node $v_j$ is an \textit{in-neighbor} of $v_i$ if the network's edge set includes a directed edge from node $v_j$ to node $v_i$ (i.e., $e_{j,i}\in E$) \cite[p.\,58]{VanSteen2010}. 
The \textit{in-degree} $k_i^{\textrm{in}}$ of node $v_i$ is the number of edges that end in $v_i$ \cite[p.\,130]{Newman2018}. The nodes \textit{out-degree} $k_i^{\textrm{out}}$ of $v_i$ is the number of edges that start in $v_i$ \cite[p.\,130]{Newman2018}.

\subsection{Subgraphs}\label{sec:rev:subnetworks}

A \textit{subgraph} (i.e., \textit{subnetwork}) of a network $G=(V,E)$ is a network $G_S=(V_S,E_S)$ with node set $V_S \subseteq V$ and edge set $E_S\subseteq E$ \cite{Bollobas2013}. In the remainder of this subsection, we consider special types of subgraphs. The large yellow box in the top half of Fig.\,\ref{fig:graphs} includes various examples of subgraphs of the example network in the lower-left box.

\subsubsection{Paths.}\label{sec:paths}

A \textit{path} is a subgraph $(V_P,E_P)$ with node set $V_P=\{v_i\}_{i=1,\ldots,{\ell+1}}\subseteq V$ and edge set $E_P=\{e_{i,i+1}\}_{i=1,\ldots,\ell}$, where an edge $e_{i,j}$ starts at node $v_i$ and ends at node $v_j$ and $v_{1},\ldots,v_{{\ell+1}}$ are distinct nodes \cite[p.\,4]{Bollobas2013}.\footnote{In directed networks, one can distinguish between \textit{directed paths} and \textit{undirected paths}. The definition of a directed path in a directed network is equivalent to the definition of a path in an undirected network. An undirected path in a directed network is a subgraph $(V_P,E_P)$ with $V_P=\{v_i\}_{i=1,\ldots,{\ell+1}}$ such that for each pair of nodes $v_i,v_{i+1} \in V_P$, the edge set $E_P$ includes either $e_{i,i+1}$ or $e_{i+1,i}$.} We say that nodes $v_i,v_j\in V$ are \textit{connected by a length-$\ell$ path} if there exists a path with $v_1 = v_{i}$ and $v_{{\ell+1}} = v_j$. 

An orange box in the top half of Fig.\,\ref{fig:graphs} includes four examples of paths that are subgraphs of the example network in the lower-left box.

The $\ell$-\textit{hop neighborhood} of a node $v_i \in V$ is the node set that consists of all nodes $v_j \in V$ that are connected to $v_i$ by a length-$\ell$ path \cite[p.\,4]{Bollobas2013}.

\subsubsection{Node-induced subgraphs.}

A \textit{node-induced subgraph} (or simply an \textit{induced subgraph}) of a graph $G=(V,E)$ on a node set $V_S$ is a subgraph $G_S=(V_S,E_S)$ of $G$, where $E_S$ includes all edges in $E$ that connect pairs of nodes in $V_S$ \cite[p.\,5]{Thulasiraman2011}.
A light orange box in the top half of Fig.\,\ref{fig:graphs} includes nine examples of node-induced subgraphs of the example network in the lower-left box.

\subsubsection{Connected subgraphs and connected components.}

In an undirected network $G=(V, E)$, a \textit{connected} subgraph $G_S=(V_S,E_S)$ is a subgraph such that for each pair of nodes $v_i,v_j \in V_S$, there exists a path from $v_i$ to $v_j$ in $G_S$.
A light orange box in the top half of Fig.\,\ref{fig:graphs} includes twelve examples of connected subgraphs of the example network in the lower-left box.

One can partition a network into one or several connected subgraphs. The coarsest partition into connected subgraphs gives the \textit{connected components} of a network \cite[p.\,54]{Eiselt2000}. The example network in Fig.\,\ref{fig:graphs} has two connected components, which we show in a dark orange box in the top half of Fig.\,\ref{fig:graphs}.

In directed networks, one distinguishes between strongly connected and weakly connected components \cite[p.\,53]{Eiselt2000}.\footnote{A \textit{strongly connected} subgraph of a network is a subgraph $G_S=(V_S,E_S)$ such that for each pair of nodes $v_i,v_j\in V_S$, there exists both a directed path from $v_i$ to $v_j$ and a directed path from $v_j$ to $v_i$ in $(V_S,E_S)$ \cite[p.\,43]{Tou2012}. A \textit{weakly connected} subgraph of a network is a subgraph $G_S=(V_S,E_S)$ such that for each pair of nodes of $v_i,v_j\in V_S$, there exists an undirected path between $v_i$ and $v_j$ in $G_S$. A strongly connected subgraph of a network is necessarily also a weakly connected subgraph, but a weakly connected subgraph does not need to also be a strongly connected subgraph.}

A \textit{connected network} has exactly one connected component. A network is \textit{fragmented} if it is not connected. The \textit{size} of a connected component is the number of nodes in it. Many researchers have used measures of performance that are related to the number and sizes of a network's connected components \cite{Albert2000, Shargel2003, Chaverri2010, Bradonjic2011, Bilal2013, Hossain2013, Trajanovski2013, Danziger2014, Priester2014, Ventresca2014, Wang2014a, Zhou2014, Ahmadi2015, Gong2015, LaRocca2015, Radicchi2015, Yang2015, Khansari2016, Ren2016, Schieber2016, Williams2016, DeDomenico2017, Mourier2017, daCunha2017, Phan2018, Kazawa2020, Mimar2022, Ferrari2023, Tomassini2023}.

\subsection{Walks and closed walks}\label{sec:rev:walks}

A \textit{walk} in a network $G=(V,E)$ is a sequence
\begin{align}
	w=(v_{1},e_{1,2},v_{2},e_{2,3},\dots, e_{{\ell-1},\ell},v_{\ell}, e_{\ell, {\ell+1}},v_{{\ell+1}})\nonumber
\end{align}
of nodes $v_{1},v_{2},\ldots,v_{\ell},v_{{\ell+1}}\in V$ and edges $e_{1,2},e_{2,3},\ldots,e_{{\ell-1},\ell},e_{\ell, {\ell+1}}\in E$ such that each edge $e_{i,j}$ starts at node $v_i$ and ends at node $v_j$ \cite[p.\,4]{Bollobas2013}.
A \textit{length-$\ell$ closed walk} is a walk where $v_1$ and $v_{l+1}$ are the same node. Closed walks are also called \textit{cycles} \cite{Newman2018}.
In walks and closed walks, neither the nodes $v_{i}$ nor the edges $e_{i,j}$ need to be distinct \cite[p.\,131]{Newman2018}. Because walks and closed walks can be sequences of non-distinct nodes and edges, they are not subgraphs in general. 

In the blue and orange boxes in the lower right-hand corner of Fig.\,\ref{fig:graphs}, we show eight examples of walks, including four examples of closed walks, that can take place on the example network in the lower-left box.

\subsection{Matrix representations of networks}\label{sec:rev:matrices}

Two ubiquitous matrix representations of networks are the adjacency matrix and the (combinatorial) Laplacian matrix \cite[p.\,142]{Newman2018}. For a network $G=(V,E)$ with $N$ nodes, the \textit{adjacency matrix} is the $N \times N$ matrix $\mathbf A=(a_{i,j})$, where \cite[p.\,110]{Newman2018}
\begin{align}
	a_{i,j} := \left\{\begin{matrix*}[l]
		1\,,&\textrm{ if $e_{j,i}\in E$}\\
		0\,,&\textrm{ otherwise}\,.\\
\end{matrix*}\right.\label{eq:adjacency_matrix}
\end{align}
The (combinatorial) \textit{Laplacian matrix} of $G$ is the $N\times N$ matrix $\mathbf L=(l_{i,j})$, where \cite[p.\,142]{Newman2018}
\begin{align}
	l_{i,j} := \left\{\begin{matrix*}[l]
		k_i\,,&\textrm{ if $i=j$}\\
		-1\,,&\textrm{ if $e_{j,i}\in E$ and $i\neq j$}\\
		0\,,&\textrm{ otherwise}\,,\\
\end{matrix*}\right.\nonumber
\end{align}
where $k_i$ is the degree of node $v_i$. 

One can represent a network with edge weights by a \textit{weighted} adjacency matrix (i.e., a \emph{weight matrix}) or a weighted (combinatorial) Laplacian matrix. In a weighted adjacency matrix, the entry $a_{i,j}$ is the weight of the edge $e_{j,i}$ if $e_{j,i}\in E$ and is $0$ otherwise. In a weighted (combinatorial) Laplacian, the nondiagonal entry $l_{i,j}$ is equal to the negative of the weight of an edge if $e_{j,i}\in E$ and is $0$ otherwise. The diagonal entry $l_{i,i}$ is equal to node $i$'s \textit{weighted degree} (i.e., its \textit{node strength}), which is equal to the sum of the weights of its incident edges.

\section{Specifying a model of a system} \label{sec:models}

In different scientific disciplines, the word `model' can refer to many different things. For example, in experimental biology, it is common to use mice as a \textit{model organism} to study various aspects of mammalian life \cite{Lee1997, Couffinhal1998, Chiu2001, Hsieh-Li2000}. In theoretical studies, models typically refer to mathematical or computational descriptions of real-world systems. Such models can take the form of a set of coupled differential equations \cite{Jacobsen2008}, an agent-based model \cite{Nair2011}, a decision tree \cite{Walsh2013}, and many others. In our review, we focus on theoretical models that include networks. For example, researchers have used road networks to model aspects of city infrastructures \cite{Li2015a} and protein-interaction networks to model cellular processes \cite{Jeong2001, Hakes2008}. 
It is also common to employ models that consist of both a network and additional features, such as edge directions, node or edge weights, node or edge labels, polyadic interactions, spatial structure, temporal structure, multilayer structure, a dynamical process, and coevolution between the network and a dynamical process. For example, popular models of the spread of disease take the form of a dynamical process that occurs on a network \cite{Pastor-Satorras2015}.

\subsection{Model specification matters}\label{sec:model_matters}

For many real-world systems, one can construct many different network models. For example, consider a system of chemical reactions. To model this system, one possibility is to construct a bipartite network in which (1) each node corresponds either to a metabolite or to a reaction and (2) an edge between a metabolite and a reaction indicates that a species takes part in a reaction \cite{Winterbach2011}. However, one can also consider a metabolite-based network in which nodes correspond to metabolites and edges indicate chemical reactions that can transform one species into another. One can also construct a reaction-based network in which nodes correspond to reactions and edges indicate that reactions share products or educts. Each of these three network constructions can yield a suitable model of a system of chemical reactions. However, the study of robustness (and many other properties) of these three different types of networks for the same system of chemical reactions can lead to different qualitative results \cite{Larhlimi2011, Winterbach2011}. In particular, the same structural property of a network can reflect different system properties when using different network constructions. For instance, consider the size $N_{\textrm{LCC}}$ of the largest connected component (LCC; see Section \ref{sec:review:components}). When using a reaction-based network to model a system of chemical reactions, the number $N_{\textrm{LCC}}$ indicates the largest number of connected chemical reactions. By contrast, in a metabolite-based network, $N_{\textrm{LCC}}$ instead indicates the number of metabolites that take part in the largest component of connected reactions. In a bipartite network model of a system of chemical reactions, $N_{\textrm{LCC}}$ is the sum of the number of metabolites and the number of reactions in the largest component of connected reactions. Using the same structural network property as a measure of system performance in different network models of a system can reflect very different notions of system performance. Consequently, studies of a system's robustness that use different network models may lead to very different results, even when all other specifications of the robustness problem are identical.

This discussion of network models of metabolic networks demonstrates that network models of the same system can be very different. It also demonstrates that one can construct different network models of the same system by choosing nodes to represent different entities and/or by choosing edges to represent different types of interactions between entities. Network models can also differ in the types of details that one includes in them. For many real-world systems, an appropriate characterization requires more information than binary data on pairwise interactions between components. Researchers have proposed many network models that can include additional knowledge and data about a system~\cite{Porter2020, Bick2023}.

\subsection{Network models with additional information}\label{sec:networks_plus}

\begin{table}[ht!]
\begin{center}
\begin{tabular}{p{3.75cm} p{3.75cm} p{5cm}}
\toprule
Network model & Application areas & References \\
\midrule
simple network & general systems & \citecell{0.33}{Callaway2000, Ventresca2014, Albert2000,Iyer2013,Beygelzimer2005,Hong2017,Schneider2011} \relvspace\\
& microbiological systems & \citecell{0.33}{Jeong2001,Li2006,Dartnell2005, Wang2018} \relvspace\\
& ecological systems & \citecell{0.33}{Nakayama2019}  \relvspace\\
& social systems & \citecell{0.33}{Dodds2003,Agreste2016}  \relvspace\\
& infrastructure systems & \citecell{0.33}{DeDomenico2017, Cohen2000, 
Wang2014a,Rosas-Casals2007, Cardenas2010, Bilal2013, Gao2017}  \bigrelvspace \\
\midrule
directed network & general systems & \citecell{0.33}{Pu2012,Boguna2005} \relvspace \\
& microbiological systems & \citecell{0.33}{Winterbach2011,Larhlimi2011,Takemoto2012}  \relvspace\\
& ecological systems & \citecell{0.33}{Dunne2002,Dunne2004}  \relvspace\\
& social systems & \citecell{0.33}{Cross2001, Caudell2015, Baggio2016, Zheng2015}  \relvspace\\
& infrastructure systems & \citecell{0.33}{Tang2016,Sun2017,Nair2011,Feng2009}  \bigrelvspace \\
\midrule
network w. node weights & general systems & \citecell{0.33}{Wang2014,Wang2015}  \relvspace\\
& social systems & \citecell{0.33}{Mourier2017}  \relvspace\\
& infrastructure systems & \citecell{0.33}{Galias2017,Liu2017,Chen2017,Nair2011}  \bigrelvspace \\
\midrule
network w. edge weights & general systems & \citecell{0.33}{VanMieghem2005}  \relvspace\\
& microbiological systems & \citecell{0.33}{West2012}  \relvspace\\
& ecological systems & \citecell{0.33}{Vieira2015,Baehner2017,Lance2017,Schleuning2016}  \relvspace\\
& social systems & \citecell{0.33}{Mourier2017, Matakos2020, Gou2017, Freire2011}  \relvspace\\
& infrastructure systems & \citecell{0.33}{Hossain2013,
Sun2017, Phillips2012, Akoglu2012, Li2015a}  \bigrelvspace \\
\midrule
multilayer network & general systems & \citecell{0.33}{Gao2015,Morino2011}  \relvspace\\
& microbiological systems & \citecell{0.33}{Radicchi2015}  \relvspace\\
& ecological systems & \citecell{0.33}{Dattilo2016}  \relvspace\\
& social systems & \citecell{0.33}{Baggio2016}  \relvspace\\
& infrastructure systems & \citecell{0.33}{Zhou2019, Ren2016}  \bigrelvspace \\
\midrule
polyadic interactions & general systems & \citecell{0.33}{Cooley2018, Bianconi2018} \relvspace\\
\midrule
spatial network & general systems & \citecell{0.33}{Shao2015, Berezin2015} \relvspace\\
& microbiological systems & \citecell{0.33}{Alstott2009}  \relvspace\\
& infrastructure systems & \citecell{0.33}{Neumayer2011, Bernstein2014, Deng2019, Dong2018, Bernstein2014}  \bigrelvspace \\
\midrule
temporal network & social systems & \citecell{0.33}{Trajanovski2012, Scellato2011, Sur2015, Sur2016}  \relvspace\\
& infrastructure systems & \citecell{0.33}{Trajanovski2012, Feng2016, Feng2019}  \bigrelvspace \\
\midrule
process on network & general systems & \citecell{0.33}{Wang2014,Wang2015} \relvspace\\
& microbiological systems & \citecell{0.33}{Winterbach2011,Larhlimi2011,Jacobsen2008}  \relvspace\\
& ecological systems & \citecell{0.33}{Dunne2002,Dunne2004,Baehner2017,Dattilo2016,Vieira2015,Lance2017,Schleuning2016,Santamaria2016}  \relvspace\\
& social systems & \citecell{0.33}{Youssef2011}  \relvspace\\
& infrastructure systems & \citecell{0.33}{Galias2017,Liu2017,Chen2017,Nair2011}  \bigrelvspace \\
\bottomrule
\end{tabular}
\end{center}
\vspace{-0.5cm}
\caption{Examples of studies of the robustness of network models in different application areas.}
\label{tab:networks}
\end{table}

There are various ways to incorporate information beyond binary interaction data in network models.
In Table \ref{tab:networks}, we sort various publications about network robustness by application area and network-model features. We include references appear in multiple locations if they include either multiple network models or a single network model that combines two or more of the listed features. The references in Table \ref{tab:networks} are a tiny small sample of the thousands of publications about network robustness across numerous research areas. We consider five application areas: general systems, microbiological systems, ecological systems, social systems, and infrastructure systems. Network models of microbiological systems are models of organisms or cellular systems. In such networks, nodes correspond to cells, genes, proteins, protein complexes, or chemical species that are relevant for metabolism. Ecological networks are models of ecosystems. In such networks, nodes correspond to species of plants, animals, fungi, or unicellular organisms. Edges can indicate various relationships between species (e.g., predator--prey relationships, mutualistic relationships, or competitive relationships). In social networks, edges indicate some type of social relationship or interaction (e.g., friendships, acquaintanceships, collaborations, and economic transactions) and nodes represent people, animals, and other entities that can participate in social interactions. Infrastructure networks are models of human-built systems, such as internet networks, power grids, and traffic systems. In the category `general systems', we list theoretical studies and studies that have considered networks from multiple application areas.

\subsubsection{Directed networks.} 

The inclusion of edge directionality is a very common extension of simple networks \cite{Newman2018}. A directed network can be an appropriate to include in a model of a system if the two participants of a pairwise interaction have different roles. For example, food webs are ecological networks in which an interaction indicates that one species eats another species. Each interaction thus includes one predator species and one prey species, and one can use the direction of an edge to indicate which species is the predator and which species is the prey in the associated relationship \cite{Dunne2002, Dunne2004}. In chemical reaction networks and metabolic networks, one can use directed edges to indicate educts and products of a chemical reaction \cite{Larhlimi2011, Winterbach2011, Takemoto2012}. Many pairwise social interactions are asymmetric, as one participant is a giver and the other participant is a receiver of information, advice, goods, supplies, money, or something else. One can thus use edge directions to incorporate such information in a network model of a system \cite{Cross2001, Caudell2015, Zheng2015, Baggio2016}. In a network model of the mobility of humans, animals, or goods, edge directions can indicate movement directions \cite{Nair2011, Tang2016, Sun2017}. In a road network, one can use directed edges to indicate one-way streets \cite{Feng2009}. 

One can represent a directed network using an asymmetric adjacency matrix (see Section \ref{sec:rev:matrices}). Many (but not all) performance measures (see Section \ref{sec:performance_measures}) and node and edge centrality measures (see Sections \ref{sec:node_centralities} and \ref{sec:edge_centralities}) are functions of a network's adjacency matrix, and one can compute them in the same way for simple networks and directed networks.

\subsubsection{Weighted networks.}

One can include additional information about nodes or edges in a network via node weights or edge weights
In a network with node weights, the weights can encode measurable node properties \cite{Antoniou2008, Phillips2012, Gou2017, Mourier2017, Matakos2020}. Examples include the exposure of persons to conservative or liberal media \cite{Matakos2020}, a person's susceptibility to an infection \cite{Gou2017}, the probability of a shark to be caught \cite{Mourier2017}, and the capacity of a power plant \cite{Phillips2012}. In a network with edge weights, the weights can encode various aspects of interactions, such as speed \cite{Li2015a}, duration \cite{Akoglu2012}, intensity \cite{Freire2011}, or probability \cite{Raimondo2021}.

One can represent a network with edge weights using a weighted adjacency matrix (i.e., a `weight matrix'), whose nonzero elements can take any real scalar value, rather than only $1$. Many (but not all) performance measures (see Section \ref{sec:performance_measures}) and node and edge centrality measures (see Sections \ref{sec:node_centralities} and \ref{sec:edge_centralities}) are functions of a network's adjacency matrix, and one can compute them for simple networks and for networks with positive edge weights in the same way. Many researchers have adapted centrality measures and performance measures that are based on geodesic distance (see Section \ref{sec:review:paths}) for the study of robustness of networks with edge weights \cite{Zhang2013b, Zhang2013b, Bellingeri2018, Zhou2019a, Zhang2021, Qian2022, Zheng2022}. Bellingeri et al.~demonstrated that incorporating edge weights in performance measures and target strategies can substantially change the results of a network's robustness \cite{Bellingeri2018}. For a set of transportation networks, Zhang et al.~reported that, even when measuring performance in a edge-weight-agnostic way (i.e., via a performance measure that does not take edge weights into account), the impact of node removal tends to be larger when one targets nodes using an edge-weight-dependent centrality measure than when one targets nodes using an edge-weight-agnostic centrality measure \cite{Zhang2021}.

For networks with node weights, it is sometimes sensible to define edge weights as the sum, product, or other function of two node weights \cite{Ravichandran2019}. In other examples of the use of node weights in network models of systems, researchers design performance measures and centrality measures that are specific to their system and depend explicitly on node weights \cite{Phillips2012, Gou2017, Mourier2017, Matakos2020}.

\subsubsection{Annotated networks.}\label{sec:anno}

One can include annotations (i.e., labels) for nodes and edges. Annotations allow one to encode different types of entities and relationships (and interactions) \cite{faust1994, Light2020}. For example, node annotations can help distinguish between power producers and power consumers in a power grid \cite{Albert2004}. One can use edge annotations to distinguish between different types of social relationships and interactions (e.g., friendship, communication, trade, and so on) when constructing a network model of a social system~\cite{faust1994}. When edge annotations take numerical values, it is convenient to represent the annotation information using a weighted adjacency matrix. It is also common to use a multilayer-network framework (see Section \ref{sec:multilayer_networks}) to incorporate node or edge annotations in a network model \cite{Kivela2014}.

\subsubsection{Multilayer networks.}\label{sec:multilayer_networks}

When a system includes multiple types of interactions, interactions at multiple different points in time, or interactions in multiple places, a popular approach is to model it as a \textit{multilayer network} \cite{Kivela2014, Porter2018, dedomenico2022, dedomenico2023}. In a multilayer network, each node is associated with one or more layers; edges can either connect nodes in the same layer or connect nodes in different layers. For example, in investigations of interconnected infrastructure systems, researchers have used multilayer networks with various constraints to study system robustness. These efforts include the analysis of {`multiplex networks'} \cite{Baggio2016, Bianconi2016, Hackett2016, Zhao2016, Kazawa2017, Osat2017, Radicchi2017, Kazawa2018, Kazawa2020}, {`multilevel networks'} \cite{Cetinkaya2013, Cetinkaya2015, Lordan2017}, {`coupled networks'} \cite{Schneider2013, Wang2013, Gong2015, Liu2016b, Wang2016}, {`interconnected networks'} \cite{Siami2013, Siami2013a, Tan2013, Siami2014, Zhang2015a, Zhao2015, Garas2016, Xia2016, Zhang2016, Zhu2016, Murakami2017}, {`interdependent networks'} \cite{Buldyrev2010, Huang2011, Gao2012, Zhou2012, Dong2013a, Huang2013a, Parandehgheibi2013, Zhang2013a, Wang2014c, Watanabe2014, Chattopadhyay2015, Shahrivar2015, Tan2015, Ji2016, Liu2016a, Chattopadhyay2017, Chen2017, Rueda2017, Cui2018, Gao2018, Wang2018a, Kazawa2020}, {`networks of networks'} \cite{Dong2013, Bianconi2014, Shekhtman2014, Gao2015, Khoury2015, Danziger2016, Shekhtman2016, Roth2017, Shekhtman2018}, and many other multilayer structures. 

One can represent a multilayer network using an adjacency tensor \cite{Kivela2014}. By projecting the nodes and edges of a multilayer network into a single layer, one obtains a single-layer (i.e., `monolayer') network, for which it is possible to compute performance measures and centrality measures that are defined for simple networks. However, in doing so, one loses the information that the multilayer structure captures. To use the information that is captured by a multilayer structure in a study of network robustness, one can either employ centrality measures and network properties that researchers have defined for multilayer networks \cite{Erlebach2006, deDomenico2013, Klimm2014, Basaras2017, Zhang2017, Estrada2019, Sole-Ribalta2016} or define appropriate new measures. For surveys of research on robustness of multilayer networks, see \cite{Bachmann2020, Mahabadi2023}.

\subsubsection{Networks with polyadic interactions.}

If a model includes interactions between more than two entities, one can encode such interactions as a \textit{hypergraph} (which sometimes is also called a \textit{hypernetwork}) \cite{Ramadan2004, Ghoshal2009, Klamt2009, Zlatic2009, Zhang2010, Young2020} and, under a suitable assumption (specifically, by satisfying a closure condition for which interactions exist), as a simplicial complex \cite{Otter2017}. See \cite{Lambiotte2019, Battiston2020, Torres2020, Battiston2021, Bianconi2021, Battiston2022, Bick2023, Boccaletti2023} for review of such \emph{polyadic} (i.e., \emph{higher-order}) networks. 
 
It is possible to convert a hypergraph $H$ into a simple (or directed or weighted) bipartite network $G=(V,E)$ in which each hyperedge of $H$ corresponds to a node of $G$ and two nodes $v_i$, $v_j$ of the network $G$ are adjacent via an edge $e_{i,j}$ if node $v_i$ corresponds to a node of $H$ and node $v_j$ corresponds to a hyperedge of $H$ that is incident to $i$. One can use performance measures and node centrality measures on the network $G$ to define notions of node importance and edge importance for $H$. However, it can often be important to interpret these measures differently for the simple network and hypergraph descriptions \cite{Benson2019}. 

The study of networks with pairwise interactions has led to development of various concepts and computational tools \cite{Newman2018}. Developing useful variants of these concepts for networks with polyadic interactions can be a challenging endeavor \cite{Torres2020, Battiston2021, Bianconi2021, Battiston2022}. Several researchers have proposed centrality measures that are designed specifically for hypergraphs \cite{Bonacich2004, Kapoor2013, Benson2019}. Cooley et al.~\cite{Cooley2018} proposed a definition of the LCC, which is a popular measure of performance (see Section \ref{sec:LCC}), for hypergraphs. Bianconi et al.~\cite{Bianconi2018} studied the robustness of the largest $k$-connected component\footnote{A \emph{$k$-connected component} of a network $G$ is the largest subgraph of $G$ that includes $k$ distinct paths between every pair of its nodes. In other words, a $k$-connected component remains a connected component (i.e., a 1-connected component) when one removes any set of $k$ edges from it \cite{Diestel2017}[p.\,11].} of simplicial complexes with respect to structural perturbations. 

The study of robustness of networks with polyadic interactions is an active research area \cite{Peng2022, Do2023, Liu2023, Peng2023, Zhao2023, Peng2024, Zhao2024}. Several researchers have used percolation-theoretic approaches to polyadic interaction systems to approach the study of robustness of hypergraphs \cite{Liu2023} and simplicial complexes \cite{Peng2023} to the removal of nodes. These and other researchers have also used computational node-removal experiments to explore the robustness of undirected hypergraphs \cite{Peng2022, Do2023, Peng2024}, directed hypergraphs \cite{Zhao2023}, interdependent undirected hypergraphs \cite{Peng2024} , and interdependent directed hypergraphs \cite{Zhao2024}.

\subsubsection{Spatial networks.}

In a \textit{spatial network}, the nodes and possibly also the edges have associated positions in some space, which can either be a physical space or a latent space \cite{barthelemy2022}. It can be very important to include spatial information in a network model, including for anatomical systems (e.g., the human brain \cite{Alstott2009, Aerts2016}), natural distribution systems (e.g., mycelial systems \cite{Bebber2007, Papadopoulos2018}, vasculature systems \cite{Papadopoulos2018}, and ant colonies \cite{Cook2014}), human-made distribution systems (e.g., power grids \cite{Bernstein2014, Dong2018}), transportation systems (e.g., road systems \cite{Sakakibara2004, Deng2019}), and communication systems (e.g. fiber networks \cite{Neumayer2011, Deng2019}). Other spatial systems involve latent spaces, such that nodes occupy a position in that space and the probability there is an edge between two nodes depends on the distance between them in the latent space~\cite{boguna2021}. For example, for many social networks, one can find suitable latent spaces that capture a portion of a person's attributes --- like their age, interests, values, economic status, cultural background, and geographical location --- because two people who are similar in some of these attributes tend to be more likely to form a social tie than two people who do not have anything in common \cite{Light2020}. Barthelemy~\cite{barthelemy2022} reviewed centrality measures and network properties that depend both on network structure and the spatial positions of the nodes of a network. When studying the robustness of spatial networks, properties that depend on spatial information can be suitable performance measures. For example, centrality measures that depend on spatial data can be useful to design strategies that target or protect a network's nodes (or other subgraphs). For example, researcher have considered {`localized attacks'} \cite{Berezin2015, Shao2015, Vaknin2017, Dong2018, Deng2019} and {`geographically correlated failures'} \cite{Bernstein2014} on spatial networks. In such perturbations, one specifies a region and then removes the nodes and/or edges whose positions lie in that region \cite{Dong2018}. Using localized attacks and geographically correlated failures, researchers can study simplistic models of phenomena such as the effect of lesions in the brain \cite{Alstott2009} and natural disasters and deliberate attacks on power grids \cite{Bernstein2014, Dong2018} and communication networks \cite{Neumayer2011, Deng2019}.

\subsubsection{Temporal networks.}

A \emph{temporal network} is a network with a time-dependent structure~\cite{Holme2012, holme2015, Lambiotte2016, Holme2019}. In a temporal network, the nodes or edges may be present or active at some times but missing or inactive at other times. Time-dependent changes of a network's structure can occur on many different time scales. For example, road networks can change through the construction of new roads and the destruction of existing roads, and these changes can take months or even years. By contrast, some social interactions, such as meetings or phone calls, can be short-lived (perhaps lasting only a few minutes or even only a few seconds). These examples illustrate that the structure of a network of social interactions can change very rapidly. 

Researchers have used a variety of approaches to formalize temporal networks \cite{Holme2012, holme2015, Lambiotte2016, Holme2019, Porter2020}. When information about the presence or activity of nodes is available at discrete times, one can represent a temporal network as multilayer network (see Section \ref{sec:multilayer_networks}) in which each layer corresponds to a point in time \cite{Porter2020}. One can also represent temporal information using annotated networks (see Section \ref{sec:anno}) by annotating each node or edge with a set of time points or time intervals during which it is present or active \cite{Holme2012}. For example, Kempe et al.~\cite{Kempe2002} used an {event-based representation} in which each edge is an event and each event has an associated starting time and duration.

In studies of network robustness, it can be crucial to consider temporal changes in network structure. Some popular measures of performance are based on walks or paths (see Section \ref{sec:performance_measures}). Importantly, some walks and paths that seem to be present when ignoring temporal information about a network's structure are not \textit{time-respecting} walks and paths (i.e., walks and paths that exist in a network when one accounts for temporal information) \cite{Konschake2013, Holme2019}, and they affect walk-based and path-based performance measures. When focusing on time-respecting walks or paths, researchers have extended various centrality measures \cite{Pan2011, Estrada2013, Rocha2014, Praprotnik2015, Lambiotte2016, Takaguchi2016, Taylor2017, Holme2019, Lv2019} and various other network properties \cite{Nicosia2013, Lambiotte2016, Holme2019} to temporal networks. For example, Scellato et al.~\cite{Scellato2011} studied the robustness of a mobile communication network by calculating the `efficiency' (see Section \ref{sec:efficiency}) of time-respecting paths as a performance measure. Several subsequent studies of the robustness of temporal networks have also used this performance measure \cite{Trajanovski2012, Sur2015, Feng2016, Sur2016, Feng2019}.

\subsubsection{Processes on networks.} 

One can include processes on networks in a model by considering dynamical systems or stochastic processes \cite{Porter2016}. Popular examples include compartmental models of the spread of infectious diseases \cite{Pastor-Satorras2015}, agent-based models of the adoption of behaviors or opinions in social groups \cite{noor2020}, Boolean dynamics of gene regulation and other systems \cite{Albert2004a, Schwab2020} ,
and differential equations that describe reaction kinetics \cite{Winterbach2011}, the load distribution in a power grid \cite{Cuadra2015}, predator--prey interactions in ecological systems \cite{Jordan2004}, and many other phenomena. 

Researchers have incorporated dynamical systems in studies of network robustness in a variety of ways. Many researchers have used dynamical systems that are mathematically or computationally tractable to develop dynamics-based centrality measures \cite{Freeman1991, Brandes2005, Newman2005, Konschake2013} for targeted node or edge removal or dynamics-based performance measures \cite{Wu2010, Ellens2011, Schwab2020}. 
See Section \ref{sec:model_specific_performance} for additional discussion.

It can be extremely difficult to mathematically study complicated dynamical systems (which may be high-dimensional and may include many parameters) that aim to be realistic models of real systems, and one may need extensive
 computational resources to simulate. Nevertheless, it is often valuable to 
 study complicated dynamical systems because, for many applications, it is not practical or even impossible to conduct perturbation experiments on real systems. For example, such experiments are infeasible for large infrastructure systems (such as power grids)
 and are ethically unjustifiable (e.g., for food webs and other ecosystems). 
For such applications, researchers often use dynamical-systems theory \cite{May1972, Allesina2012, Medeiros2021} and computational simulations of dynamical systems to study system robustness \cite{Carreras2002, Nedic2006, Carreras2013, Rezaei2014, Henneaux2015, Song2015} or to compare a model with detailed information about processes on networks with less-detailed models \cite{Winterbach2011, Cuadra2015}, which one can either derive (e.g., through mean-field approximations or other methods) or simply write down~\cite{Porter2016}. For example, Winterbach et al.~\cite{Winterbach2011}
used stoichiometric modeling to examine biomass production in a metabolic system. In the study of power grids, many researchers have developed intricate models of
cascading failures \cite{Carreras2002, Nedic2006, Carreras2013, Rezaei2014, Cuadra2015, Henneaux2015, Song2015}.

\subsubsection{Combining network models.}

There are many more possibilities for constructing elaborate network models of real-world systems than we can possibly enumerate in the present review. Our list above is far from complete. Moreover, many combinations of the approaches that we discussed above yield further network models. Examples include `spatiotemporal' networks \cite{Williams2016}, dynamics on spatial networks \cite{Danziger2014}, and combinations of dynamics on networks and dynamics of networks. Models in which dynamics on networks are coupled to dynamics of networks are called `coevolving networks' or `adaptive networks' \cite{Gross2007, Sayama2013, berner2023adaptive, sawicki2023}, and it can be very important to study them.

\subsubsection{Interchangeabeability of network models.}

There are various approaches to constructing a network model when incorporating additional information. For instance, one can categorical data about nodes or edges with labels in a monolayer network \cite{faust1994, Light2020} or layer affiliations in a multilayer network \cite{Kivela2014}. Numerical values associated with nodes or edges can be included as weights or continuous-valued labels. Temporal networks might be modeled as multilayer networks, with each layer representing a time stamp or aggregated time period. Additionally, a hypergraph can be expressed as a bipartite graph with node labels.

Given additional information that one may wish to incorporate, there are often several approaches to constructing a network model. Examples include, but are not limited to, categorical and numerical data associated with nodes or edges, temporal interaction data, and non-dyadic interaction data. One can include categorical information about nodes or edges via node or edge labels in a monolayer network \cite{faust1994, Light2020} or via layer affiliations in a multilayer network \cite{Kivela2014}. For numerical values that are associated with nodes or edges, one commonly uses node weights or edge weights \cite{Antoniou2008} to include this data in a network model, but one can also incorporate this information as numerical node or edge labels \cite{Malhotra2021}. For temporal interaction data, one can choose a temporal network as one's network model, but it is sometimes also possible to represent this information in a multilayer network in which each layer corresponds to a time stamp or aggregated information over some time period \cite{Kivela2014}. One can represent polyadic interaction data using a bipartite network \cite{Asratian1998, Pavlopoulos2018}. A special case of such a network model are Petri nets, which are directed bipartite networks in which one set of nodes represents places and another set of nodes represents transitions \cite{Reisig2012}. Two alternative polyadic-interaction models that have gained popularity are hypergraphs and simplicial complexes \cite{Lambiotte2019, Battiston2020, Torres2020, Battiston2021, Bianconi2021, Battiston2022, Bick2023, Boccaletti2023}. Several scholars have considered directed hypergraphs as an alternative model for applications of Petri nets \cite{Ausiello1986, Restivo2001, Ritz2014}.

\subsection{Popularity and applicability of network models}

In general, one can expect different network models to yield different results when studying a system's robustness. The effect that the choice of a network model can have on the outcome of a study of robustness indicates that researchers need to be very careful when choosing network models. In this subsection, we discuss the tradeoff between simple-network models and network models with additional information. 
In Section \ref{sec:choose_model}, we give practical advice for choosing an appropriate network model.

\subsubsection{Model specification matters for models with additional features.}

In Section \ref{sec:model_matters}, we discussed that the way that one abstracts a system into a simple network (i.e., choosing what aspects of a system are encoded nodes and edges) can greatly affect the results of studies of the system's robustness. For network models with additional information, another choice that can greatly affect the results of a study of system robustness is the choice of what aspects of a system are represented by those additional features. For example, one can choose to use a weighted network to represent a city's road system. Depending on the focus of one's study and the availability of data, it can be reasonable to use road length, mean travel time, mean traffic, age, or other attributes of roads as edge weights. Two edge-weighted network models of the same city's road system can have very different structural properties if the edge weights correspond to different features of roads.

\subsubsection{Strengths and weaknesses of simple-network models.} 

To study system robustness, many researchers have used simple networks as models \cite{Albert2000, Callaway2000, Iyer2013, Ventresca2014} (also see Table \ref{tab:networks}). A benefit of simple networks is that one can build simple-network models from many types of data from many different application areas. This can facilitate comparisons of network robustness across application areas. 
Many researchers have conducted comparative studies of network robustness across domains to uncover common design principles \cite{Sole2002, Demetrius2005, Wang2008, Wang2012a, Shimada2014} that may have evolved independently in different biological systems \cite{Hintze2008, Zhang2016a} and in nature and technology \cite{Iyer2013, Ventresca2014}. However, the characterization of the robustness of some or all of the compared systems often requires analyzing a model that is more detailed than a simple network. Network characteristics that may seem to be related to the robustness of systems across domains can be artifacts of the methods that one uses to construct a simple network \cite{Cantwell2020}. It is important to treat the results of  comparative studies of simple-network models with caution and validate them using network models with additional information when possible.

\subsubsection{Strengths and weaknesses of network models with additional information.}

Whenever one expects an aspect of a system to affect its robustness, it is relevant and often extremely important to incorporate that information into one's model. For example, when one studies the spread of information on a system of mobile communication, including information about the temporal structure of phone calls and text messages (and other things) can be crucial to identify nodes that can communicate messages to a large portion of the network \cite{Scellato2011}. 

In practice, however, it is essentially impossible to construct a model that includes all salient information. The necessary data is often not even available. The availability of insufficient data can often justify the use of a simple-network model. Using network models with additional information can also make it difficult to conduct a meaningful comparative study of robust network structures across domains. Nevertheless, when it is not possible to construct a simple-network model of a system without omitting information that is important for a system's robustness, it is unclear whether or not the results of studies of simple-network models of them (and thus a comparative study of simple-network models across domains) can be meaningful at all. When one aims to fit a network model to data, the inclusion of many forms of additional information in the model can yield an overspecified model (i.e., a model that includes redundant variables \cite{Mohr2021}) which have the potential to overfit patterns in the data.

\section{Specifying a measure of performance}\label{sec:performance_measures}

When studying a robustness problem, a crucial point is the specification of a performance measure. In this section, we discuss challenges in identifying an appropriate performance measure for various scenarios. We also discuss several structural properties of networks that researchers have used as performance measures.

\subsection{Performance measure and system function}

When studying the robustness of a system, a comment question is whether or not it can continue to perform some function after it has experienced some perturbation(s). In a study of system robustness, one thus typically examines model properties that hopefully measure the extent to which a system can perform some function. In this subsection, we discuss several model properties that researchers have employed.

The specification of a performance measure is a crucial part of specifying a robustness problem. In general, the insensitivity of one model property to some perturbation does not imply that some other property is similarly insensitive. Consequently, studies of robustness that use different performance measures can differ considerably in their conclusions \cite{Ventresca2014}.

Whether or not a model property is an appropriate measure of performance depends on the system that one studies and the model that one uses to study it. Moreover, the choice of a performance measure also reflects the system function (or functions) in which a researcher is interested. Again consider a system of chemical reactions. Perhaps one is interested in the system's ability to produce a specific metabolite (e.g., a metabolite, such as adenosine triphosphate (ATP), that is important for a cell's survival) or its ability to degrade a specific metabolite (e.g., a toxin that can hinder the survival of a cell or organism). Alternatively, one may be interested in the system's ability to turn one specific metabolite into another specific metabolite (e.g., the production of ATP from glucose). It is likely that different model properties are appropriate performance measures for these different system functions.

In network models with additional features, one can use the information from these features to help guide choices of appropriate performance measures \cite{LaRocca2015}. Simple-network models do not have such features, so it can be difficult to identify relevant performance measures. For example, consider protein--interaction networks (PINs), which are popular models in the study of cellular processes. These models include only information about protein--protein interactions and exclude, e.g., information about the interactions of proteins with other molecules. It is largely unclear which properties of a PIN's structure can measure the viability (i.e., the ability to survive) of a cell or an organism \cite{Hakes2008}. When it is difficult to identify appropriate performance measures, it is common to examine the robustness of several structural properties of a network model \cite{Winterbach2011, Bilal2013, Ventresca2014, Yamashita2019} that one suspects to be either positively or negatively correlated with system performance.

As in most previous work, we focus on scalar performance measures. However, one can also examine more complicated performance measures. For example, Mohorosi~\cite{Morohosi2010} used the distribution of geodesic shortest-path lengths as a performance measure studied how it changed when removing edges from networks. We expect that such distribution-based and other non-scalar performance measures can be appropriate when a researcher is interested in a system's ability to perform several functions. 

Many researchers have studied the robustness of the output of algorithms for community detection \cite{Karrer2008, Subelj2011, Li2015, Shizuka2016} and node ranking \cite{Costenbader2003, Borgatti2006, Frantz2009, Niu2015, Segarra2015} to changes in network structure. The motivations of studies of algorithm robustness tend to be different than those of studies of system robustness, but studies of node-ranking and community-detection robustness on networks can have a similar character (see Fig.\,\ref{fig:flowchart}) to studies of robustness of non-scalar performance measures.

For some applications, it can be reasonable to suppose that a network $G$ has the uniquely optimal structure for performing a required function (or a set of required functions) and that network structures that are very different from the structure of $G$ are unlikely to perform the required function(s) well. For example, for many biological systems, it is common to assume that evolution produces phenotypes that are optimal or near-optimal for the survival of an organism in a given environment \cite{Kitano2004}. Such applications motivate the use of graph distances to measure the impacts of structural perturbations. Some researchers have used graph distances instead of performance measures to compare network structures before and after structural perturbations \cite{Boldi2011, Schieber2016}. In such studies, one can calculate a graph distance $\Delta(\cdot,\cdot)$ and measure the impact of a perturbation by calculating the distance $\Delta(G',G)$ between a perturbed network $G'$ and the original network $G$. This approach eliminates the need to specify a performance measure. Instead, it requires a researcher to specify a graph distance that is sensitive to the structural changes of interest. See Donnat et al.~\cite{Donnat2018} for a review of the use of graph distances for different applications.

In Table \ref{tab:performance}, we list a variety of performance measures and studies of them. Some studies of network robustness do not clearly distinguish between performance measures and measures of robustness (see Section \ref{sec:performance_concept}). It is thus difficult to partition the set of structural network properties that researchers have studied in the context of network robustness into performance measures and robustness measures. We include a network property as a performance measure in Table \ref{tab:performance} if researchers have studied changes of this structural property under structural perturbations of networks. Table \ref{tab:performance} thus includes some structural properties that researchers have also used as measures of robustness. In Section \ref{sec:measures}, we discuss the use of these structural properties as robustness measures. In the remainder of this subsection, we give definitions of the structural properties in Table \ref{tab:performance} and we discuss their applications as performance measures.

\begin{table}[p!]
\begin{center}
\begin{tabular}{p{6.cm} p{6.5cm}}
\toprule
Performance measure & References \\
\midrule
Number of nodes & \citecell{0.45}{Dunne2002, Dunne2004, Allesina2009, Kaiser-Bunbury2010, Mello2011, Mello2011a, Takemoto2012, Wang2012a, Zhang2013c, Wang2015, Schleuning2016, DeDomenico2017, Chen2017, Sajjad2017, Lance2017, Kim2022} \relvspace\\
Number of edges & \citecell{0.45}{Baggio2016, Sun2017, Wang2013, Wang2014c} \relvspace\\ 
Mean degree & \citecell{0.45}{Sarma2012, Bilal2013, Lordan2014, Colladon2017} \bigrelvspace \\
\midrule
Number of components & \citecell{0.45}{Chaverri2010, Bradonjic2011, Hossain2013, Bilal2013, Zhang2015b} \relvspace\\
Mean component size & \citecell{0.45}{Albert2000, Shargel2003, Chaverri2010} \relvspace\\
Size of the largest connected component & \citecell{0.45}{Albert2000, Cohen2000, Holme2002, Shargel2003, Barrett2004, Bollobas2004, Paul2004, Motter2004, Tanizawa2005, Wang2006, Rosas-Casals2007, Sole2008, Cohen2010, Chaverri2010, Herrmann2011, Nair2011, Zeng2012, Louzada2013, Sha2013, Lordan2014, Wang2014, Wang2014a, Nie2015, Wang2015, Gao2015, Agreste2016, Sun2017, Yamashita2019, Deng2019} \relvspace\\
Relative LCC size & \citecell{0.45}{Cohen2000, Dodds2003, Beygelzimer2005, Estrada2006, Sole2008, Schneider2011, Wu2011, Borge-Holthoefer2012, Zeng2012, Zhou2012, Bilal2013, Hossain2013, Iyer2013, Sha2013, Zhang2013a, Danziger2014, Trajanovski2013, Priester2014, Ventresca2014, Wang2014a, Zhou2014, Ahmadi2015, Gong2015, LaRocca2015, Radicchi2015, Yang2015, Zhang2015b, Khansari2016, Ren2016, Schieber2016, Williams2016, daCunha2017, DeDomenico2017, Mourier2017, Phan2018, Kazawa2020, Mimar2022, Ferrari2023, Tomassini2023} \relvspace\\
Reachability & \citecell{0.45}{Ventresca2014, Hossain2013, Bilal2013, Pinnaka2015, Zhang2013} \bigrelvspace\\
\midrule
Mean local clustering coefficient & \citecell{0.45}{Feyessa2011, Winterbach2011, Lordan2014, Nie2016} \relvspace\\
Global clustering coefficient & \citecell{0.45}{Winterbach2011, Nair2011, Bilal2013} \relvspace\\
Graph entropy & \citecell{0.45}{Ventresca2014, Schieber2016, Baehner2017, Jiang2019}\bigrelvspace \\
\midrule
Network diameter & \citecell{0.45}{Sarma2012, Zhang2013, LaRocca2015, Ren2016, Schieber2016, Colladon2017}
\relvspace\\
Mean shortest-path length & \citecell{0.45}{Albert2000, Holme2002, Shargel2003, Beygelzimer2005, Dartnell2005, Demetrius2005, Li2006, Cardenas2010, Chaverri2010, Bradonjic2011, Nair2011, Bilal2013, Sha2013, Lordan2014, Agreste2016, Ren2016, Colladon2017} \relvspace\\
Efficiency & \citecell{0.45}{Holme2002, Beygelzimer2005, Latora2005, Feyessa2011, Scellato2011, Trajanovski2012, Hossain2013, Trajanovski2013, Ventresca2014, Lordan2014, Wang2014a, LaRocca2015, Nie2015, Feng2016, Peng2016,  Ren2016, Williams2016, Gao2017, Zhou2021, Ferrari2023, Li2023b}\bigrelvspace \\
\midrule
Natural connectivity & \citecell{0.45}{Wu2010, Wu2010a, Bradonjic2011, Zhang2013b, Peng2016, Gao2017, Wang2018, Yamashita2019, Nakayama2019, Deng2019} \relvspace\\
Resistance distance & \citecell{0.45}{Yamashita2019, Ellens2011,Wang2014b, Wang2018}\bigrelvspace \\
\midrule
Model-specific performance measures & \citecell{0.45}{Albert2004, Wilhelm2004, Eubank2004, Scott2006, Erlebach2006, Nagurney2007, Behre2008, Larhlimi2011, Winterbach2011, Nair2011, Youssef2011, Witthaut2012, Doerr2013, LaRocca2015, Baggio2016}\vspace{-0.35cm}\\
\bottomrule
\end{tabular}
\end{center}
\vspace{-0.5cm}
\caption[Performance measures in studies of network robustness.]{Examples of performance measures in studies of network robustness. Horizontal lines separate the performance measures that we discuss in Section \ref{sec:review:numbers}, Section \ref{sec:review:components}, Section \ref{sec:review:triangles}, Section \ref{sec:review:paths}, Section \ref{sec:review:spectral}, and Section \ref{sec:model_specific_performance}.}
\label{tab:performance}
\end{table}

\subsection{Performance measures that are based on the number of nodes or the number of edges}\label{sec:review:numbers}

A basic way to measure the impact of a perturbation on a network is to determine `how much of it is left' after the perturbation. The number of nodes (i.e., the network size) and the number of edges are two straightforward measures of how much of a network is left.

\subsubsection{Number of nodes (i.e., network size).}

When considering the robustness of a simple network's number of nodes, it is trivial to calculate the impact of node-removal perturbations. The removal of $n$ nodes from a network with $N_0$ nodes yields a network with $N' = N_0 - n$ nodes. For node-removal perturbations of a simple network, the number of nodes is not very interesting. However, it is much more interesting to consider dynamics on a network that model successive failures of nodes or edges following the removal of an initial set of nodes or edges. Models that include such dynamics are often used to examine \textit{cascading failures} or \textit{cascading behavior} \cite{Crucitti2004a, Karimi2013, Heath2014, Porter2016, Borge-Holthoefer2013}. For example, it is common to study {trophic cascades} in food webs \cite{Dunne2002, Dunne2004, Vieira2015, Dattilo2016, Baehner2017, Lance2017}. The extinction of one species in an ecosystem can lead to the extinction of additional species that, for example, rely on the first species as prey \cite{Dunne2002}. In studies of power grids, it is common to consider perturbations that can cause power blackouts for entire cities or countries \cite{Cuadra2015}. 
Many researchers have modeled such blackouts as  {cascading failures} in a power grid \cite{Cuadra2015, Chen2017, Galias2017, Liu2017}. In a network model of a power grid, nodes correspond to power generators or consumers and edges correspond to power transmission lines. Each generator node contributes to the overall power generation of a power grid, and each edge contributes to the distribution of power to consumers. If a node or edge fails, the contributions (i.e., `loads') of other nodes and edges tend to increase to meet consumer demand. Such load redistribution can cause the failure of additional nodes or edges, and it can thus trigger a cascading failure. The number of nodes that remain after a cascade (or, alternatively, the `cascade size', which is the number of nodes that `become extinct' or `fail' during a cascade) depends on a network's structure \cite{Cuadra2015}. It is thus a nontrivial measure of performance for network models with cascades.

\subsubsection{Number of edges and mean degree.}

One can also consider the number of edges as a performance measure.
Baggio et al.\:\cite{Baggio2016} proposed `interconnectedness' as a performance measure in a multilayer network of ecological and social relationships between households. In their study, interconnectedness is proportional to the number of interlayer edges of a multilayer network.
Sun et al.~\cite{Sun2017} used the number of edges in an air transportation network as a performance measure. Lordan et al.~\cite{Lordan2014} also studied air transportation networks. They and others \cite{Sarma2012, Colladon2017} have calculated the mean degree, which is the number of edges of a network divided by the number of nodes of a network, as a performance measure.

\subsection{Performance measures that are based on connected components}\label{sec:review:components}

\subsubsection{Number of connected components and mean component size.} 

It is common to assume that the ability of many real-world systems to perform a function requires interactions between its parts. Under this assumption, a necessary (but not sufficient) condition for good system performance is that most nodes of an associated network are part of the network's LCC. The number $n_c$ of components is an example of a performance measure that is likely to be negatively correlated with a system's ability to perform a function, because a large value of $n_c$ indicates that a network consists of many components that do not interact with each other. The mean component size $\overline{N}_c$ is another network property that researchers have studied frequently in the context of network robustness. It is related to $n_c$ by
\begin{align}
	\overline{N}_c = \frac{N}{n_c}\,.\nonumber
\end{align}
Albert et al.~\cite{Albert2000} and Shargel et al.~\cite{Shargel2003} calculated $\overline{N}_c$ as a measure of system performance. Hossain et al.~\cite{Hossain2013} examined $n_c$ under node removal and edge removal. Chaverri~\cite{Chaverri2010} calculated $n_c$ and $\overline{N}_c$ as performance measures in a social network of bats.

\subsubsection{Absolute and relative size of the largest connected component.}\label{sec:LCC}

A very popular measure of system performance is the (absolute) size $N_{\textrm{LCC}}$ of the LCC~\cite{Albert2000, Cohen2000, Holme2002, Shargel2003, Barrett2004, Bollobas2004, Motter2004, Paul2004, Tanizawa2005, Wang2006, Sole2008, Chaverri2010, Cohen2010, Herrmann2011, Nair2011, Zeng2012, Louzada2013, Sha2013, Lordan2014, Wang2014, Wang2014a, Gao2015, Nie2015, Wang2015, Agreste2016, Sun2017, Deng2019, Yamashita2019, Rosas-Casals2007}. The relative size of the LCC is the proportion $\Frag := N_{\textrm{LCC}}/N$ of the number $N_\textrm{LCC}$ of nodes of a network's LCC to the number $N$ of nodes of the network.
The size of the LCC is an integer in $[0,N]$. The relative LCC size is a real number in [0,1], which facilitates comparing relative LCC sizes for networks of different sizes. One can use results from percolation theory \cite{Kesten1982, Sahini1994, Stauffer2018} to calculate expected values of the absolute and relative LCC sizes under several types of node-removal or edge-removal perturbations \cite{Cohen2000, Boguna2005, Cohen2010, Sahimi2023}.

For network models with additional features, one can define many LCC variants that account for such features. For example, in a directed networks the largest weakly connected component and the largest strongly connected component give different performance measures. For multilayer networks, one can define performance measures based on the largest \textit{mutually connected component}, which is a network component in which each node is connected to each in every layer by a path \cite{Cellai2013}. For infrastructure networks in which some nodes are labeled as `critical facilities', Dong et al.~\cite{Dong2019} defined a `robust component' as a component in which each node is connected to at least one node that is a critical facility. The largest robust component of a network can thus include several of the network's connected components. Dong et al.~used the size of the largest robust component as a performance measure of infrastructure systems \cite{Dong2019}.

\subsubsection{Reachability.} 

The \textit{reachability} $r$ of a network is proportional to the number of node pairs that are connected by at least one path \cite{Rai1990, Neumayer2010, Bilal2013}. It is given by
\begin{align}
	r := \frac{1}{2}\binom{N}{2}^{-1}\sum_{\substack{i,j=1\\ j\neq i}}^N \Xi_{i,j}\,,\label{eq:reachability}
\end{align}
where $\Xi_{i,j} = 1$ if there is a path from $v_i$ to $v_j$ and $\Xi_{i,j} = 0$ otherwise.
In an undirected network, one determines a network's reachability from the sizes of its connected components by calculating \cite{Ventresca2014}
\begin{align}
	r = \frac{1}{2}\sum_{i} N_{c_i}(N_{c_i}-1)\,,\nonumber
\end{align}
where $N_{c_i}$ is the size of the $i$th component of a network. Many researchers have studied reachability, but they have used a variety of different names for it (e.g., `pairwise connectivity' \cite{Ventresca2014}, `flow robustness' \cite{Zhang2013, Pinnaka2015, Alenazi2016}, and `average two-terminal reliability' \cite{Neumayer2010, Bilal2013}).

\subsection{Performance measures that are based on node degree and triangles}\label{sec:review:triangles}

In this subsection, we discuss performance measures that are based on properties of nodes and small subgraphs. We consider (1) an entropy measure that is a function of a network's degree distribution and (2) the mean local clustering coefficient and the global clustering coefficient, which measure the frequency of triangles in a network.

\subsubsection{Graph entropy.} 

For a network with degree distribution $p(k)$ and maximum degree $k_{\textrm{max}}$, one can define a notion of graph entropy\footnote{The base $b$ of the logarithm in \eq{first_log} determines the units of entropy. If one chooses $b = 2$, one measures entropy values in bits. If one chooses $b = e$, one measures entropy values in nats.} as \cite{Wang2006}
\begin{align}
	H = -\sum_{k=1}^{k_\textrm{max}}p(k)\log_b(p(k))\,.\label{eq:first_log}
\end{align}
Wang et al.~\cite{Wang2006} used $H$ to measure the degree heterogeneity of a network and demonstrated that it is related to the robustness of a network's LCC to removing nodes uniformly at random. This relationship motivated Ventresca and Aleman~\cite{Ventresca2014} to consider $H$ as a performance measure. B\"ahner et al.~\cite{Baehner2017} examined `interaction evenness', which is proportional to a network's graph entropy, as a measure of performance.  

Several researchers have proposed calculating entropy of various distributions to characterize a network's performance or robustness. For example, Demetrius et al.~\cite{Demetrius2005} calculated a notion of network entropy using the steady-state distribution of a stochastic process 
and argued that such an entropy is a good measure of the robustness of a network. 
Schieber et al.~\cite{Schieber2016} used the entropy of the distribution of geodesic distances (see Section \ref{sec:review:paths}) between nodes as a performance measure. Gao et al.~\cite{Gao2017} proposed a node-centrality measure that they called the `critical degree' and defined a performance measure via the entropy of the distribution of the corresponding centrality values of nodes. Jiang et al.~\cite{Jiang2019} considered graph entropy, the entropy of the distribution of node betweenness-centrality values, and the entropy of the distribution of products of node degree and node betweenness-centrality values as measures of performance. Zingg et al.~\cite{Zingg2019} used random-graph models to describe social organizations and proposed using the entropy of the random-graph model to measure the robustness of a social organization. As these studies illustrate, there are many different distributions --- including degree distribution, a distribution of node or edge centralities, a distribution of distances, the steady-state distribution of a dynamical process, probability distributions from a random-graph model, and many others --- that one can associate with a networked system. These different distributions capture different aspects of these systems, and their entropies do not need to be correlated with each other. For example, a network's degree distribution having a large entropy does not apply that the steady-state distribution of some stochastic process also has a large entropy.

\subsubsection{Global clustering coefficient.}

The \textit{global clustering coefficient} (i.e., \textit{transitivity}) of a network $G$ is \cite{Newman2018}
\begin{align}
	T := 3\times\frac{\textrm{number of triangles in $G$}}{\textrm{number of connected triples in $G$}}\,,\nonumber
\end{align}
where the `number of connected triples' is the number of length-2 paths in $G$ and the `number of triangles' is the number of subgraphs that consist of a length-2 path and an edge that connects the starting node and ending node of that path.

\subsubsection{Mean local clustering coefficient.}

Watts and Strogatz~\cite{Watts1998} defined the \textit{local clustering coefficient} $C_i$ of a node $v_i$ in a network $G=(V,E)$ as 
\begin{align}
	C_i := \left\{\begin{matrix}2\frac{|E^{(1)}_{i}|}{k_i(k_i-1)}\,, & \textrm{if } k_i \geq 2 \\
		0\,, & \textrm{if } k_i < 2\,,\end{matrix}\right.\label{eq:lcc}
\end{align}
where $k_i$ is the degree of node $v_i$ and $|E^{(1)}_{i}|$ is the number of edges in the induced subgraph on the 1-hop neighborhood of $v_i$ (i.e., the number of edges between neighbors of $v_i$). 
For a network with $N$ nodes, the \textit{mean local clustering coefficient} is $C = \sum_{i=1}^NC_i/N$.

The mean local clustering coefficient and the global clustering coefficient are measures of the frequency of triangles of a network, relative to the number of connected triples.
Several researchers have linked the number of triangles of a network to system performance \cite{Ogras2006, Liu2013}. Triangles in a network are related to a notion of `redundancy' \cite[p.\,186]{Newman2018} because a triangle in an undirected network indicates that there are at least two paths that connect each pair of nodes in the triangle. The concept of redundancy is closely related to a system's ability to robustly perform a function \cite{Schreiber2002, Zhou2004, Wagner2005, Meepetchdee2007, Masel2009, Whitacre2010, Randles2011, Yazdani2011}. Additionally, many researchers have proposed that large values of $C$ and $T$ are characteristic of many real-world networks \cite{Watts1998, Newman2003, Fagiolo2007, Radicchi2017}. Based on this proposition, some have argued that $C$ and $T$ can in some scenarios be suitable proxies for a system's ability to fulfill a function \cite{Alon2007, Liu2013}. For directed networks, several scholars have argued that triangular subgraphs, such as the 3-node feedback loop and the 3-node feedforward loop, are components of important functional units in biological systems \cite{Kitano2004, Alon2007, Kwon2008}.

\subsection{Performance measures that are based on geodesic distance}\label{sec:review:paths}

The geodesic distance from a node $v_i$ to a node $v_j$ in a network $G$ is the length of a shortest path from $v_i$ to $v_j$. The geodesic distance thus indicates the minimum number of edges that a message, person, good, or other thing needs to traverse to go from an origin node $v_i$ to a destination node $v_j$.

\subsubsection{Network diameter.}

The \textit{network diameter} $D$ indicates the maximum geodesic distance between two nodes in a network $G=(V,E)$~\cite[p.\,133]{Newman2018}. That is,
\begin{align}
	D := \max_{i,j} \, d_{i,j}\,,\label{eq:diameter}
\end{align}
where $d_{i,j}$ is the geodesic distance from $v_i \in V$ to $v_j \in V$. Fragmented networks have node pairs, $v_i$ and $v_j$, that are not connected by a path. To avoid the difficulty of defining a distance between such nodes, one can define $D$ as the maximum geodesic distance between a a two nodes in the LCC of $G$. Alternatively, one can use the convention that the distance between such nodes is infinite.

\subsubsection{Mean shortest-path length.}

The \textit{mean shortest-path length} (i.e., \textit{mean geodesic distance}) between two nodes of a network is
\begin{align}\label{eq:mspl}
	L := \frac{1}{2}\binom{N_{\textrm{LCC}}}{2}^{-1}\hspace{0.1cm}\sum_{\substack{i,j=1, \\ j\neq i}}^{N_{\textrm{LCC}}} d_{i,j}\,.
\end{align}
To avoid the difficulty of defining a distance between disconnected nodes in a fragmented network, one can consider only nodes in the LCC of a network. In some publications, researchers have referred to the mean shortest-path length \eq{mspl} as the `average shortest-path length', `average path length', and `diameter' \cite{Albert2000, Shargel2003, Li2006}. Unfortunately, using the term `diameter' for the mean shortest-path length is inconsistent with established terminology in network theory, and it can lead to confusion about whether a study considers the performance measure \eq{diameter} (i.e., the actual diameter of a network) or \eq{mspl}.

\subsubsection{Efficiency.}\label{sec:efficiency}

One can define the \textit{efficiency} $\effi$ as the mean of the inverse geodesic distances between the pairs of nodes of a network \cite{Beygelzimer2005} or the node pairs in its LCC \cite{Goni2013}. That is,
\begin{align}
	\effi := \frac{1}{2}\binom{N'}{2}^{-1}\sum_{\substack{i,j=1\\j\neq i}}^{N'} d_{i,j}^{-1}\,,\label{eq:effi}
\end{align}
where $N'$ is either $N$ \cite{Beygelzimer2005} or $N_{\textrm{LCC}}$ \cite{Goni2013}. For $N' = N$, the inverse geodesic distance between nodes in different connected components is 0 \cite{Beygelzimer2005}. A common way to incorporate edge weights in definition of efficiency is to replace the inverse geodesic distances $d_{i,j}^{-1}$ in \eq{effi} with the smallest harmonic mean of edge weights along a path between $i$ and $j$ \cite{Latora2001, Bellingeri2018}.\footnote{Some researchers additionally require the considered paths to be shortest paths in the corresponding unweighted network \cite{Zhang2013b, Zhou2019a, Zhang2021, Qian2022}.}
For temporal networks, it is common to study \textit{temporal efficiency}, which one can compute from \eq{effi} by replacing $d_{i,j}$ with the distances of shortest time-respecting paths \cite{Scellato2011}. In networks with node labels, paths that connect nodes with similar or different labels can contribute differently to a system's performance.
 For example, LaRocca et al.~\cite{LaRocca2015}
proposed a definition of efficiency in power grids that only includes paths from power-producing nodes to power-consuming nodes.

The performance measure in \eq{effi} has been given various names by different researchers. Examples include `average inverse geodesic length' \cite{Holme2002, Hossain2013, Wang2014a, Nie2015}, `network efficiency' \cite{Feyessa2011, LaRocca2015, Zhou2021, Li2023b}, `efficiency of communication' \cite{Beygelzimer2005}, `information transmission efficiency' \cite{Gao2017},  and `global efficiency' \cite{Lordan2014}.

The network diameter, mean shortest-path length, and the efficiency $\effi$ are especially popular measures of performance for infrastructure and communication networks, which have the function of transporting and distribute people, goods, messages, or other things between various sites \cite{Cardenas2010, Nair2011, Hossain2013, Zhang2013, Lordan2014, LaRocca2015, Ren2016, Colladon2017}. In such systems, it is possible that people or algorithms with complete information about a network's structure guide agents or objects to take a shortest path to go from one site to another. For example, when traveling from one location in a city to another, it is likely that people plan their route using GoogleMaps or other route planning services. These services seek either a shortest path (according to distance or some other measure of cost) or, more typically, a very short path that is not necessarily a shortest one. For many other systems, the assumption of shortest-path-based routing seems ill-conceived \cite{Lee2012}. For example, in a metabolic system, metabolites do not choose their reactions and thus do not attempt to choose shortest paths in a metabolic network. Even in systems in which humans or algorithms aim to select a shortest path, it can be impossible to reliably find shortest paths if only incomplete information about a network's structure is available \cite{Kleinberg2000}. For example, in the foundational experiments on the `six degrees of separation' in human networks, Milgram and his collaborators conducted experiments in which they asked people to deliver a parcel to a person by mailing it to one of their acquaintances \cite{Milgram1967, Milgram1969}. (For review of these experiments, see \cite{Newman2018, Baek2021}.) Although the participants of these experiments can potentially seek shortest paths to help deliver their letters to the target individual, one should expect that many letters traversed social networks on paths other than the shortest ones, because the participants did not have complete information about those networks. Erlebach et al.~\cite{Erlebach2006} argued that the assumption of shortest-path-based routing is unrealistic even for the internet because of restrictions that peer-to-peer protocols impose on the paths that data packages can traverse.

\subsection{Performance measures that are based on spectral network properties}\label{sec:review:spectral}

For some directed networks and for all undirected networks, it is possible to diagonalize $\bf A$ and $\bf L$ and thus represent these networks by a set of eigenvalues and a corresponding set of eigenvectors. Spectral graph theory (i.e., the study of eigenvalues and eigenvectors of matrix representations of networks) has uncovered various relationships between a network's structural properties and its eigenvalues and eigenvectors \cite{VanMieghem2023}. Using these relationships, one can construct mathematical functions of eigenvalues and eigenvectors of $\bf A$ and $\bf L$ that can measure system performance or system robustness.

\subsubsection{Natural connectivity.}

In several studies, researchers have used properties that are based on the exponential of a network's adjacency matrix as measures of network functionality \cite{Estrada2005, Estrada2008, Estrada2012a}. 
Exponentials of ${\bf A}$ and ${\bf L}$ are closely linked to linear dynamics on networks \cite{estrada2012review} because the system
\begin{align} \label{eq:linear_system}
	\frac{d{\bf x}}{dt} = {\bf M}{\bf x}
\end{align}
of linear differential equations has the solution ${\bf x}(t) = \exp({\bf M}t){\bf x}_0$ for a diagonalizable matrix ${\bf M}$ and initial conditions ${\bf x}_0$. If ${\bf M} = -{\bf L}$, one can consider \eq{linear_system} as a model of a linear diffusion process on a network \cite{Mirzaev2013}. When ${\bf M} = {\bf A}$, one can consider \eq{linear_system} as a model of a spreading process on a network \cite{Masuda2017}. 
In particular, entry $(i,j)$ of $\exp({\bf A}t)$ is given by a weighted sum of edge-weight products along all length-$\ell$ walks from node $v_i$ to node $v_j$. It is thus related to the number of walks along which a person, object, message, or signal can travel from $v_i$ to $v_j$. Estrada and Hatano~\cite{Estrada2008}interpreted the $(i,j)$th entry of $\exp({\bf A}t)$ as the number of walks along which $v_i$ can `communicate' with $v_j$. They defined the \textit{communicability function}
\begin{align}
	\kappa(i,j) := \left(e^{\mathbf A}\right)_{i,j}\,.\label{eq:commfunc}
\end{align}
Wu et al.~\cite{Wu2010} studied the natural logarithm of the mean of `self-communicability' $\kappa(i,i)$ of a network. They considered the \textit{natural connectivity}
\begin{align}
	K := \ln\left(\sum_{i=1}^N\kappa(i,i)/N\right) = \ln\left(\trace(e^{\bf A})\right) - \ln N\nonumber
\end{align}
as a measure of system performance \cite{Wu2010}. In the last decade, natural connectivity has become a popular measure for the study of network robustness \cite{Wu2010, Wu2010a, Peng2016, Wang2018, Yamashita2019}. Different researchers have proposed slightly different interpretations of natural connectivity and its relation to robustness. Estrada et al.~\cite{Estrada2008} proposed communicability as a measure of the ability of nodes to communicate with each other. Wu et al.~\cite{Wu2010} proposed natural connectivity (see \eq{commfunc}) as a measure of redundancy in networks and argued that natural connectivity is also a good measure of network robustness. 
However, both in their paper \cite{Wu2010} and in subsequent studies of natural connectivity \cite{Shang2011, Zhang2013b, Chan2014}, examinations of changes of a network's natural connectivity under structural perturbations suggest that it can be desirable to preserve a large natural connectivity under structural perturbations of a network. We therefore include natural connectivity in our list of performance measures rather than in our list of robustness measures. 
Natural connectivity can be an appropriate measure of performance when the system function of interest is a system's ability to enable communication between nodes along many different walks. Researchers have considered natural connectivity for applications in areas such as communication systems \cite{Wu2010a, Gao2017}, transportation systems \cite{Xu2019}, and road networks \cite{Deng2019}.

\subsubsection{Resistance distance.}

Interpreting each edge of a network as a resistor in an electrical circuit, one can calculate the \textit{effective resistance}
$\omega_{i,j}$ between any two nodes of the network using Kirchhoff's laws. Klein and Randi{\'c}~\cite{Klein1993} showed that one can use $\omega_{i,j}$ as a distance on a network and thus referred to $\omega_{i,j}$ the \textit{resistance distance} between node $v_i$ and node $v_j$. Devriendt explored the connection between effective resistance and various network properties \cite{Devriendt2022}. Several researchers have considered the mean resistance distance 
\begin{align}
	\Omega:=\frac{1}{2}\binom{N}{2}^{-1}\sum_{\substack{i,j=1\\i\neq j}}^N\omega_{i,j}\,\nonumber
\end{align}
as a measure of system performance or robustness \cite{Ellens2011, Wang2014b, Tizghadam2015, Grunberg2016, Mackin2017, Coletta2019}. 
The link between resistance distance and electric circuits has contributed to the popularity of $\Omega$ as a measure of performance for power grids \cite{Grunberg2016, Mackin2017, Coletta2019}.

One can compute $\Omega$ using the formula~\cite{Mackin2017}
\begin{align}
	\Omega = N\,\textrm{tr}({\bf L}^+)\,,\nonumber
\end{align}
where ${\bf L}^+$ is the Moore--Penrose pseudoinverse \cite{Penrose1955} of ${\bf L}$.

\subsubsection{Spectral network properties and dynamics on networks.}

In our discussion of performance measures that are based on spectral network properties throughout this subsection, we pointed out links between these performance measures and dynamical processes. Spectral network properties are structural properties of a network in the sense that one can compute them from a network's structure. However, much of the motivation for investigating them in studies of performance and robustness is based on models of dynamics on networks. Wang et al.~\cite{Wang2018} use several performance measures that are based on spectral network properties to rank the robustness of a set of networks. Unsurprisingly, they found that different spectral properties lead to different rankings of networks. One likely reason is that the performance measures that they considered are related to different dynamical processes.

\subsection{Model-specific performance measures}\label{sec:model_specific_performance}

In Section \ref{sec:review:spectral}, we discussed that many spectral performance measures are closely linked to dynamical processes on networks. In using these performance measures, researchers are making assumptions  about the movement of people, goods, electricity, information, and other things on a network. In this subsection, we discuss examples of performance measures that depend explicitly on dynamical processes or other aspects of a network model. Calculating these performance measures requires one to explicitly solve dynamical system on a network, either via analytical calculations or numerical simulations.  We also consider examples of performance measures that depend on node labels and multilayer structure.

\subsubsection{Performance measures for dynamics on metabolic networks.}\label{sec:review:specific}

It is common to model a metabolic network as a chemical-reaction network with kinetic reaction fluxes \cite{Larhlimi2011}. Such a network consists of a set of chemical species with concentrations $x_1, x_2, x_3,\ldots$ and a set of reactions with fluxes $\tilde v_1, \tilde v_2, \tilde v_3, \ldots$. The stoichiometric matrix ${\bf S}$ encodes the stoichiometric information  (i.e., the number of molecules of a species that a reaction consumes or produces) for the chemical reactions \cite{Larhlimi2011}. For a closed and well-mixed chemical system, the concentrations of species are related to the reaction fluxes by
\begin{align}
	\frac{d{\bf x}}{dt}={\bf S}\tilde{\bf v}\,,\nonumber
\end{align}
where ${\bf x}:= (x_1, x_2, x_3,\ldots)$ is a \textit{concentration vector} and ${\tilde{\bf v}}:=(\tilde v_1, \tilde v_2, \tilde v_3, \ldots)$ is a \textit{flux vector}. Many researchers have studied steady-state flux vectors (i.e., the vectors $\tilde{\bf v}$ that satisfy ${\bf S}\tilde{\bf v} = 0$) and proposed links between steady-state flux vectors and notions of robustness of metabolic networks~\cite{Wilhelm2004, Behre2008, Winterbach2011}. For example, a popular concept in the study of metabolic-network robustness are \textit{elementary flux modes}, which are minimal\footnote{These sets are `minimal' in the sense that the removal of any reaction from an elementary flux mode yields a set of reactions that is not an elementary flux mode.} subsets of reactions that have a steady-state flux vector \cite{Wilhelm2004, Behre2008}.\footnote{This definition of elementary flux modes applies only to closed systems. Many researchers have used a definition of elementary flux modes that also facilitates the study of elementary flux modes in open systems \cite{Schuster2000, Zanghellini2013}.} By contrast, Winterbach et al.~\cite{Winterbach2011} measured the performance of a metabolic system by its maximum\footnote{Formally, the maximum rate $\mu$ of biomass production is $\max_{\tilde{\bf v}}(\bf{c}^T\tilde{\bf v})$, where one takes the maximum over all steady-state flux vectors $\tilde{\bf v}$. Each element of $\bf c$ encodes the contribution of a `key reaction' to biomass production \cite{Winterbach2011}.} `rate of biomass production', which is a linear combination of the steady-state fluxes of `key reactions' for biomass production.

\subsubsection{Performance measures for dynamics on transportation networks.}

Scott et al.~\cite{Scott2006} considered a congestion measure for studying the robustness of highway networks to edge removal. They used a game-theoretic model \cite{Wardrop1952, Correa2010} for the decision process of drivers as they determine which route to take. This model has a so-called `Wardrop equilibrium' \cite{Correa2010}, at which no driver can improve their travel time by unilaterally changing routes \cite{Scott2006}. Scott et al.~\cite{Scott2006} used the Wardrop equilibrium to assign road-congestion values to highways, and they used these congestion values to compute their performance measure. Ying et al.~considered customer mobility measured by the total mean queue time in a supermarket as a measure of performance and sought network structures that optimize this measure \cite{Ying2023}.

\subsubsection{Performance measures for networks with node labels or multilayer structure.}

Network models can include node labels \cite{Dickinson2003} and/or affiliations of nodes to layers \cite{Kivela2014} (see Section \ref{sec:models}). For many applications, performance measures that use node labels and layer affiliations have clearer links to system performance than performance measures that ignore node labels and layer affiliations. For example, if one supposes that the delivery of electricity from generators to consumers depends on a power grid, then paths from generators to consumers are more relevant for performance than paths between generators. Albert et al.~\cite{Albert2004} proposed to use a version of reachability (see \eq{reachability}) as a performance measure for the North American power grid. This version of reachability includes paths only from generators to distribution substations. Paths from generators to generators and paths from substations to substations do not affect this version of reachability. Erlebach et al.~\cite{Erlebach2006} modeled the internet as a network in which nodes are customers or providers. They suggested viewing the providers as forming an `upper' layer of a two-layer network and consumers as forming a `lower' layer of the network. They argued that common routing policies prevent data packages from taking `valley paths' (i.e., producer--consumer--producer paths) and subsequently proposed a performance measures that is based on the path lengths of `valley-free paths'. 

\subsubsection{Perspectives on a system's function}
The examples that we considered in this subsection demonstrate that one can use node labels, layer affiliations, and aspects of dynamics on networks to construct performance measures with closer links to system performance than performance measures that depend only on a simple network's structure. Additionally, there is often more than one way to incorporate aspects of dynamics on networks when constructing a performance measure. For example, for reaction kinetics on metabolic networks, we discussed two different approaches to construct performance measures; One approach is based on elementary flux modes, and the other is based on the production rates of a set of `key metabolites'. The different approaches reflect different perspectives on a system's function.

\subsection{Popularity and applicability of performance measures}

The myriad performance measures that researchers have considered in studies of network-robustness problems entail several challenges in the further study of such problems. Should one expect different performance measures to lead to similar results in a study of robustness? If not, why are the results different? Which performance measures should one use? In this subsection, we discuss these and further questions. For practical guidance on choosing appropriate performance measures for a study of robustness in networks, see Section \ref{sec:choose_performance}.

\subsubsection{Can one expect different performance measures to lead to similar robustness results?}

In general, the answer is to this question `No!' \cite{Erlebach2006, Alderson2010, Winterbach2011, Ventresca2014, Cuadra2015, Wang2018, Yamashita2019}. Wang et al.~\cite{Wang2018} examined different spectral network properties to assess the robustness of a set of networks. They then ranked networks by their robustness. They found that different spectral network properties yield different robustness rankings of networks. Winterbach et al.~\cite{Winterbach2011} and Yamashita et al.~\cite{Yamashita2019} considered several performance measures and examined correlations between them. Yamashita et al. reported that many, but not all, considered performance measures were positively correlated with each other. Winterbach et al. computed a performance measure from reaction kinetics in a chemical-reaction network. They compared this performance measure to several measures that one can calculate from a network's structure. They reported weak positive correlations between the performance measure that they calculated using reaction kinetics and the performance measures that they calculated from network structure. They concluded that these correlations were too weak to justify using the structure-based performance measures as proxies for the performance of reaction kinetics on a metabolic network.

\subsubsection{Why do some performance measures lead to similar robustness results?}
\label{sec:well-connectedness}

In general, one cannot expect two performance measures to be positively correlated or to lead to similar results in network-robustness studies. However, some performance measures are positively correlated with each other for some sets of networks. For example, Yamashita et al.~\cite{Yamashita2019} found that several widely-used performance measures are positively correlated on a set of networks of various sizes and various densities. A possible explanation of these positive correlations is that many structural network properties that researchers have considered as performance measures are measures of similar things. For example, many structural properties in Table \ref{tab:performance} measure some notion of `connectedness'. Reachability, mean component size, and the absolute and relative sizes of an LCC are positively correlated with the probability that there is some path that connects a pair of nodes that one chooses uniformly at random. The number of components is negatively correlated with such connectedness. Efficiency is related to a notion of `\textit{well}-connectedness'; it is positively correlated with the likelihood that there is a \textit{short} path that connects a pair of nodes that one chooses uniformly at random. The mean shortest-path length of a network is negatively correlated with this notion of well-connectedness. The mean local clustering coefficient and global clustering coefficient are positively correlated with the probability that there are two short paths (specifically, one length-1 path and one length-2 path) that connect a pair of nodes that one chooses uniformly at random. Natural connectivity is positively correlated with the number of closed walks in a network that connect a node to itself.

Measuring connectivity using any of these structural properties tends to increase as one increases the number $m$ of edges in a network. 
In studies of sequential node or edge removal (see Section \ref{sec:sets_and_sequences}), it is common to compare performance measures on networks with different values of $m$. In such studies, positive correlations between two performance measures can result from positive correlations between the performance measures and $m$. For a set of networks with identical values of $m$, Wang et al.~\cite{Wang2018} found that different performance measures lead to different robustness rankings.

\subsubsection{Is connectedness a necessary condition for good performance?}

\begin{figure}[t]
\centering
\includegraphics[trim={0cm 0cm 0cm 0cm},clip,width=1\textwidth]{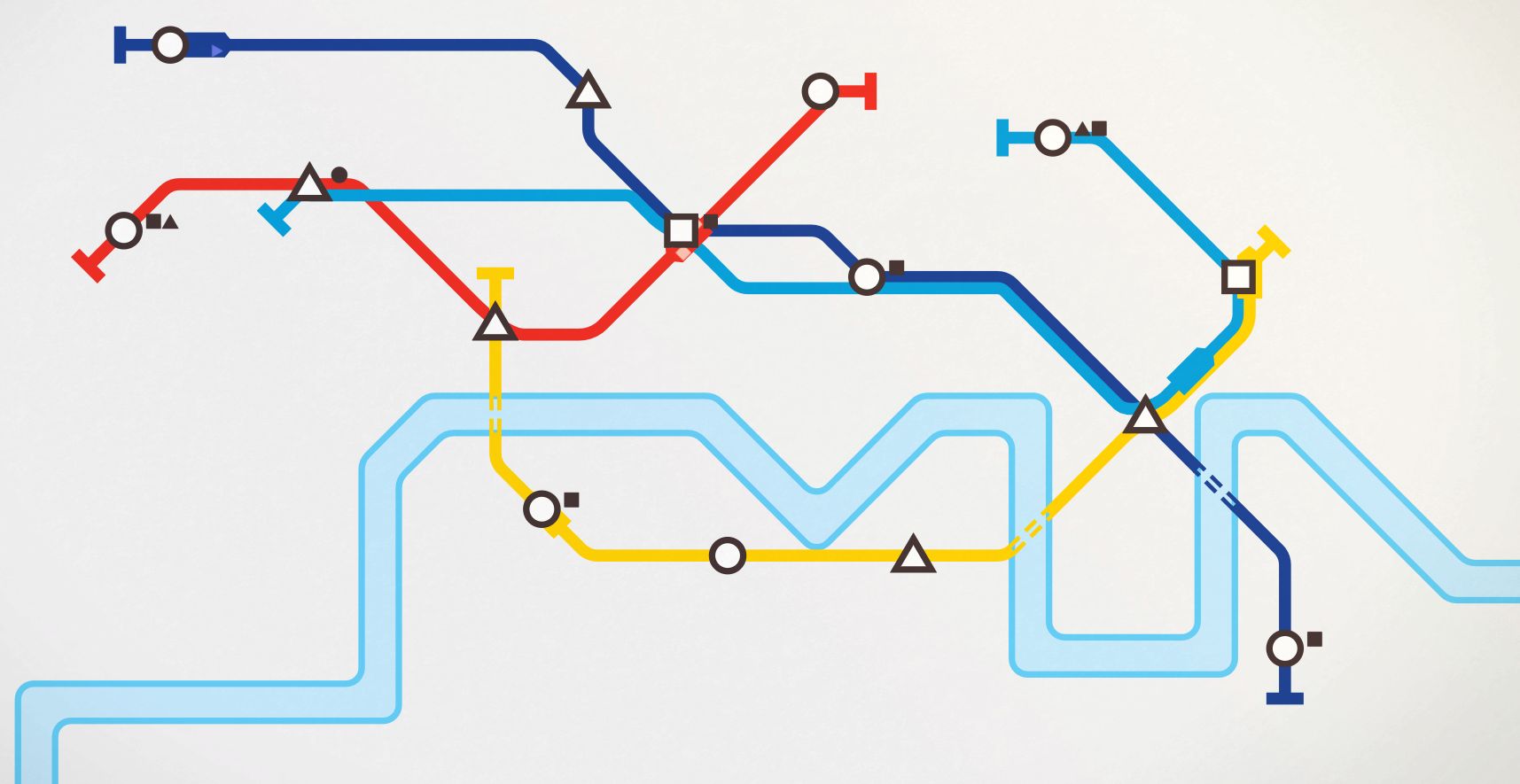}
\caption[A map of a connected transportation network from the video game \textit{Mini Metro}.]{{\bf A map of a connected transportation network from the video game \textit{Mini Metro}.} White symbols with black borders indicate train stations. Small black symbols indicate passengers who are waiting to embark on a train. The shape of a station indicates its type. The shape of a passenger indicates their desired destination type. (Copyright \textcopyright\:2024 Dinosaur Polo Club.  Mini Motorways and Mini Metro are trademarks of Dinosaur Polo Club. All rights reserved.)
} 
\label{fig:mini}
\end{figure}

\begin{figure}[t]
\centering
\includegraphics[trim={0cm 0cm 0cm 0cm},clip,width=1\textwidth]{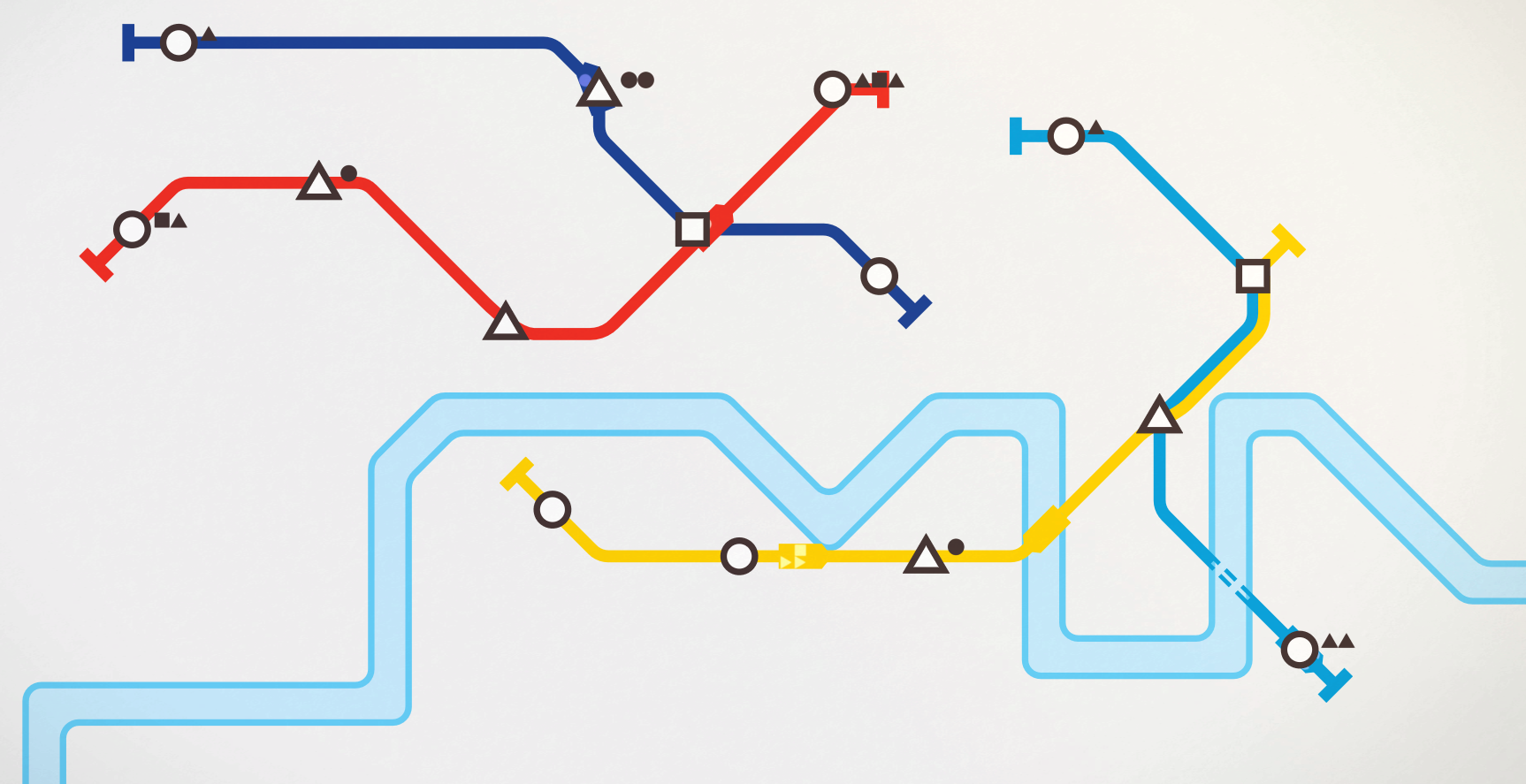}
\caption[A map of a fragmented transportation network from the video game \textit{Mini Metro}.]{{\bf A map of a fragmented transportation network from the video game \textit{Mini Metro}.} 
White symbols with black borders indicate train stations. Small black symbols indicate passengers who are waiting to embark on a train. The shape of a station indicates its type. The shape of a passenger indicates their desired destination type. (Copyright \textcopyright\:2024 Dinosaur Polo Club.  Mini Motorways and Mini Metro are trademarks of Dinosaur Polo Club. All rights reserved.)
} 
\label{fig:mini2}
\end{figure}

The answer to this question is `yes' in some cases but `no' in others. Performance measures that are related to some notion of connectedness are very popular in the study of robustness problems. (See Table \ref{tab:performance} for references.) 
Their popularity reflects the widely-applied assumptions that only a  connected network or only a well-connected network is a well-performing network. Whether or not this is an appropriate assumptions for a given network-robustness problem depends, among other things, on the focus of a particular study. For example, if a relevant function of an urban transportation system is to transport people from any origin to any destination, then the associated network needs to be connected to fulfill this function \cite{Farahani2013}. However, if one is interested in some other function of a transportation system, then connectedness may not be necessary. As a toy example, consider the maps from the video game \textit{Mini Metro} \cite{MiniMetro} in Figs.\,\ref{fig:mini} and \ref{fig:mini2}. In this game, a player's task is to connect stations (white symbols with black borders) on a map by train lines so that all passengers (small black symbols) can travel to their desired destination type, which is designated by a passenger's symbol shape. One can construct a connected network of train lines that fulfills this task (see Fig.\,\ref{fig:mini}). Because each type of symbol occurs at least twice in this example, one can also construct a fragmented network that achieves the same function (see Fig.\,\ref{fig:mini2}). Another example of a system function for which network connectedness is not a performance requirement is the rate of biomass production of a metabolic system.
A well-functioning metabolic network can have several distinct connected components that represent functional units. These functional units can produce one or several biomolecular species without interacting with other functional units \cite{Ma2013}. A metabolic network can also include isolated nodes that do not contribute to its function. In such cases, a large number of disconnected components can suggest that model reduction is necessary, rather than indicating the fragility of the system. Winterbach et al.~\cite{Winterbach2011} studied biomass production rate of a metabolic system and demonstrated that this rate is not strongly correlated with several performance measures that relate to some notion of connectedness.

\subsubsection{Is connectedness a sufficient condition for good performance?}

The answer to this question also differs across studies, but it is `no' in many situations. A network's connectedness from LCC size, relative LCC size, mean component size, and reachability is related to the probability that a pair of nodes that one chooses uniformly at random is connected by a path. 
There are many reasons why such paths can be largely (or even almost entirely) irrelevant in practice. For example, a system can have restrictions on interactions that depend on node labels and edge labels, regardless of whether one includes node labels or edge labels in a model of the system. Erlebach et al.~\cite{Erlebach2006} explained that peer-to-peer protocols impose restrictions on the paths that data packages can traverse on the internet. Whether or not a path is relevant for data transmission on the internet depends on the labeling of a path's nodes as consumers and producers. Paths that do not satisfy a protocol's criteria are irrelevant for the transmission of data packages because they are the `wrong type' of path.

\subsubsection{How is connectedness related to performance?}
When one assumes that all nodes contribute in a similar way to system performance and that all paths are equally relevant to a system function, measures of connectedness that give equal weight to the existence of a path between any pair of nodes tend to be relevant performance measures. When additional information indicates that some parts of a network are more important for maintaining system functions than others, the such measures of connectedness tend to be unsuitable performance measures.
Node labels and edge labels can indicate that (1) some nodes and edges contribute differently to performance than other nodes and edges and that (2) some paths on a network are more relevant than other paths to system functions.
In a \textit{Mini Metro} network, passengers select a node type (instead of an individual node) as their destination. Because a player understands this aspect of the game's dynamics and because each node type occurs at least twice on a map, the player can construct a fragmented transportation network that fulfills the required function. For the example of biomass production in a metabolic system, one assumes that some nodes that correspond to key metabolites are more relevant than other nodes to biomass production. Many studies assume that measures of connectedness are relevant for system performance because they consider networks that do not have node labels or edge labels, either because there exists only one node type and one edge type or because information about node labels and edge labels is not available. In such studies, it is common to assume that all nodes contribute in a similar way to system performance and that all paths are equally relevant to a system function and to consequently consider the LCC size or other performance measures that are related to some notion of connectedness as suitable performance measures.

\section{Specifying a set of perturbations}\label{sec:perturbations}

Formulating a robustness problem requires the specification of a set of perturbations that a system experiences. There are various possibilities for modeling both internal and environmental (i.e., external) perturbations. In a parameter-dependent model (e.g., weighted networks or dynamical processes on networks), one can consider changes in model parameters as perturbations \cite{Stelling2004, Walsh2013}. However, when using networks to model real-world systems, one is often interested in the effects of changes in network structure on a system's ability to perform a specified function. It is possible to consider edge rewirings \cite{Karrer2008} or the addition of nodes or edges \cite{Witthaut2012, Wang2014b} as structural perturbations. In edge-weighted networks, one can study the effect of increasing or decreasing edge weights on a network's robustness \cite{Zhang2013b}. The most widely-studied structural perturbations are node removals and edge removals \cite{Cohen2010, Gao2015}. \footnote{These perturbations correspond to substantial changes to a network's adjacency matrix. The removal of an edge corresponds to changing an element of the adjacency matrix from 1 to 0. The removal of an edge corresponds to removing a row and a column from the adjacency matrix. These structural changes are thus conceptually different from matrix perturbations in a mathematical sense because they cannot be infinitesimally small. The operation that describes the change of an adjacency matrix when an edge is removed is called a `matrix modification' \cite{Higham2003, Liu2003}. (For an introduction to the mathematical framework for studying matrix modifications, sett \cite{Higham2003}.) Our use of the word `perturbation' for the structural changes that we consider in this review is consistent with common terminology in network science and the engineering sciences.} In this section, we consider several approaches to defining sets or sequences of node or edge removals to study network robustness.

\subsection{Sets and sequences of perturbations}\label{sec:sets_and_sequences}

When one is interested in a system's response to a structural perturbation (e.g., the removal of nodes or edges), the research question that guides the study design can take many different forms. For example, one can ask 'What is the average outcome of perturbing a system with a perturbation selected according to some probability distribution from a set of perturbations?' 
Alternatively, one can ask 'What is the most severe (or least severe) outcome of perturbing a system with a perturbation out of a set of perturbations?' Some researchers have focused on the distribution of performances after a perturbation. Others have examined the progression of a system's performance to complete failure via a sequence of perturbations. These different foci of studies of system robustness require different measures of the insensitivity of performance to perturbation (see Section \ref{sec:measures}). These foci also guide how one selects sets or sequences of perturbations to consider.

If one is interested in the average, largest, or smallest impact of the removal of a single node, it is common to consider a set $\{p_v\}_{v\in V}$ of $N$ perturbations, where the perturbation $p_v$ corresponds to the removal of node $v$ \cite{Wilhelm2004, Wang2014, Tang2016}. One then computes the average\footnote{In principle, the average can be a mean that is weighted by the probabilities of perturbations if the perturbations occur with different probabilities. However, we are not aware of a study in which researchers have taken this approach. Studies typically consider a median or an unweighted mean to determine an average.}, maximum, or minimum of the resulting impacts $\{I_v\}_{v\in V}$ (see Section \ref{sec:scalar_robustness}). 
If one is interested in the average, largest, or smallest impact of the removal of a single edge, one considers the set $\{p_e\}_{e\in E}$ of $m$ perturbations and the set of resulting impacts \cite{Latora2005}.

Alternatively, if one is interested in the impact of simultaneously removing several nodes from a network, one can consider a set $\{p_{V_S}\}_{{V_S}\subset V}$ of perturbations, where the perturbation $p_{V_S}$ corresponds to the removal of the subset $V_S \subset V$ of nodes from the network $G=(V,E)$. Similarly, if one is interested in the impact of simultaneously removing several edges from a network, one can consider a set $\{p_{E_S}\}_{{E_S}\subset E}$ of perturbations, where the perturbation $p_{E_S}$ corresponds to the removal of the subset $E_S\subset E$ of edges from the network $G$.) In studies of a network's robustness to simultaneous node removal, it is common to limit comparisons of the impacts of perturbations to the removal of node sets $V_S$ of the same size $n$. For example, when $n = 3$, one considers the removal of three nodes. However, even for small $n$, it can be computationally challenging to investigate the impact of all size-$\binom{N}{n}$ subsets of nodes of a network. In studies in which one is interested in the average impact of perturbations, many researchers have thus sampled $\{I_{V_S}\}_{V_S\subset V}$ \cite{Tang2016} or used techniques from probability theory and statistical mechanics to calculate (or estimate) expected values of the impact of the perturbation $p_{V_S}$ \cite{Cohen2000, Boguna2005, Cohen2010}. 
In studies that consider the maximum impact of perturbations, researchers have used heuristics to identify sets of nodes or edges that, when removed, tend to lead to a particularly large impact on a network's performance. Examples of such heuristic methods include the use of centrality measures (see Sections \ref{sec:node_centralities} and \ref{sec:edge_centralities}), measures of \textit{collective influence} of node sets and edge sets \cite{Morone2015, Lue2016, Morone2016, Pei2019}, and optimal or almost optimal \textit{dismantling sets} \cite{Morone2015, Braunstein2016, daCunha2017, deAbreu2018}. 

Many studies of network robustness examine the progression of declining system performance until a system reaches a complete loss of function(s) \cite{Albert2000, Zeng2012, Hossain2013, Iyer2013, Ventresca2014}. In such studies, one can consider a sequence $(p_{V_1}, p_{V_2}, p_{V_3}, \ldots)$ of node-removal perturbations with progressively more nodes (i.e., $|V_1|<|V_2|<|V_3|< \cdots \leq N$). It is particularly common to consider sequences of node-removal perturbations in which the first set of nodes to remove consists of a single node (i.e., $V_1=\{v_1\}$) and each subsequent node set $V_k$ includes all nodes from the set $V_{k-1}$ along with one additional node \cite{Albert2000, Iyer2013, Ventresca2014}. For such a study, one can construct a sequence of perturbations from a sequence $(v_1, v_2, v_3, \ldots)$ of nodes that one removes from a network in the order of the sequence. 
Albert et al.~\cite{Albert2000}, Iyer et al.~~\cite{Iyer2013}, and Ventresca et al.~\cite{Ventresca2014} (and many other researchers) have taken this approach. They constructed node sequences using both random sampling and using rankings that are based on node centrality measures (see Section \ref{sec:node_centralities}). They then studied the impact of the resulting sequence of perturbations on several performance measures. Hossain et al.~\cite{Hossain2013} used a similar approach to study the progressive decline of system function under sequential edge removal. Zeng et al.~\cite{Zeng2012} compared the progressive decline of performance under sequential node removal, sequential edge removal, and a mixed-attack strategy in which one samples a sequence $(x_1, x_2, x_3, \ldots)$ of entities $x_i$ to remove, where $x_i$ is the largest-degree node with probability $f$ and the largest-geodesic-betweenness edge (see Section \ref{sec:betweenness}) with probability $1 - f$.

\subsection{Random failures versus targeted attacks}

Albert et al.~\cite{Albert2000} wrote arguably the most influential paper on the topic of network robustness. They examined the robustness of the mean component size, the LCC size, and the mean shortest-path length of a variety networks under sequential node removal. They observed that the choice of the sequence $(v_1, v_2, v_3,\ldots)$ crucially affects the robustness of performance measures. When one chooses $v_1, v_2, v_3,\ldots$ uniformly at random from the node set $V$, the change of the performance measures tends to be smaller than when choosing nodes in order from largest degree to smallest degree. In various subsequent studies, researchers have reported similar observations for various types of networks and various strategies for node selection \cite{Motter2002, Herrmann2011, Lordan2014, Ventresca2014} or edge selection \cite{Hossain2013, Lordan2014}. We refer to such a strategy to select nodes or edges for removal as a \textit{targeting strategy}. The most common targeting strategies use \textit{centrality measures}, which measure the importance of a subset of a network. \textit{Node centrality measures} assign a value of  `centrality' (i.e., importance) to each node of a network \cite[p.\,159]{Newman2018}. Similarly, \textit{edge centrality measures} assign a centrality value to each edge of a network. One can rank nodes or edges by their centrality values\footnote{It is common to rank nodes or edges in decreasing order of centrality, but one can also consider rankings of nodes or edges from smallest centrality to largest centrality \cite{Dunne2002, Dunne2004, Dattilo2016}.} and use this ranking to choose a sequence of nodes or edges for removal. Several researchers have pointed out that removing a node or an edge typically changes the centrality values of the remaining nodes and edges \cite{Holme2002, girvan2002}. Holme et al.~distinguished targeting strategies that select nodes or edges based on their initial centrality values from targeting strategies that select nodes or edges based on centrality values that one updates after each node or edge removal \cite{Holme2002}. They proposed to refer to targeting strategies that use updated centrality values as {`dynamical'} targeting strategies.  However, many subsequent studies of network robustness instead referred to such strategies as \textit{adaptive} targeting strategies \cite{Herrmann2011, Zhou2014, Ma2015, Yang2015, Hong2017}. Holme et al.~\cite{Holme2002}
demonstrated that adaptive targeting strategies tend to yield larger changes in performance measures than nonadaptive targeting strategies. Subsequent studies have reported similar results \cite{Holmgren2006, Han2009, Nie2015, Lu2016, Nie2016}.

Other target strategies use additional information about nodes and/or edges to select perturbations. For example, Colladon and Vaggagini considered selecting nodes for removal based on their annotated role in enterprise intranet social networks \cite{Colladon2017}. Berezin et al.~studied robustness and vulnerability of 2D spatially embedded networks and considered `localized attacks' that correspond to removing all nodes within a circular region of the embedding space \cite{Berezin2015}.

\subsection{Node centrality measures}\label{sec:node_centralities}

In Table \ref{tab:node_targets}, we list node centrality measures and studies of network robustness that have used these centrality measures for adaptive or nonadaptive targeting strategies of node removal. In this subsection, define the centrality measures in Table \ref{tab:node_targets}.
 
\begin{table}[t!]
\begin{center}
\begin{tabular}{p{3.2cm} p{9.3cm}}
\toprule
Node centrality measure & References \\
\midrule
Degree & \citecell{0.63}{Albert2000, Jeong2001, Dunne2002, Holme2002, Shargel2003, Dunne2004, Motter2004, Paul2004, Beygelzimer2005, Dartnell2005, Tanizawa2005, Estrada2006, Sole2008, Allesina2009, Kaiser-Bunbury2010, Herrmann2011, Nair2011, Schneider2011, Wu2011, Borge-Holthoefer2012,  Pu2012, Trajanovski2012, Hossain2013, Iyer2013,  Sha2013, Trajanovski2013, Zhang2013, Zhang2013a, Lordan2014, Priester2014, Ventresca2014, Wang2014a, Zhou2014, Ahmadi2015, Gong2015, Nie2015, Pinnaka2015, Yang2015, Agreste2016, Hao2016, Dattilo2016, Khansari2016, Ren2016, Colladon2017, daCunha2017, DeDomenico2017, Mourier2017, Hong2017, Lance2017, Sun2017, Gao2017, Wandelt2018, Deng2019, Jiang2019, Nakayama2019, Zhou2019, Kazawa2020, Zhou2021,Tomassini2023} \\
Geodesic betweenness centrality & \citecell{0.63}{Holme2002, Cardenas2010, Feyessa2011, Borge-Holthoefer2012, Trajanovski2012, Hossain2013, Iyer2013, Trajanovski2013, Zhang2013, Lordan2014, Ventresca2014, Ahmadi2015, Gong2015, Nie2015, Pinnaka2015, Yang2015, Agreste2016, Khansari2016, Williams2016, daCunha2017, Sun2017,  DeDomenico2017, Gao2017, Hong2017, Mourier2017, Wandelt2018, Deng2019, Jiang2019, Zhou2019, Zhou2021} \\
Geodesic closeness\\ centrality & \citecell{0.63}{Motter2004, Feyessa2011, Trajanovski2012, Iyer2013, Trajanovski2013, Zhang2013, Ventresca2014, Ahmadi2015, Pinnaka2015, Yang2015, Zhang2015b, Agreste2016, Khansari2016, Williams2016, Sun2017} \\
Eigenvector centrality & \citecell{0.63}{Feyessa2011, Iyer2013, Trajanovski2013, Zhang2013, Ahmadi2015, Pinnaka2015, Khansari2016, Gao2017, Sun2017} \\
Katz centrality & \citecell{0.63}{Zhang2015b, Khansari2016, Wandelt2018} \\
PageRank centrality & \citecell{0.63}{Ventresca2014, Gong2015, Khansari2016, Wandelt2018} \\
Subgraph centrality & \citecell{0.63}{Khansari2016, Pu2018, Deng2019} \\
Damage centrality & \citecell{0.63}{Latora2005, Lordan2014, Wang2014a, Khansari2016, Sun2017} \vspace{-0.4cm}\\
\bottomrule
\end{tabular}
\end{center}
\vspace{-0.5cm}
\caption[Examples of centrality measures for node removal.]{Examples of node centrality measures that researchers have used as targeting strategies in studies of network robustness.}
\label{tab:node_targets}
\end{table}

\subsubsection{Degree centrality.} 

A very popular centrality measure is \emph{degree centrality} (or simply \emph{degree}) \cite{Albert2000, Jeong2001, Dunne2002, Holme2002, Shargel2003, Dunne2004, Motter2004, Paul2004, Beygelzimer2005, Dartnell2005, Tanizawa2005, Estrada2006, Sole2008, Allesina2009, Herrmann2011, Nair2011, Schneider2011, Wu2011, Pu2012, Trajanovski2012, Hossain2013, Iyer2013, Sha2013, Trajanovski2013, Zhang2013, Zhang2013a, Lordan2014, Priester2014, Ventresca2014, Wang2014a, Zhou2014, Ahmadi2015, Gong2015, Nie2015, Pinnaka2015, Yang2015, Agreste2016, Dattilo2016, Hao2016, Khansari2016, Ren2016, Colladon2017, DeDomenico2017, Gao2017, Hong2017, Lance2017, Mourier2017, Sun2017, daCunha2017, Deng2019, Jiang2019, Nakayama2019, Zhou2019, Kazawa2020, Zhou2021, Tomassini2023, Borge-Holthoefer2012, Kaiser-Bunbury2010}. One ranks nodes by their degree and considers nodes with large degree to be more important than nodes with small degree.

\subsubsection{Betweenness centrality.} \label{sec:betweenness}

The \emph{(geodesic node) betweenness centrality} of a node is 
a measure of how many shortest paths in a network traverse the node. Formally, the betweenness centrality $\textrm{BC}_i$ of a node $v_i$ of a network $G$ is~\cite[p.\,152]{Estrada2015} 
\begin{align}
	\textrm{BC}_i := \sum_{\substack{v_j, v_k\in V\backslash\{i\}\,,\\j\neq k}}\frac{n^{\textrm{SP}}_{i}(j,k)}{n^{\textrm{SP}}(j,k)}\,,\nonumber
\end{align}
where $n^{\textrm{SP}}(j,k)$ the number of shortest paths from $v_j$ to $v_k$ and $n^{\textrm{SP}}_i(j,k)$ is the number of shortest paths from $v_j$ to $v_k$ that traverse $v_i$ \cite{Freeman1977}. One can define other types of betweenness centrality, such as ones that are based on random walks rather than on strictly shortest paths \cite{Newman2005}.

\subsubsection{Closeness centrality.} 

Many researchers have referred to a family of node centrality measures that uses geodesic distances between nodes as \textit{(geodesic) closeness centrality}. Centrality measures of this family use the geodesic distances $d_{i,j}$ between a node $v_i$ and nodes $v_{j}\in V$ to assign a centrality value $\textrm{CC}_i$ to $v_i$. It is very common to define closeness centrality as the inverse of the mean of geodesic distance between a node and all nodes in $V$ \cite[p.\,171]{Newman2018}. That is,
\begin{align}
	\textrm{CC}_i:=\left(\frac{1}{N}\sum_{j=1}^Nd_{i,j}\right)^{-1}\,.\label{eq:ccdef1}
\end{align}
This definition of $\textrm{CC}_i$ has several issues. For example, if a network has more than one connected component, no node of the network is connected to all other nodes. For each node $v_i$, there exists at least one node $v_{j}$ with $j \neq i$ such that there is no path from $v_i$ to $v_j$. If one sets $d_{i,j} = \infty$ for such node pairs, one obtains $\textrm{CC}_i = 0$ for all $v_i \in V$. As pointed out by Newman~\cite{Newman2018},
the alternative definition
\begin{align}
	\textrm{CC}_i:=\frac{1}{N-1}\sum_{j\neq i}^Nd_{i,j}^{-1} \nonumber
\end{align}
avoids this and other issues of \eq{ccdef1} \cite[p.\,172]{Newman2018}. Many studies of network robustness have considered targeting strategies that are based on the closeness centrality in \eq{ccdef1} \cite{Zhang2013, Ahmadi2015, Pinnaka2015, Agreste2016, Sun2017} or on other notions of closeness \cite{Feyessa2011, Iyer2013, Ventresca2014}.

\subsubsection{Eigenvector centrality.}\label{sec:eigenvector_centrality}

One can suppose that a node's importance depends on the importance of its neighboring nodes. Eigenvector centrality is the simplest notion of centrality that captures such a notion of importance. One way to obtain the eigenvector centralities of the nodes of a network employs an illustrative iterative procedure. One sets the initial centrality of each node to $x_i^{(0)} = 1$ and subsequently updates the centrality of each node to a value that is proportional to the sum of the centrality values of its neighbors, That is,
\begin{align}
	x_i^{(1)}=\frac{1}{\lambda_1}\sum_ja_{i,j}x_j^{(0)}\,,\nonumber
\end{align}
where one uses the principal eigenvalue  $\lambda_1$ of the network's adjacency matrix as a normalization constant. Updating the centrality values over and over again leads to an iterative procedure that one can describe formally by writing
\begin{align}
	x_i^{(t+1)} = \frac{1}{\lambda_1}\sum_j a_{i,j}{x_j^{(t)}}\,, \quad
	x_i^{(0)} = 1\,.\label{eq:ev-iteration}
\end{align}
As $t \rightarrow \infty$, the quantity $x_i^{(t)}$ converges to the eigenvector centrality $\textrm{EC}_i$ (where $i \in\{1,\dots,N\}$) of the $i$th node of a network. A more efficient (but arguably less illustrative) way to calculate the eigenvector centralities of a network's nodes uses the principal eigenvector of the network's adjacency matrix \cite[p.\,169]{Newman2018}. In this approach, one writes
\begin{align}
	\textrm{EC}_i = {\bf x}_i\,,\nonumber
\end{align}
where ${\bf x}_i$ is the $i$th entry of the principal eigenvector of the network's adjacency matrix.\footnote{Directed networks have a set of left eigenvectors and a set of right eigenvectors, which can differ from each other. To decide whether the principal left eigenvector or the principal right eigenvector gives a more relevant ranking of nodes, one needs to establish whether importance in a given network spreads from nodes to their out-neighbors or to their in-neighbors \cite[p.\,161--162]{Newman2018}. When using \eq{adjacency_matrix} to define the entries of the adjacency matrix of a directed network, the entries of the adjacency matrix's principal right eigenvector are the eigenvector centralities of the nodes when importance spreads from nodes to their out-neighbors and the entries of the adjacency matrix's principal left eigenvector are the eigenvector centralities of the nodes when importance spreads from nodes to their in-neighbors.} For a review of interpretations of eigenvector centrality, see \cite{Baek2021}.

One can consider eigenvector centrality to be an extension of degree centrality \cite[p.\,330]{Newman2018}. Degree centrality uses the number of neighbors of a node as a measure of the node's importance. The degree centrality of a node is proportional to the values $x_i^{(1)}$ in the first iteration of the iterative procedure in \eq{ev-iteration}. Eigenvector centrality uses a normalized sum of the importances of neighboring nodes as a measure of a node's importance \cite{Bonacich1972}.

\subsubsection{Katz centrality.}\label{sec:Katz}

In directed networks that are not strongly connected, many nodes likely have an eigenvector centrality of 0. Specifically, only nodes in a strongly connected component and nodes that are connected to a strongly connected component via a directed path to those nodes have nonzero eigenvector centralities \cite[p.\,163]{Newman2018}.
To obtain nonzero centrality values for the other nodes of a network, one can adapt the iterative procedure in \eq{ev-iteration} in Section \ref{sec:eigenvector_centrality} by adding a constant term. This yields the iterative procedure
\begin{align}
	x_i^{(t+1)} = 1 + \alpha\sum_j a_{i,j}{x_j^{(t)}}\,, \quad x_i^{(0)} = 1\,.\label{eq:katz}
\end{align}
For $\alpha\in(0,1/\lambda_1)$, the quantity $x_i^{(t)}$ converges to a finite value as $t \rightarrow \infty$. Because Katz~\cite{Katz1953} first proposed to use these limit values as a measure of node centrality, many researchers refer to this centrality measure as \textit{Katz centrality}. (Other researchers have used the names `alpha centrality` \cite{Ghosh2011, Ide2013}, `beta centrality' \cite{Soheili2014, Jennings2019}, and --- referring to the subsequent reinvention of this centrality measure by Bonacich \cite{Bonacich1987} --- `Bonacich centrality' \cite{Mizruchi1981, Ballester2006, Fujiyama2020}, `Bonacich power centrality' \cite{Smith2014, Koufteros2021}, and `Katz--Bonacich centrality' \cite{Matveenko2019, Dasaratha2020, Bloch2023}.)
The Katz centrality $\textrm{KC}_i$ of node $v_i$ is
\begin{align}
	\textrm{KC}_i = {\bf x}^{(\infty)}_i\,,\nonumber
\end{align}
where 
\begin{align}
	{\bf x}^{(\infty)} = ({\bf I}-\alpha{\bf A})^{-1}{\bf 1} \,,\nonumber
\end{align}
with matrix $\bf I$ is the $N\times N$ identity matrix and $\bf 1 = (1,1,\ldots,1)^T$ an $N$-dimensional vector whose entries are all equal to $1$ \cite[p.\,163]{Newman2018}.

\subsubsection{PageRank centrality.}

For some applications, it can be reasonable to assume that nodes with few out-neighbors affect their out-neighbors' importance more strongly than nodes that have many out-neighbors. For example, in a social network of mentor--mentee relationships, a mentor who has many mentees may have less time for one-on-one mentorship meetings with each of their mentees on average than a mentor with few mentees. To give greater weight to the relationships that nodes with a small out-degree have with their out-neighbors than nodes with a large out-degree, one can weight the entries of the adjacency matrix in the iterative procedure in \eq{ev-iteration} in Section \ref{sec:Katz} by the respective node's out-degree $k_j$. One then writes
\begin{align}
	x_i^{(t+1)} = 1+\alpha\sum_j \frac{a_{i,j}}{k_j}{x_j^{(t)}}\,, \quad x_i^{(0)} = 1\,.\label{eq:pagerank}
\end{align}
For  $\alpha\in(0,1/\lambda_1)$, the quantity $x_i^{(t)}$ converges to a finite value as $t \rightarrow \infty$. This finite value is the \textit{PageRank centrality} of the $i$th node of a network \cite{Gleich2015}. The PageRank centrality $\textrm{PRC}_i$ of node $v_i$ is \cite[p.\,166]{Newman2018}
\begin{align}
	\textrm{PRC}_i = {\bf x}^{(\infty)}_i \nonumber
\end{align}
where 
\begin{align}
	{\bf x}^{(\infty)} = ({\bf I}-\alpha{\bf A}{\bf D}^{-1})^{-1}{\bf 1} \,,\nonumber
\end{align}
and $\bf D$ is the degree matrix with elements
\begin{align}
	d_{i,j}:= \left\{\begin{matrix*}[l]
		k_i\,,&\textrm{ if $i = j$}\\
		0\,,&\textrm{ otherwise}\,.\\
	\end{matrix*}\right. \nonumber
\end{align}
Brin and Page proposed PageRank centrality as part of their development of the Google web search engine \cite{Brin1998}. The used PageRank centrality to model the final distribution of a large number of web users who browse the World Wide Web by following a link on each webpage that they land on uniformly at random. In that context, one can interpret $\alpha$ in \eq{pagerank} as the parameter that tunes the probability of a web user interrupting their clickstream and `teleporting' to any page on the World Wide Web. (Specifically, small values of $\alpha$ correspond to frequent teleporting of web users.) 

Some extensions of PageRank allow non-identical constant terms in \eq{pagerank}, which correspond to a non-uniform distribution of teleportation destinations over network nodes \cite{Gleich2015}. Other extensions include formulations of PageRank for temporal networks and multilayer networks \cite{Taylor2021}.

\subsubsection{Subgraph centrality.}

Estrada and Rodriguez-Velazquez~\cite{Estrada2005} proposed a centrality measure that is based on the number of closed walks in which a node participates. The \emph{subgraph centrality} of a node $v_i$ is the weighted sum of the numbers $n^{\textrm{CW}}_i(\ell)$ of closed walks of length $\ell$ that traverse $v_i$,
\begin{align}
    \textrm{SC}_i:=\sum_{\ell=0}^\infty c_\ell n^{\textrm{CW}}_i(\ell)\,,\nonumber
\end{align}
with weights $c_\ell$. When one selects $c_\ell=\alpha^{-\ell}$, subgraph centrality is identical to Katz centrality \cite{estrada2012review}. A common choice of weighting is $c_\ell=(\ell!)^{-1}$, i.e., 
\begin{align}
    \textrm{SC}_i:=\sum_{\ell=0}^\infty\frac{n^{\textrm{CW}}_i(\ell)}{\ell!}\,.\label{eq:subgraph-centrality}
\end{align}
One can calculate the subgraph centrality in \eq{subgraph-centrality} of node $v_i$ using the eigenvalues $\lambda_i$ and eigenvectors ${\boldsymbol\eta}_i$ of a network's adjacency matrix ${\bf A}$. Specifically, 
\begin{align}
	\textrm{SC}_i=\sum_{j=0}^N\left(\eta_j^{(i)}\right)^2e^{\lambda_j}\,,\nonumber
\end{align}
where $\eta_j^{(i)}$ is the $i$th component of the eigenvector $\boldsymbol\eta_j$.

\subsubsection{Damage centrality.} 

For any performance measure $X$, one can define a node centrality measure as the change of $X$ under the removal of a node.  That is,
\begin{align}
	\textrm{DC}_i := X_{(V,E)}-X_{(V\setminus\{v_i\},E')}\,,\nonumber
\end{align}
where $X_{(V,E)}$ denotes the performance measure for the initial network and $X_{(V\setminus\{v_i\},E')}$ denotes the performance measure for the network after removing node $v_i$. Latora and Marchiori~\cite{Latora2005} introduced this family of centrality measures and referred to them as \textit{damage centrality} (or simply \textit{damage}). Many researchers have considered targeting strategies that are based on the damage to the LCC size (see Section \ref{sec:review:components}) of a network \cite{Lordan2014, Wang2014a}. Sun et al.~\cite{Sun2017} examined the impact on efficiency (see Section \ref{sec:review:paths}) of targeting nodes by damage centrality. It is possible to view other centrality measures as a damage-centrality measure. For example, in a simple network, one can view the degree of a node as a node's damage centrality to the number of edges.

\subsection{Edge centrality measures}\label{sec:edge_centralities}

In Table \ref{tab:edge_targets}, we list edge centrality measures and studies of network robustness that have used these centrality measures for adaptive or nonadaptive targeting strategies for edge removal. In this subsection, we give definitions of the centrality measures in Table \ref{tab:edge_targets}.

\begin{table}[h!]
\begin{center}
\begin{tabular}{p{5.5cm} p{7.cm}}
\toprule
Edge centrality measure & References \\
\midrule
Geodesic edge betweenness & \citecell{0.5}{Holme2002, He2009, Zeng2012, Hossain2013} \\
Edge clustering coefficient & \citecell{0.5}{Kaiser2004, Zeng2012} \\
Degree sum & \citecell{0.5}{Tomassini2023} \\
Degree product & \citecell{0.5}{Holme2002, Kaiser2004, Zeng2012, Zhang2013b} \\
Degree difference & \citecell{0.5}{Kaiser2004}\\
Products of node centrality measures & \citecell{0.5}{Zheng2014, Hao2020}\\
Epidemic importance & \citecell{0.5}{Matamalas2018, Jhun2021}
\vspace{-0.4cm}\\
\bottomrule
\end{tabular}
\end{center}
\vspace{-0.5cm}
\caption[Examples of centrality measures for edge removal.]{Examples of edge centrality measures that researchers have used as targeting strategies in studies of network robustness.}
\label{tab:edge_targets}
\end{table}

\subsubsection{Edge betweenness.}

The \emph{(geodesic) edge betweenness centrality} of an edge $e_{i,j}$ is a measure of how often a shortest path in a network traverses the edge. That is \cite{girvan2002},
\begin{align}
	\textrm{BC}_{i,j} := \sum_{\substack{k,l=1\,,\\j\neq i}}^N\frac{n^{\textrm{SP}}_{i,j}(k,l)}{ n^{\textrm{SP}}(k,l)}\,,\nonumber
\end{align}
where $n^{\textrm{SP}}(k,l)$ the number of shortest paths from $v_k$ to $v_l$ and $n^{\textrm{SP}}_{i,j}(k,l)$ is the number of shortest paths from $v_k$ to $v_l$ that traverse $e_{i,j}$. Edge betweenness centrality is a popular edge centrality measure for studies of network robustness \cite{Holme2002, He2009, Zeng2012, Hossain2013}.

\subsubsection{Edge clustering coefficient.}

In an undirected network, the edge clustering coefficient of an edge $e_{i,j}$ between two nodes, $v_i$ and $v_j$, with degrees $k_i, k_j \geq 2$ is \cite{Radicchi2004}
\begin{align}
	C_{i,j} := \frac{|V_{(i,j)}|}{\min\{k_i,k_j\}-1}\,,\nonumber
\end{align}
where $V_{(i,j)}\subset V\setminus\{v_i,v_j\}$ is the set of nodes that are adjacent to both $v_i$ and $v_j$. Kaiser and Hilgetag~\cite{Kaiser2004} and Zeng and Liu~ \cite{Zeng2012}  used $C_{i,j}$ in studies of network robustness.

\subsubsection{Degree sum, degree product, and degree difference.}

Several researchers have assigned centrality values to edges using the degrees $k_i$ and $k_j$ of the nodes that are attached to edge $e_{i,j}$~\cite{Holme2002, Kaiser2004, Zeng2012}. For example, one can use the degree sum $k_i + k_j$ or the degree product $k_ik_j$ as edge centrality measures \cite{Holme2002, Wang2008b, Tomassini2023}. Both of these edge centrality measures are large for edges that connect two nodes with large degrees. Another example is (the absolute value of) the degree difference, $|k_i - k_j|$, which is an edge centrality measure that assigns large centrality values to edges that connect large-degree nodes to small-degree nodes \cite{Kaiser2004}. 

The distribution of degree products and the distribution of degree differences in a network are related to the network's degree--degree assortativity. In a \textit{degree--degree assortative} network, edges tend to connect nodes with similar degrees \cite{Newman2002}. Conversely, in a degree--degree disassortative network, edges tend to connect nodes with different degrees. Several researchers have explored links between degree--degree assortativity and robustness in networks \cite{Newman2002, Newman2003b, Herrmann2011, Zhou2012, Trajanovski2013, Shizuka2016}.

\subsubsection{Edge centrality measures that are based on node centrality measures.}

In addition to using the sum, product, and absolute value of difference of node degrees as edge centrality measures, one can also consider the sum, product, and absolute value of difference of various node centrality measures that are not degree centrality. For example, researchers have used the product of node betweenness centrality \cite{Zheng2014}, the product of node closeness centrality \cite{Hao2020}.

Another way to build edge centrality measures based on node centrality measures on a network $G(V,E)$ is to use node centrality measures on the corresponding line graph \cite{Brohl2019}. 

\subsubsection{Edge centrality measures that are based on processes on networks.}
One can construct edge centrality measures that are based on processes on networks. For example, Matamalas et al.~proposed an edge centrality measure that measures an edge's potential for disease transmission in an SIS model of the spread of infectious diseases on a network \cite{Matamalas2018}. Their  edge centrality measure, which they called \textit{epidemic importance}, assigns each edge $e_{i,j}$ a positive number that is proportional to the probability that one of the two nodes $v_i$, $v_j$ is infected while the other node is not infected.

\subsection{Targeting strategies and applications}

In this subsection, we review similarities and differences in the impact of different targeting strategies on performance measures. We subsequently discuss motivations for considering different targeting strategies in studies of network robustness.

\subsubsection{Correlations between centrality measures.}

Many researchers have compared the impacts on network performance measures of uniformly random node and edge removal to the impacts of different targeting strategies for removing nodes and edges \cite{Motter2002, Herrmann2011, Hossain2013, Iyer2013, Lordan2014, Ventresca2014}. These results suggest that (1) many strategies for targeted node and edge removal affect performance measures more strongly than removing nodes or edges uniformly at random and that (2) the difference between the impacts of targeted and uniform random removal is much greater than the difference between the impacts of different targeting strategies. More specifically, some researchers have reported positive correlations between centrality measures when comparing targeting strategies for node removal, and they have argued that the similar impacts of different targeting strategies arise from the strong positive correlations between different centrality measures \cite{Holme2002, Iyer2013, Lordan2014}. 

Many popular centrality measures are related to each other, so one should expect to observe positive correlations between their values \cite{Estrada2005, Newman2018}. For example, one can derive eigenvector centrality from an infinite iterative process that updates the centrality of a node by replacing the node's centrality by the sum of the centralities of its neighboring nodes in each iteration (see Section \ref{sec:eigenvector_centrality}). In the first iteration, one obtains degree centrality. The iterative procedures that converge to eigenvector centrality, Katz centrality, and PageRank centrality have many similarities, and these similarities can lead to similar centrality values for nodes in undirected networks and nodes in a strongly connected component or `downstream' of it (i.e., connected via a directed path that originates from a strongly connected component) of a directed network. 
A node with many edges tends to be connected to many other nodes by short paths, which suggests that one can expect a positive correlation between degree and closeness centrality in many networks \cite[p.\,171]{Newman2018}. Subgraph centrality assigns more importance to closed walks of short length $\ell$ than to closed walks of lengths that are larger than $\ell$. If one considers only closed walks of length $\ell = 2$, subgraph centrality and degree centrality result in identical node rankings in an undirected, unweighted network. If one considers only closed walks of length $\ell = 3$, subgraph centrality indicates the number of triangles that are associated with a node in a simple network, and it is thus correlated with the numerator of the local clustering coefficient (see \eq{lcc}).

However, in comparing the results of different studies, one finds that the strength of the correlations between different centrality measures varies significantly across networks \cite{Wacker2020}. The strength of the correlation and the functional form of the relationship between two centrality measures depends on network structure \cite{Holme2002, Valente2008}. Uncovering the effects of network structure on correlations of centrality measures is an active research area \cite{Chung2012, Iyer2013, Zhang2013, Jackson2014, Nomikos2014, Li2015b, Niu2015, Bloch2017, Oldham2019, Wacker2020, Bloch2023}.

\subsubsection{Should one consider more than one targeting strategy?}

If one tends to observe similar impacts on performance measures for different targeting strategies, does one need to consider more than one targeting strategy in a study of network robustness? The answer to this question depends on a study's aim. For example, if one seeks to demonstrate that the impact of sequential node removal can be very different for different sets of removed nodes, it is likely sufficient to compare the impact of removing nodes uniformly at random to the impact of targeting nodes by largest degree (see, for example, Ref.\,\cite{Albert2000}). However, a study's aim may be to find a good heuristic approach to identifying a fixed-size set of nodes that, if removed, leads to the largest possible change of a performance measure. For such a study, it can be useful to compare a variety of different targeting strategies. For many performance measures, neither degree nor any other known node centrality measure can reliably identify fixed-size sets of nodes or edges whose removal leads to the largest change in a performance measure \cite{Braunstein2016}. Several studies have demonstrated that, for some networks, targeting nodes by largest betweenness centrality and by other centrality measures can lead to larger impacts on performance measures than targeting nodes by largest degree \cite{Holme2002, Iyer2013, Lordan2014, Ventresca2014}. 

Considering different centrality measures can also be relevant when modeling adversarial attacks on networked systems. For example, if one is interested in the impact of a hacker's attack on the internet, one can examine centrality measures that hopefully reflect a hacker's strategy for selecting target nodes. As another example, consider extinctions in an ecological network. If one assumes that a particular centrality measure reflects the likelihood of extinction for 
species, then it should be reasonable to use that centrality measure to construct realistic node-removal perturbations. In some studies, researchers have used observations on real systems to rank nodes or edges \cite{Li2015, Mourier2017}. Mourier et al.~\cite{Mourier2017} used data from a catch-and-release program to assign a value of `catchability' blacktip reef sharks. Li et al.~\cite{Li2015a} used measurements of traffic flow to assign `velocities' to the edges of a road network. They then examined the change of the LCC size when removing edges from largest to smallest velocity.

\subsubsection{Which targeting strategy leads to the largest impacts?}

Suppose that one is interested in reliably identifying fixed-size sets of nodes that, if removed, lead to the largest possible change of a performance measure. To do this, it may not be possible to reliably identify a node centrality measure (or a combination of node centrality measures). In general, one cannot calculate the impact of the removal of a set of nodes from the impacts of removing individual nodes. Researchers refer to the impact of removing a set of nodes as the set's \textit{collective influence} \cite{Morone2015, Lue2016, Morone2016, Pei2019}. The study of collective influence is an active research area \cite{Braunstein2016, Lue2016, Mugisha2016, Zhu2018, Boltz2019}. If one is interested in the impact of removing a set of nodes on a network's LCC size, one can consider strategies for \textit{network dismantling} \cite{Braunstein2016, Zdeborova2016, Ren2019}. In a network-dismantling problem, one seeks a minimal set of nodes that, if removed, reduces the LCC size to some prescribed value \cite{Braunstein2016}. Such sets of nodes are known as `optimal dismantling sets' \cite{Braunstein2016, daCunha2017, deAbreu2018}. Many researchers have proposed heuristic algorithms to find almost optimal dismantling sets on various networks \cite{Morone2015, Braunstein2016, daCunha2017, deAbreu2018}.

\section{Specifying a measure of robustness}\label{sec:measures}

In this section, we review common approaches to calculate measures of robustness --- or, alternatively, measures of vulnerability --- for networked systems. In Section \ref{sec:impact}, we explain how the specification of a system or model of a system, a performance measure, and a set or sequence of perturbations yields a set or sequence of perturbation impacts, which one can use to compute different notions of robustness. In Section \ref{sec:scalar_robustness}, we review several scalar robustness measures that yield a single real-valued number (i.e., a real scalar) that summarizes the impact of perturbations on system performance of a system.
In Section \ref{sec:scalar_robustness_lcc}, we review several scalar robustness measures that one can compute directly from a network's adjacency matrix or Laplacian matrix without conducting computational node-removal or edge-removal experiments when the chosen performance measure is the network's LCC size.
In Section \ref{sec:choosing_robustness}, we discuss some considerations that we anticipate to be relevant when choosing an appropriate robustness measure for a given study.

\subsection{The impact of a perturbation}\label{sec:impact}

When examining the effect of a perturbation $p$ of a network on a performance measure $X$, it is common to calculate the value $X_0$ of the performance measure for the unperturbed network and the value $X_p$ of the performance measure after the perturbation. One can consider the \textit{impact} measure $I$ of a perturbation to be the change
\begin{align} \label{eq:impact_as_change}
	I = X_0 - X_p
\end{align}
of a performance measure or the relative change
\begin{align} \label{eq:impact_as_relative_change}
    I = \frac{X_0 - X_p}{X_0} 
\end{align}
of that performance measure \cite{Albert2004, Latora2005, Nagurney2007, Behre2008, Dattilo2016, Lance2017, Liu2017}. To ensure that impacts are positive, the measure of impact can be the absolute value 
\begin{align} \label{eq:impact_as_abs_change}
    I = |X_0 - X_p|
\end{align}
of the performance difference or the absolute value 
\begin{align} \label{eq:impact_as_abs_relative_change}
    I = \frac{|X_0 - X_p|}{X_0}
\end{align}
of the relative change of performance.
After selecting one of the definitions of impact in \eqrange{impact_as_change}{impact_as_abs_relative_change} (or perhaps some other definition of impact), one can compute a set or sequence of impacts from a set or sequence of perturbations (see Section \ref{sec:sets_and_sequences}).

\subsubsection{Sets and sequences of impacts.}\label{sec:impact_sets}

In studies of network robustness, it is useful to examine the impacts of perturbations using a variety of perturbations. For example, one can study the robustness of networks to single-node removal or single-edge removal. In a study of single-node removal, one calculates the set $\{I_v\}_{v\in V}$ of impacts for the removal of nodes $v \in V$ \cite{Wilhelm2004, Wang2014, Tang2016}. Similarly, in a study of single-edge removal, one calculates the set $\{I_e\}_{e\in E}$ of impacts $I_{e}$ for the removal of edges $e \in E$ \cite{Latora2005}.

One can also study the robustness of networks to the removal of multiple nodes or multiple edges. We distinguish between studies of \textit{simultaneous} node/edge removal and studies of \textit{sequential} node/edge removal. In a study of simultaneous node removal, one examines the impact of removing a set $V_S\subset V$ of nodes from a network. A set of $N$ nodes has $\binom{N}{n}$ subsets $V_S$ of size $n$, so it is almost always impractical to conduct an exhaustive computation of the full set $\{I_{V_S}\}_{V_S\subset V}$ of impacts of removals of size-$n$ node sets. Researchers have circumvented this computational difficulty either by sampling from the set $\{I_{V_S}\}_{V_S\subset V}$ \cite{Tang2016} or by using theoretical results from probability theory and statistical mechanics to calculate expectations of $I_{V_S}$ \cite{Cohen2000, Boguna2005, Cohen2010}. 

In a study of sequential node removal, one constructs one sequence or several sequences $(v_1,v_2,v_3,\ldots)$ of nodes. For each sequence of nodes, one then calculates a corresponding sequence $(I_{v_1},I_{v_2},I_{v_3},\ldots)$ of impacts, where $I_{v_i}$ is the impact of removing the set $\{v_1,\ldots,v_i\}$ of nodes from a network \cite{Albert2000, Iyer2013, Ventresca2014}. In a study of sequential edge removal, one calculates a similar sequence of impacts for the removal of a sequence of edges \cite{Zeng2012, Hossain2013}.

\subsection{Scalar measures of robustness and vulnerability}\label{sec:scalar_robustness}

Sets or sequences of impacts for structural perturbations give detailed information about the robustness (i.e., insensitivity) and vulnerability (i.e., sensitivity) of a performance measure to a set of perturbations. 
However, to facilitate simple comparisons of the robustness of different networks, researchers have developed several approaches to summarize this information in a single real number as a scalar measure of robustness \cite{Latora2005, Schneider2011, Wang2014}. In this subsection, we review such scalar robustness measures that one can calculate from sets or sequences of perturbation impacts. 

In Table \ref{tab:R}, we list the robustness measures that we review in this subsection. We include references to publications that have used these measures.

\begin{table}[h!]
\begin{center}
\begin{tabular}{p{6.1cm} p{6.4cm}}
\toprule
Robustness measure & References \\
\midrule
Mean impact of node or edge removal & \citecell{0.45}{Wilhelm2004, Wang2012a, Zhang2013c, Wang2013, Wang2014, Tang2016} \\
Maximum impact of removing & \multirow{2}{*}{\citecell{0.45}{Latora2005, Wandelt2018}} \\ 
one or several nodes or edges \vspace{0.2cm}\\
Probability of achieving a performance goal & \citecell{0.45}{Henneaux2015, Henneaux2015a, Panteli2015, Phan2018} \\
Critical fraction of nodes & \citecell{0.45}{Cohen2000, Albert2002, Newman2003, Rosas-Casals2007, Sole2008, Buldyrev2010, Zhou2012, Sha2013, Danziger2014, Nie2016, Roth2017} \\
Critical fraction of edges & \citecell{0.45}{He2009, Cozzo2018, Thornhill2018, Mimar2022} \\
Schneider et al.'s $R$ index & \citecell{0.45}{Schneider2011, Wu2011, Zeng2012, Schneider2013, Iyer2013, Priester2014, Zhou2014, Ahmadi2015, Gong2015, Yang2015, daCunha2017, Hong2017, Dong2019, Li2023b, Tomassini2023}
\vspace{-0.4cm}\\
\bottomrule
\end{tabular}
\end{center}
\vspace{-0.5cm}
\caption[Examples of robustness measures.]{Examples of robustness measures that researchers have used in studies of network robustness..}
\label{tab:R}
\end{table}

\subsubsection{Mean impact and maximum impact.}

For sets of impacts of single node-removal or edge-removal perturbations, it is common to use a set's mean value \cite{Wang2014} or maximum value \cite{Latora2005} as a measure of vulnerability. Mean impacts indicate how severely one can expect the performance of a system to decline on average if one assumes that all considered perturbations are equally likely. By contrast, maximum impacts illustrate the worst-case scenario for a system's performance for a given set of possible perturbations. The choice of using the mean impact or maximum impact to measure a system's robustness thus reflects one's interest in either the average performance loss or the worst-case performance loss that a system can experience under a perturbation.

\subsubsection{Minimum perturbations and critical fractions of nodes or edges.}

The mean impact and maximum impact are measures of robustness that one can compute with respect to a given set $\mathcal P$ of perturbations. Conversely, one can consider a scenario in which one is given an impact value $I = c$ instead of a set of perturbations. One can then ask the following question: What is the minimum perturbation that can lead to the prescribed impact $c$? Such a measure of robustness is meaningful when one can rank perturbations by some notion of size. For example, when one considers simultaneous or sequential node-removal or edge-removal perturbations, one can examine the number of nodes or edges that one needs to remove to obtain an impact of at least $c$ on a system's performance.

For studies of simultaneous node-removal perturbations, a common measure of robustness is the smallest fraction $f_v$ of nodes for which the expectation of the impact $I_{V_S}$ of removing a node set $V_S$ with $|V_S| = \lceil f_vN\rceil$ is at least a specified value \cite{Cohen2000, Dunne2002, Shargel2003, Wang2008}. In many such studies, researcher have referred to $f_v$ as the {`critical fraction of nodes'} \cite{Cohen2000, Albert2002, Buldyrev2010, Danziger2014}. Similarly, researchers have use the term {`critical fraction of edges'} for the smallest fraction $f_e$ for which the expectation of the impact $I_{E_S}$ of removing a node set $E_S$ with $|E_S| = \lceil f_em\rceil$ is at least a specified value \cite{Cozzo2018}. 

When one uses the LCC size of a network that one draws from an Erd\H{o}s--R\'enyi (ER) random-graph ensemble as a measure of a system's performance, one can obtain a value of $f_v$ from the ER percolation threshold $p_c = 1/\langle k\rangle$ in the $N \rightarrow \infty$ limit. Removing a proportion $1 - p_c$ nodes from an ER random graph tends to cause a steep decline of the relative LCC size \cite[p.\,573]{Newman2018}.

\subsubsection{Probability of achieving a performance goal.}

Another measure of robustness is the probability that an impact is smaller than or larger than a prescribed tolerable impact $I^*$ \cite{Henneaux2015, Henneaux2015a, Panteli2015, Phan2018}. Let $P(I<I^*)$ denote the probability that an impact is smaller than $I^*$, and let $P(I>I^*)$ denote the probability that an impact is larger than $I^*$.
These probabilistic robustness measures are useful when it is important that a system maintain a minimum performance $X^*$ after a perturbation. That minimum performance corresponds to the impact $I^*$.
In such a case, the probability $P(I<I^*)$ is a measure of how likely it is for a system to maintain an acceptable performance after a perturbation.

\subsubsection{Schneider et al.'s $R$ index.} 

For studies of sequential node or edge removal, Schneider et al.~\cite{Schneider2011} used the mean of the relative performances $X_p/X_0$ as a measure $R$ of robustness and the mean of the relative changes $|X_0 - X_p|/X_0$ as a measure $V$ of vulnerability. For both cases, one takes the mean over the impacts of a set $\{p_1,p_2,p_3, \ldots, p_{N-1}\}$ of progressively more severe perturbations, where $p_i$ corresponds to the removal of $i$ nodes from a network. 
Schneider et al. referred to their measure of robustness (i.e., the mean of the relative performances $X_p/X_0$) as the `$R$ index'. When the selected performance measure is the relative LCC size, then the $R$ index takes values in the interval $[1/N^2, 0.5]$ (and it takes values in the interval $[1/N,0.5]$ if the network is initially connected) \cite{Schneider2011}. 

The Schneider et al.~'s $R$ index measures how rapidly a system's performance declines as one sequentially removes a network's nodes according to a specified removal sequence. A large value of the $R$ index indicates that the system is able to maintain a good performance as one removes progressively more nodes. Conversely, a small value of 
the $R$ index indicates that a system's performance drops strongly after the removal of even a few nodes. Changing the order of the sequence of nodes can greatly impact the $R$ index. Applying different targeting strategies when selecting a sequence of sequential node-removal perturbations can thus lead to very different values of the $R$ index for the same system and performance measure.

Many researchers have used Schneider et al.'s $R$ index in studies of network robustness \cite{Herrmann2011, Mello2011, Mello2011a, Schneider2011, Zeng2012, Iyer2013, Louzada2013, Ventresca2014, Zhou2014, Ahmadi2015, Hong2017}. Although most of these studies refer to this measure of robustness as `the $R$ index', we caution that this terminology can lead to confusion with `$r$-robustness' (see Section \ref{sec:r-robustness}) and several other existing robustness indices \cite{Wilhelm2004, Gorban2007, Fang2023}.

\subsection{Scalar robustness measures for the size of a network's largest connected component}\label{sec:scalar_robustness_lcc}

In mathematical studies, it is common to investigate simple graphs and directed graphs independently of their ability to model real-world systems. In graph-theoretical studies of network robustness, researchers have developed a variety of robustness measures that use the LCC size as a performance measure and that one can compute without conducting a set of computational perturbation experiments \cite{Hoory2006, Shang2012, Zheng2022}.\footnote{Lou et al.~proposed to call robustness measures that one can compute without conducting a set of computational perturbation experiments `a priori robustness measures' \cite{Lou2023}. They proposed to call robustness measures that require computational or other perturbation experiments `a posteriori' robustness measures.} In this subsection, we review such robustness measures. Some of the measures that we discuss have also been used as performance measures in computational studies \cite{Eubank2004, Wu2010a, Zhang2013b, Wang2014b, Cozzo2018, Yamashita2019}. We discuss these ambiguities in Section \ref{sec:choosing_robustness}.

\begin{table}[h!]
\begin{center}
\begin{tabular}{p{4cm} p{8.5cm}}
\toprule
Robustness measure & References \\
\midrule
Node connectivity & \citecell{0.45}{VanMieghem2005, Erlebach2006, Abbas2018} \relvspace\\
Edge connectivity & \citecell{0.45}{VanMieghem2005, Erlebach2006, Shang2012} \bigrelvspace \\
Algebraic connectivity (i.e., the Fiedler value) & \citecell{0.45}{VanMieghem2005, Yazdani2011, Bilal2013, Shahrivar2015, Cheng2017, MoradiShahrivar2017, Wang2018, Yamashita2019} \relvspace\\
Node expansion & \citecell{0.45}{Bagchi2004, Barrett2004, Eubank2004} \relvspace\\
Edge expansion & \citecell{0.45}{Bagchi2004, Wang2014d, Cheng2017, MoradiShahrivar2017, Oehlers2021} \relvspace\\
Spectral gap of $\bf A$ & \citecell{0.45}{Yazdani2011, Wang2014d, Wang2018, Yamashita2019} \relvspace\\
Spectral radius of $\bf A$ & \citecell{0.45}{Yamashita2019, Bilal2013, Jamakovic2006, Wang2018} \relvspace\\ 
$r$-robustness & \citecell{0.45}{Zhang2012, LeBlanc2012, Shahrivar2015, Zhang2015, MoradiShahrivar2017, Usevitch2017, Abbas2018, Ishii2022, Usevitch2020} \relvspace\\
$(r,s)$-robustness & \citecell{0.45}{LeBlanc2012, Usevitch2017, Ishii2022, Usevitch2020} 
\vspace{-0.4cm}\\
\bottomrule
\end{tabular}
\end{center}
\vspace{-0.5cm}
\caption[Examples of robustness measures for the largest connected component (LCC) of a network.]{Examples of robustness measures that researchers have used to study the robustness of the size of the largest connected component (LCC) of a network.}
\label{tab:lcc_robustness_measures}
\end{table}

\subsubsection{Node connectivity and edge connectivity.}\label{sec:ne_connectivity}

In a connected network, one can examine how many nodes or edges it is necessary to remove to fragment the network into multiple components. If removing a set of nodes or set of edges causes a network or its LCC to become fragmented, then that set is a \textit{node cut set} or \textit{edge cut set}, respectively. A network's \textit{node connectivity} $\kappa_v$ is the size of a minimal node cut set of a network \cite{Shang2012, Bollobas2013}. A network's \textit{edge connectivity} $\kappa_e$ is the size of a minimal edge cut set \cite{Shang2012, Bollobas2013}.  The node connectivity, edge connectivity, and minimum degree $k_{\min}$ of a simple network are related by the inequality \cite[p.\,207,]{Bollobas2013}
\begin{align}
    \kappa_v\leq \kappa_e\leq k_{\min}\,.\nonumber
\end{align}

\begin{figure}[t]
\centering
\includegraphics[trim={2.7cm 0.5cm 2.2cm 0.5cm},clip,width=1\textwidth]{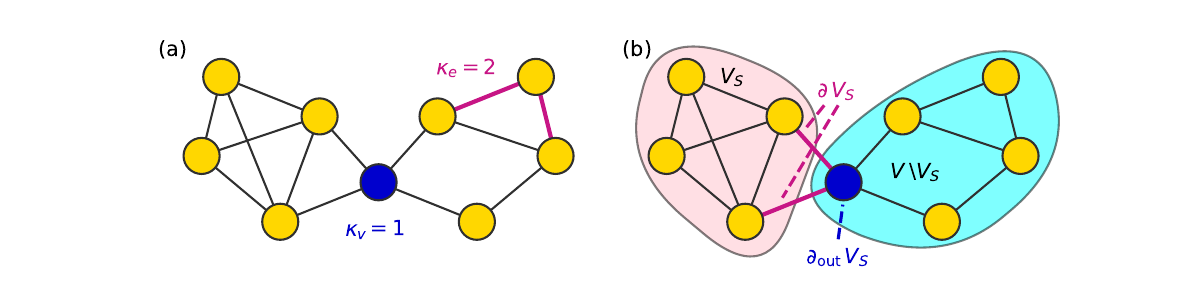}  
\caption{An illustration of node connectivity, edge connectivity, and the node sets and edge sets for the node and edge expansions of an example network. (a) An example network with 9 nodes and 14 edges. The minimal node cut set has a single node (in blue). The network thus has node connectivity $\kappa_v = 1$. The network has several minimal edge cut sets; each of them include two edges. We indicate one of these edge cut sets in magenta. The network's edge connectivity is $\kappa_e = 2$. (b) Partitioning the network into two subnetworks with node sets $V_S$ and $V \setminus V_S$ leads to an edge boundary $\partial S$ of two edges (in purple) and an outer node boundary $\partial_{\textrm{out}}S$ of one node (in blue). The network thus has edge expansion $h_e = 1/2$ and node expansion $h_v = 1/4$.} 
\label{fig:kappa}
\end{figure}

In Fig.\,\ref{fig:kappa}(a), we show an example of a network's minimal node cut set, minimal edge cut set, and the resulting values of node connectivity and edge connectivity.

What insights can one gain about the robustness of a network's LCC by knowing $\kappa_v$ and $\kappa_e$? The number $\kappa_e - 1$ indicates the number of edges that one can remove from a network without decreasing its LCC size.
The number $\kappa_v - 1$ indicates the number of nodes that one can remove from a network without increasing the number of nodes that are not part of its LCC. (This slight difference in the interpretation of $\kappa_e$ and $\kappa_v$ arises because a node cut set includes nodes in the LCC, so the LCC size can change when removing any subset of a node cut set.) These relationships between $\kappa_v$, $\kappa_e$, and a network's LCC help motivate the study of node connectivity \cite{VanMieghem2005, Erlebach2006} and edge connectivity \cite{VanMieghem2005, Erlebach2006, Shang2012} in the context of network robustness.

\subsubsection{Algebraic connectivity.}\label{sec:algebraic-connectivity}

The \textit{algebraic connectivity} (i.e., \textit{Fiedler value}) of a network is the second-smallest eigenvalue of its combinatorial Laplacian matrix $\mathbf L$ \cite{Estrada2015}. The algebraic connectivity of a network is positive if and only if the network is connected \cite{Jamakovic2007}. In that case, one can view a network's algebraic connectivity as a measure of how well it is connected \cite{deAbreu2007}. The algebraic connectivity $a$ is linked to the node connectivity $\kappa_v$ and the edge connectivity $\kappa_e$ by the relation \cite{Cvetkovic1980, VanMieghem2005}
\begin{align}
	2\kappa_e\left(1-\cos\left(\pi/N\right)\right)\leq a\leq \kappa_v\,.\nonumber
\end{align}
This relationship helps motivates the study of algebraic connectivity in various contexts \cite{Fiedler1973, Mohar1991, Jamakovic2007, Jamakovic2007a, Spielman2007, deAbreu2007}. For example, several researchers have used $a$ as a performance measure \cite{Yazdani2011, Bilal2013, Yamashita2019} or robustness measure \cite{VanMieghem2005, Jamakovic2007, Jamakovic2007a, Bigdeli2009, Martin2014, Wang2018, VanMieghem2023}.

\subsubsection{Node and edge expansions.}

Consider a bipartition of a network into two node sets $V_S$ and $V\setminus V_S$. The number of edges that connect nodes in $V_S$ to nodes in $V$ is the \textit{edge boundary} $\partial V_S:= \{e_{i,j}\in E:\,\,v_i \in V_S \land v_j \in V\setminus V_S\}$. The \textit{edge expansion} $h_e$ (which is also called \textit{isoperimetric constant} or \textit{Cheeger constant} \cite{VanMieghem2023}) of a network $G=(V,E)$ with $N$ nodes is the smallest value of the fraction of the edge boundary and the size of $V_S$ \cite{VanMieghem2023}. In formal language, one writes \cite[p.\,152]{VanMieghem2023}
\begin{align}
	h_e := \min_{0<|V_S|\leq\frac{N}{2}}\frac{|\partial V_S|}{|V_S|}\,.\nonumber
\end{align}
Informally, one can think of $h_e$ as a measure of `botteneckedness' in networks \cite{Laszka2012}. Researchers have investigated the relationship between several notions of a network's robustness and its expansion properties \cite{Estrada2006, Estrada2007}. The edge expansion of a network is closely related to several notions of connectivity (see Sections \ref{sec:ne_connectivity} and \ref{sec:algebraic-connectivity}). For example, for an undirected $k$-regular\footnote{An undirected network is \textit{$k$-regular} if each of its nodes has degree $k$ \cite[p.\,2]{Akiyama2011}.} network, the Cheeger inequality \cite[p.\,153]{VanMieghem2023} links $h_e$ to a network's algebraic connectivity through the inequality
\begin{align}
	\frac{a}{2}\leq h_e\leq\sqrt{2ka}\,.\label{eq:cheeger}
\end{align}
This relationship between $a$ and $h_e$ helps motivate the study of $a$ in the context of network robustness. This complements the motivations that we discussed in Section \ref{sec:algebraic-connectivity}. 

The \textit{outer node boundary} of $V_S$ is the set of nodes outside of $V_S$ that are connected to at least one node in $V_S$ by an edge. Denoting the outer node boundary by $\partial_{\textrm{out}} V_S$, one writes $\partial_{\textrm{out}} V_S := \{v_i\in V\setminus V_S:\,\,e_{i,j} \in E \land v_j \in V_S\}$.
The \textit{node expansion} $h_v$ 
of a network is the smallest value
of the fraction of the outer node boundary and the size of $V_S$.
In formal language, the node expansion of a network $G=(V,E)$ with $N$ nodes is \cite[p.\,20]{Gao2019} 
\begin{align}
	h_v := \min_{0<|V_S|\leq\frac{N}{2}}\frac{|\partial_{\textrm{out}} V_S|}{|V_S|}\,,\nonumber
\end{align}
where $V_S\subset V$ and $\partial_{\textrm{out}} V_S := \{v_i\in V\setminus V_S:\,\,e_{i,j} \in E \land v_j \in V_S\}$ is the \textit{outer node boundary} of $V_S$. 
Several researchers have studied theoretical and empirical relationships between a network's node expansion and its robustness \cite{Bagchi2004, Barrett2004} and relationships between $h_v$ and properties of dynamics on networks \cite{Eubank2004, Sauerwald2011, Giakkoupis2012, Doerr2013}.

\subsubsection{Spectral gap of ${\bf A}$.}

\Eq{cheeger} also helps motivate the study of the spectral gap of ${\bf A}$ (i.e., the difference between the largest and second-largest eigenvalues of $\bf A$ \cite[p.\,88]{VanMieghem2023}). In an undirected regular network, the spectral gap is equal to the algebraic connectivity $a$ \cite{Hoory2006, Stanic2013, VanMieghem2023}. 
Yazdani and Jeffrey~\cite{Yazdani2011} and Yamashita et al.~\cite{Yamashita2019} used the spectral gap of ${\bf A}$ as a performance measure in studies of network robustness. Wu and Holme~\cite{Wu2011} used the spectral gap of $\bf A$ as a measure of network robustness. One can also interpret the spectral gap of an undirected network as a measure of the convergence time of a random walk on a network \cite[p.\,109]{VanMieghem2023}. This relationship of the spectral gap to dynamical processes on networks is a potential avenue to link notions of structural robustness to notions of robustness of dynamical systems on networks \cite{Li2013}.

\subsubsection{Spectral radius of ${\bf A}$.}

The spectral radius $\rho(\cdot)$ of a matrix is the largest absolute value of its eigenvalues \cite[p.\,195]{Lax2002}. For many models of dynamics on networks, the spectral radius $\rho({\bf A})$ of a network's adjacency matrix is an important quantity \cite{Barnett2009, Pastor-Satorras2001}. Consider compartmental models, such as an susceptible--infected--susceptible (SIS) model, of the spread of infectious diseases on a network \cite{Pastor-Satorras2015}. For several random-graph models (e.g., ER random networks \cite[p.\,343]{Newman2018}), one can find a critical threshold $\tau_c$ for the effective spreading rate \cite{Boguna2002, Miller2009, VanMieghem2009}.\footnote{The \textit{effective spreading rate} is the ratio of the recovery rate (i.e., the rate at which an infected node recovers) and the infection rate (i.e., the rate at which a node that is adjacent to an infected node becomes infected)~\cite{Pastor-Satorras2015}.} 
When the effective spreading rate is smaller than $\tau_c$, infections die out. However, when the effective spreading rate is larger than $\tau_c$, infections can persist and a positive fraction of a network's nodes remain infected \cite{VanMieghem2009}.
Using a mean-field approach \cite[p.\,37]{Porter2016}, van Mieghem et al.~\cite{VanMieghem2009} showed that $\rho({\bf A})^{-1}$ is a lower bound of $\tau_c$. 
When considering a compartmental model of the spread of an infectious disease, the spectral radius of the adjacency matrix is thus related to a system's ability to resist an epidemic. Several researchers have used $\rho({\bf A})$ as a measure of performance or robustness in studies of network robustness \cite{Jamakovic2006, Bilal2013, Yamashita2019}.

\subsubsection{$r$-robustness and $(r,s)$-robustness.}\label{sec:r-robustness}

Motivated by a linear dynamical model of information spread on a network with malicious nodes, Zhang and Sundaram~\cite{Zhang2012} introduced a notion of robustness that they called $r$-robustness. A network $G=(V,E)$ is `$r$-robust' if for any pair of disjoint, nonempty subsets $V_{S_1}$ and $V_{S_2}$ of $V$, at least one of the two subsets includes a node with at least $r$ neighbors outside of that subset 
\cite{Zhang2012}. 
For a linear model of information spread, Zhang and Sundaram~\cite{Zhang2012} proposed an information-processing algorithm that can ensure the convergence of node states to a homogeneous steady state on an $r$-robust network, even if that network includes a set of malicious nodes that try to steer the dynamics away from that homogeneous steady state.

LeBlanc et al.~\cite{LeBlanc2012} introduced the notion of robustness that they called $(r,s)$-robustness. A network $G=(V,E)$ is `$(r,s)$-robust' if any pair of disjoint, nonempty subsets $V_{S_1}$ and $V_{S_2}$ of $V$ satisfies the following properties:
\begin{enumerate}
    \item the union $V_{S_1} \cup V_{S_2}$ includes at least $s$ nodes such that each of those nodes has at least $r$ neighbors outside its associated subset; 
    \item each node in $V_{S_1}$ has at least $r$ neighbors in $V\setminus V_{S_1}$; or
    \item each node in $V_{S_2}$ has at least $r$ neighbors in $V\setminus V_{S_2}$. 
\end{enumerate} 
Any subset $V_{S_i}$ of size $|V_{S_i}| = 1$ is a nonempty node set that includes at most one node with $r$ neighbors outside $V_{S_i}$. Conditions (2) and (3) of the definition of $(r,s)$-robustness are thus necessary to for networks to potentially be $(r,s)$-robust with $s > 1$.
For a linear model of opinion dynamics, LeBlanc et al.~\cite{LeBlanc2012} proposed an information-processing algorithm that can ensure the convergence of node states to a homogeneous steady state on an $(r,s)$-robust network, even if that network has a set of malicious nodes that try to steer the dynamics away from that homogeneous steady state.

Many researchers have used $r$-robustness and $(r,s)$-robustness in studies of network robustness \cite{Shahrivar2015, Zhang2015, MoradiShahrivar2017, Usevitch2017, Abbas2018, Usevitch2020, Ishii2022}. The two concepts have been particularly popular in studies of network robustness in the engineering sciences \cite{Abbas2018, Ishii2022}. Outside of the engineering sciences, other researchers have called other robustness measures `$r$-robustness' \cite{Gorban2007}.

\subsection{Popularity and applicability of robustness measures}\label{sec:choosing_robustness}

In this subsection, we discuss connections between the scalar robustness measures that we discussed in Sections \ref{sec:scalar_robustness} and \ref{sec:scalar_robustness_lcc}. In Section \ref{sec:robustness_dist}, we discuss some advantages and disadvantages of using scalar robustness measures. We review empirical findings on correlations between different measures of robustness in Section \ref{sec:robustness_correlations}, and we propose relevant use cases for different robustness measures in Section \ref{sec:robustness_applications}. In Sections \ref{sec:performance_ambiguity} and \ref{sec:robustness_ambiguity}, we offer explanations of why several performance measures also arise as robustness measures (and vice versa) in various studies.

\subsubsection{Do I need a scalar measure of robustness?}\label{sec:robustness_dist}

Scalar robustness measures are popular in studies of the robustness of networked systems. They facilitate straightforward comparisons between the robustness of different systems both within \cite{Dodds2003, Ahmadi2015, Li2023b} and across \cite{Schneider2011} application domains, and they also are helpful for studying the effects of various model parameters on robustness \cite{Scellato2011, Wu2011, Zhang2013}. Nevertheless, many studies of the robustness of networked systems do not use scalar robustness measures \cite{Albert2000, Trajanovski2013, Manzano2014, Henneaux2015, Henneaux2015a, Phan2018, Ferrari2023}. Some studies instead report their results either in the form of distributions of performance values or impact values over a set of perturbations \cite{Henneaux2015, Henneaux2015a, Panteli2015, Phan2018} or in the form of the trajectory of performance or impact as perturbations occur sequentially \cite{Albert2000, Iyer2013, Ventresca2014, Ren2016, Kazawa2020, Ferrari2023}, or both \cite{Trajanovski2013, Zanin2018, Sun2021}.

\begin{figure}[p]
\centering
\includegraphics[trim={1.3cm 2.2cm 2.1cm 2.6cm},clip,width=1\textwidth]{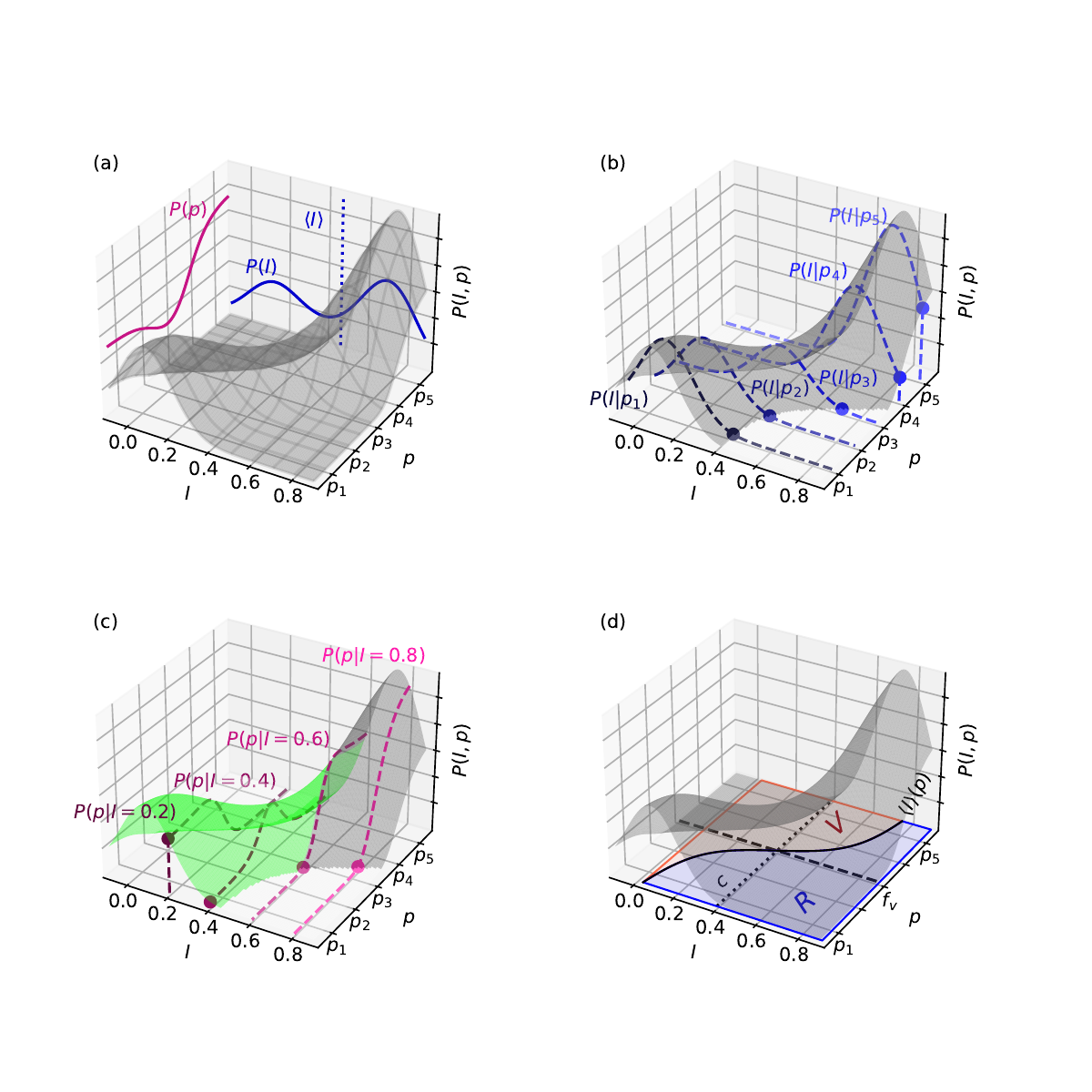}  
\caption{{\bf Relationship between the joint distribution $P(I,p)$ of perturbations $p$ and impacts $I$ and several scalar measures of robustness.} (a) The joint distribution $P(I,p)$ (gray surface) and the marginal distributions $P(p)$ (solid magenta curve) and $P(I)$ (solid blue curve). One can compute the mean impact $\langle I \rangle$ of perturbations from the marginal distribution $P(I)$. (b) The conditional probability distributions $P(I|p_i)$ of impacts given a perturbation $p_i$ are proportional to one-dimensional (1D) slices of $P(I,p)$ along the $I$ axis. (See the dashed curves in various shades of blue.) For each curve, a disk of the same shade indicates the maximum impact for the perturbation. To improve visibility, we show the gray surface that indicate $P(I,p)$ only in locations where $P(I,p) > 0$. (c) The conditional probability distributions $P(p|I_i)$ of impacts given a perturbation $I_i$ are proportional to 1D slices of $P(I,p)$ along the $p$ axis. (See the dashed curves in various shades of purple, magenta, and pink.) 
For each curve, a disk of the same shade indicates the minimum perturbation for which it is possible to observe the impact $I_i$. To improve visibility, we show the surface that indicates $P(I,p)$ only in locations where $P(I,p)>0$. The volume under the green part of the surface indicates the probability $P(I<I^*)$ of meeting a performance goal $I<I^*$. (d) A solid black curve indicates the mean impact $\langle I\rangle (p)$ as a function of a perturbation $p$. 
If the perturbations correspond to removing a proportion of the nodes, the critical fraction $f_v$ (see the dashed black curve) is the perturbation size for which $\langle I\rangle (p)$ leads to a specified impact $I = c$ (dotted black curve). The shaded pink region between $\langle I\rangle (p)$ and $\langle I\rangle = 0$  indicates Schneider et al.'s vulnerability index $V$. The shaded blue region between $\langle I\rangle (p)$ and $\langle I\rangle =I_{\textrm{max}}$ is Schneider et al.'s robustness index $R$.}
\label{fig:pip}
\end{figure}

When a perturbation $p$ occurs with a probability $P(p)$ and each $p$ leads to an impact $I$ with probability $P(I|p)$, the joint probability distribution $P(I,p)$ includes sufficient information to compute various scalar robustness measures. In Fig.\,\ref{fig:pip}, we illustrate the joint distribution $P(I,p)$ and its relationships to scalar robustness measures. The scalar robustness measures that one can calculate from $P(I,p)$ or samples of $P(l,p)$ include all of the robustness measures that we discussed in Section \ref{sec:scalar_robustness}:
\begin{itemize}
    \item{The mean impact of the removal of a single node is $\sum_I P(I)I$, where $P(I) = \sum_{p} P(I,p)$ is the marginal probability of impacts over all single-node-removal perturbations $p$. See the dotted blue curve in Fig.\,\ref{fig:pip}(a).}
    \item{The maximum impact of the removal of a single node is the largest value of $I$ for which $P(I,p) > 0$ for any single-node-removal perturbation $p$. See the blue disks in Fig.\,\ref{fig:pip}(b).} 
    \item{The probability of meeting a performance goal $I^*$ is $P(I<I^*) = \sum_{I<I^*}P(I)$, where $P(I)$ is the marginal probability $P(I) = \sum_{p} P(I,p)$ that one obtains by summing $P(I,p)$ over all considered perturbations $p$. See the volume under the green surface in Fig.\,\ref{fig:pip}(c).}
    \item{The critical fraction $f_v$ of nodes (see the dashed black curve in Fig.\,\ref{fig:pip}(d)) is the smallest value of $f_v$ such that the removal of a node set $V_S$ with $\lceil f_vN \rceil$ nodes leads to an expected impact $\langle I_{V_S} \rangle \geq c$ for some constant $c$. (See the dotted black curve in Fig.\,\ref{fig:pip}(d).) One can thus express $f_v$ via
    \begin{align}
	    \min_{f_v} f_v \hspace{0.3cm}\textrm{subject to}\, \sum_{p \in \mathcal P_f}\sum_I P(I,p)I \geq c\,,
    \label{eq:min-frac}\end{align}
    where $\mathcal P_f = \{V_S\subset V| \, |V_S|=\lceil f_vN\rceil\}$. Similarly, one can express the critical fraction $f_e$ of edges as the solution of the optimization problem in \eq{min-frac} with $\mathcal P_f=\{E_S\subset E| |E_S| = \lceil f_em \rceil\}$.}
    \item{Schneider et al.'s $R$ index is the mean impact of a sequence $(p_1, p_2, \ldots, p_{N-1})$ of perturbations, where $p_i$ corresponds to removing a node set $V_i$ of size $i$. The node sets $V_i$ are nested, i.e., $V_i\subset V_{i+1}$ for all $i=1,\dots,N-2$.
    Because the order of the perturbations in this sequence does not change the value of the mean, one can replace the sequence by the set $\mathcal P = \{p_1, p_2, \ldots, p_{N-1}\}$ of perturbations. One can then express Schneider et al.'s $R$ index as the area between the mean-impact function $\langle I \rangle (p) := \sum_I P(I|p) I$ and the $\langle I \rangle=I_{\textrm{max}}$ line: 
    \begin{align}
   	 R = \frac{1}{|\mathcal{P}|}\sum_{p\in\mathcal{P}}\left(I_{\textrm{max}}-\langle I \rangle (p)\right)\,.\label{eq:r-from-mean-impact}
    \end{align}
    If one normalizes the impacts such that $I_{\textrm{max}} = 1$, one can simplify \eq{r-from-mean-impact} to obtain
    \begin{align}
  	  R = 1 - \frac{1}{|\mathcal{P}|}\sum_{p\in\mathcal{P}}\langle I \rangle (p)\,,
    \end{align}
    which one can express in terms of conditional probabilities $P(I|p)$ by writing
    \begin{align}
 	   R = 1 - \frac{1}{|\mathcal{P}|}\sum_{p\in\mathcal{P}}\sum_I P(I|p) I
    \end{align}
    or in terms of the unconditional probabilities $P(I,p)$ by writing
    \begin{align}
   	 R = 1 - \frac{1}{|\mathcal{P}|}\sum_{p\in\mathcal{P}}\sum_I \frac{P(I,p) I}{\sum_I P(I,p)}\,.\nonumber
    \end{align}
    }
    \item{In several studies of the structural robustness of networks, researchers have reported a sequence of performance values or impact values \cite{Albert2000, Iyer2013, Ventresca2014, Ren2016, Kazawa2020, Ferrari2023} or their distribution \cite{Henneaux2015, Henneaux2015a, Panteli2015, Phan2018}. One can view these results as expected values or samples that one draws from the probability distribution $P(I,p)$. For example, when researchers report the progressive decline of performance with sequential node removal, the reported data has the form of a sequence of performance values or impact values $(I_1,I_2,\ldots)$, where each impact $I_i$ is either a sample from the conditional distribution $P(I|p_i)$ for some sequence $(p_1,p_2, \ldots)$ of perturbations or the expected value $\sum_{p\in{\mathcal P}_i} P(I,p) I$ for some sequence $({\mathcal P}_1,{\mathcal P}_2, \ldots)$ of perturbation sets.}
\end{itemize}

Because the joint probability distribution $P(I,p)$ includes sufficient information to compute many popular scalar measures of robustness, it is a rather informative non-scalar robustness measure. We are not aware of any studies in which researchers have used $P(I,p)$ to characterize a system's robustness. (Several studies have, however, included an exploration of confidence bands of $I(p)$ for sequential random node-removal perturbations, which they have called `robustness envelopes' \cite{Trajanovski2013, Zanin2018, Sun2021}.) We anticipate that it is difficult for many systems and models of systems to obtain sufficient observational, experimental, or computational data to construct $P(I,p)$. Another potential issue with using $P(I,p)$ or other non-scalar measures of robustness to characterize system robustness is that they do not facilitate straightforward comparisons of robustness between two or more systems. When two systems have different distributions $P(I,p)$, it is unclear how one should decide which of those systems is more robust. However, we argue that the difficulty that arises when trying to compare two or more distributions $P(I,p)$ is emblematic of the difficulty of ranking systems by their robustness without first fully specifying the robustness problem of interest.

\subsubsection{How are different robustness measures related?}\label{sec:robustness_correlations}

For a given network, different robustness measures do not need to be correlated with each other. For example, a small mean impact from single-node-removal perturbations of a network does not indicate that the same network has a small value of Schneider et al.'s $R$ index. These two robustness measures capture different notions of what it means for a system to be robust, and they consider different perturbation scenarios.

If one wishes to rank a set of networked systems by their robustness, the choice of robustness measure is a crucial decision in the study design. Ranking a set of systems by one robustness measure can lead to an entirely different ranking than what one obtains with a different robustness measure. Even when two systems have exactly the same value of one robustness measure, they can still differ in other robustness measures. 

Notable exceptions to this lack of correlations between different robustness measures occur when one can express one measure of robustness (e.g., a scalar robustness measure, such as the mean impact of single-node-removal perturbations) as a function of another measure of robustness (e.g., the joint distribution $P(I,p)$ of impacts and perturbations). If two systems share a probability distribution $P(I,p)$, one can expect robustness measures (such as the mean impact of single-node-removal perturbations) that are deterministic functions of $P(I,p)$ to have identical values for the two systems. Furthermore, one can expect stochastic measures of robustness (e.g., sampled impacts of removing a node uniformly at random) to be correlated for systems that have the same distribution $P(I,p)$.

\subsubsection{Interpreting scalar measures of robustness.}\label{sec:robustness_applications}

When characterizing a system's robustness by a scalar measure, it is common to focus on a specific notion of robustness and a specific set or sequence of perturbations. One's choice of a scalar measure of robustness should reflect one's interest in specific aspects of a system and its robustness. Scalar robustness measures that are maxima over a set of impacts carry information about the loss of performance in a worst-case scenario. By contrast, scalar robustness measures that are averages of a perturbation's impact or a system's performance after experiencing a perturbation tend to convey the performance loss that one can expect on average. 

Whether one is interested in a worst-case scenario or an average scenario depends on the implications of a performance loss in a system. For example, when the system of interest is a colony of bacteria and the robustness study concerns the survival of a single bacterium, it is likely that one is more interested in the average bacterial survival than in the most `unlucky' bacterium in the colony because the death of a single bacterium does not crucially affect the colony's survival. Conversely, when the system of interest is a human body (or a model of a human body), a severe loss of function can indicate a person's death, and one may be motivated to prevent this worst-case scenario, even if it is has a small probability.
Generally, if a severe loss of performance in a system is associated with very severe consequences that one hopes to prevent, those severe consequences can provide a strong motivation to study worst-case scenarios of the system's robustness.

Using the probability of achieving a prescribed performance goal as a robustness measure can be relevant when one can set a meaningful performance goal and wants to know how likely it is for a system to achieve that goal. For example, when modeling the internet as a communication network of computers, a relevant performance measure may be the average speed of data packages on this network, and a relevant performance goal may be a prescribed minimum value of the average speed of data packages on the internet. 
In this context, the probability of achieving the prescribed performance goal is potentially an important quantity. It has the potential to be more informative than the mean or maximum impact of the loss of a single computer in the network.

When it is possible to prevent perturbations to a system at some cost, one potentially interesting question is the following: How many perturbations (or what severity of perturbations) can a system withstand before there is a drastic loss of performance? In this context, the critical fraction of nodes and the critical fraction of edges that one needs to remove to cause a prescribed performance loss are potentially relevant robustness measures. If the performance measure is the LCC size or the relative LCC size, then node connectivity, edge connectivity, and algebraic connectivity are potentially relevant measures of robustness. These robustness measures can also be relevant from an adversarial perspective, in which one is interested in how many attacks are needed or what severity of attacks is needed to break a system.

\subsubsection{Performance measures as robustness measures.}\label{sec:performance_ambiguity}

In this review, we have endeavored to clearly distinguish between performance measures and robustness measures.
However, the vast literature on network robustness includes many examples that conflict with our categorization of performance and robustness measures. For example, researchers have used the entropy of a network's degree distribution as a performance measure when studying the robustness of networks to structural perturbations \cite{Ventresca2014, Baehner2017, Jiang2019}, but other studies have used the entropy of a network's degree distribution as a measure of robustness \cite{Wang2006, Greco2012, Koc2013, Baehner2017}. There are viable conceptual and technical arguments for using performance measures as robustness measures. We present three such arguments here.

In Section \ref{sec:well-connectedness}, we explained that one can view many performance measures as measures of a network's `connectedness' or `well-connectedness'. From a conceptual perspective, when we view `connectedness' (e.g., a sufficiently large relative LCC size) as a characteristic of a well-performing system, one can sometimes use measures of `well-connectedness' (e.g., transitivity) as robustness measures.
For example, suppose that a research uses the LCC size of a network model as a performance measure for an associated system. If the network's LCC has a large transitivity, then many edges in the LCC are part of a triangle.
Removing an edge that is part of a triangle from the LCC does not change the LCC size. One can thus consider the transitivity in the LCC as a measure of robustness to a single-edge-removal perturbation.

Another conceptual argument arises from the notion of a `performance buffer'. When a performance value of $X = X_{\min}$ is an indicator of a well-functioning system and perturbations tend to decrease performance by at most $\Delta X$, a performance of $X \geq X_{\min} + \Delta X$ indicates that a system most likely has a sufficient performance buffer to tolerate a single instance of the considered perturbations. For example, suppose that one measures the performance of a system by computing the mean local clustering coefficient $C$ of an associated network model $G$, and assume that the system performs well if $C \geq 0.5$. Removing a single edge $e$ from the network $G$ can decrease $C$ by at most $(\max\{k_1,k_2\} + 1)/N$, where $N$ is the number of nodes of the network and $k_1$ and $k_2$ are the degrees of the nodes that are incident to $e$. If the maximum node degree $k_{\max}$ in the network is less than $N/5 - 1$, one can thus conclude that removing a single edge from $G$ can decrease its mean local clustering coefficient by at most $0.2$. Consequently, a mean local clustering coefficient of $C \geq 0.7$ indicates that the system is performing well and is robust to a single-edge-removal perturbation. 

When one defines the impacts of perturbations using \eq{impact_as_change}, one can also make a technical argument for using performance measures as robustness measures. With \eq{impact_as_change}, the impact of a perturbation $p$ on a system is given by $-X_p$ with a constant offset $X_0$. One can then directly infer several popular robustness measures, such as the mean impact or maximum impact over a set of perturbations, from the mean value or minimum value of $X_p$ over the set of perturbations. Studies of the mean value or minimum value of $X_p$ thus tend to lead to similar insights as studies of the aforementioned robustness measures.

\subsubsection{Robustness measures as performance measures.}\label{sec:robustness_ambiguity}

Several researchers have used computational experiments to investigate the sensitivity and insensitivity of the robustness measures in Section \ref{sec:scalar_robustness_lcc} to perturbations \cite{Eubank2004, Wu2010a, Zhang2013b, Wang2014b, Cozzo2018, Yamashita2019}. Based on these studies, one can argue that 
it is reasonable to categorize these measures as performance measures. However, one can also view these studies as considering notions of `second-order' robustness. In most studies of robustness, researchers investigate whether or not a system still performs \textit{well} when it is subjected to a perturbation $p\in\mathcal P$. In studies of `second-order' robustness, researchers investigate whether or not a system, when subjected to a perturbation $p \in \mathcal P$, also performs \textit{robustly} in the sense of being able to withstand subsequent perturbations $p' \in \mathcal P'$, where $\mathcal P$ and $\mathcal P'$ can be either the same set of perturbations or different sets of perturbations.  

One may be motivated to study `second-order' robustness with $\mathcal P = \mathcal P'$ when one expects a system to include fail-safe structures (e.g., duplicate or redundant parts in an associated network model) that perturbations need to break to cause the system's performance to decline. In such a case, a decline of a relevant robustness measure precedes the decline of system performance, so the decline of the robustness measure is
a necessary condition for a loss of performance.

Designing a study of `second-order' robustness with different perturbation sets $\mathcal P$ and $\mathcal P'$ can be relevant when studying a system that is subject to frequent perturbations $p' \in \mathcal P'$ and it is safe to assume that the system's performance will decline whenever it is not robust to such perturbations. One may be interested in studying the effects of additional perturbations $p \in \mathcal P$ on this system. For example, transportation and infrastructure networks, such as railroad networks and power grids, are routinely affected by varying weather conditions, which one can treat as the perturbation set $\mathcal P'$. One can expect that a railroad network or power grid that is not robust to the effects of common weather variations will break down quickly. The robustness of these systems to common weather variations is thus a necessary condition for their sustained performance. A research objective with possibly wide-reaching implications is to understand or forecast how these systems respond to additional perturbations of a different type (e.g., targeted attacks on power stations or busy railroads).

\section{Conclusions and discussion}
\label{sec:review:conclusion}

We conclude our review with a discussion of the relevance of different specifications of a robustness problem. We also discuss several research avenues that can further understanding of network robustness.

A major challenge in the identification of `robust network structures' is that many studies of network robustness lead to different (and sometimes seemingly contradictory) conclusions about the characteristics of robust network structures. Alderson and Doyle~\cite{Alderson2010} proposed distinguishing between different notions of robustness using several specifications. Their framework helps resolve apparent conflicts between the results of different studies of robustness because it helps one recognize that many studies consider different robustness notions, which do not need to be correlated with one another. However, their framework also leads to the challenge of choosing specifications for a robustness problem. In the present paper, we reviewed various choices for specifying a model of a system, a performance measure, a set of perturbations, and a measure of insensitivity (i.e., robustness) or sensitivity (i.e., vulnerability) of a performance measure to perturbations.

\subsection{How should one choose a model?}\label{sec:choose_model}

There are usually several reasonable approaches to constructing a network model of a system.
Different network models tend to capture different aspects of a system \cite{Winterbach2011}. 
Even when using the same combination of robustness measure, performance measure, and set of perturbations, different network models can lead to different outcomes in studies of robustness. Therefore, it is important to choose an appropriate network model that captures relevant information that can affect a system's robustness.

\subsubsection{Guidance from additional features.}\label{sec:choose_model_from_features}

When modeling a system as a simple network, it is common to assume that the nodes and edges of a network are 
inherently similar (unlike, e.g., in an annotated or multilayer network) and that the different roles that nodes and edges can play in a system are determined by the simple network's structure. In reality, however, the objects that are modeled by nodes and edges can vary greatly in many of their aspects, and some of these aspect can be relevant to a system's robustness. For example, the nodes of a social network can model people who respond differently to receiving information from peers. The edges of a road network can encode roads with different lengths, numbers of lanes, conditions, speed limits, and other attributes. Therefore, different roads are more suitable for different types of traffic. One can include various attributes in network models as additional information, including in the form of node labels, edge weights, and multilayer structures (see Section \ref{sec:models}). When one anticipates that some attributes of a system (or some attributes of the nodes and edges of an associated network model) may affect a system's robustness, it is appropriate to include such information a model of a system. 
For example, when a road system includes both multilane highways and poorly maintained dust roads, it is reasonable to suppose that a dust road cannot act as an effective replacement for a broken-down highway. A model of this road system will this benefit from including information about the type and quality of roads, so that one can generate accurate insights into how structural perturbations can affect the system. If one expects some attributes (e.g., the names of the roads in a road system) of nodes or edges to be irrelevant for the robustness of a system, then it is reasonable to exclude these attributes from a model of that system.

\subsubsection{Guidance from dynamical processes.}

One can also seek guidance for building a model from one's knowledge about a system's dynamics. When modeling a system as a network, it is common to assume that something (e.g., influence, information, goods, people, or diseases) is transmitted or transported along the edges of the network. When selecting a performance measure for a robustness problem, it can be important to consider which types of transmissions or movement trajectories are important for a system to perform well. Ideally, one should construct the network model for that robustness problem so that the model includes information on whether each of these important transmission types or movement trajectories can occur or cannot occur in the system. 

When there is sufficient information about the relevant dynamics on a network that is associated with a system, one can include a mathematical or computational model of these dynamics in one's network model. In many cases, probing the robustness of properties of the dynamical process to structural perturbations in the network can reveal deeper insights into a system's robustness than assessing its performance from a performance measure that accounts only for network structure \cite{Alderson2010, Trajanovski2013, Manzano2014, Schieber2016}.

\subsubsection{What can one learn from simple network models?}

In many studies of robustness, researchers have used simple networks as models of systems \cite{Albert2000, Callaway2000, Iyer2013, Ventresca2014} (see also Table \ref{tab:networks}). An advantage of simple networks is that one can build simple-network models from many types of data and from many different domains, thereby facilitating comparisons of network robustness across domains. Many researchers have conducted comparative studies of network robustness across domains to uncover supposedly `universal' design principles \cite{Sole2002, Demetrius2005, Wang2008, Wang2012a, Shimada2014} that may have evolved independently in different domains \cite{Hintze2008, Zhang2016a} and in nature and technology \cite{Iyer2013, Ventresca2014}. However, it is possible that the characterization of the robustness of some or all of the compared systems requires a model that is more detailed than a simple network. Network characteristics that seem to be related to the robustness of systems across domains can be artifacts of the methods that one uses to construct simple networks from data \cite{Cantwell2020}. We emphasize that one should treat the results of such comparative studies of simple-network models with caution and, when possible, carefully evaluate them using network models with additional information.

\subsection{How should one choose a performance measure?}\label{sec:choose_performance}

Choosing a relevant performance measure is a crucial and sometimes challenging step in specifying a robustness problem. Many networked systems do not have an objectively identifiable purpose or output. Such situations occur in social networks, communication networks, mutualistic ecological networks, and other areas. In specifying a performance measure for a system in this situation, a researcher imposes a view of a system's purpose or output in their study. To make a conscious and reasonable selection of a performance measure, we advise researchers to take guidance from additional features in their network models and from the walks or paths that they expect to be important for a system's performance.

\subsubsection{Guidance from additional features.} 

In many studies of robustness, researchers have used the absolute or relative LCC size as a performance measure
\cite{Albert2000, Cohen2000, Cohen2000, Holme2002, Dodds2003, Shargel2003, Barrett2004, Bollobas2004, Motter2004, Paul2004, Beygelzimer2005, Tanizawa2005, Estrada2006, Wang2006, Sole2008, Sole2008, Chaverri2010, Cohen2010, Herrmann2011, Nair2011, Schneider2011, Wu2011, Zeng2012, Zeng2012, Zhou2012, Bilal2013, Hossain2013, Iyer2013, Louzada2013, Sha2013, Sha2013, Trajanovski2013, Zhang2013a, Danziger2014, Lordan2014, Priester2014, Ventresca2014, Wang2014, Wang2014a, Wang2014a, Zhou2014, Ahmadi2015, Gao2015, Gong2015, LaRocca2015, Nie2015, Radicchi2015, Wang2015, Yang2015, Agreste2016, Khansari2016, Ren2016, Schieber2016, Williams2016, DeDomenico2017, Mourier2017, Sun2017, daCunha2017, Phan2018, Deng2019, Yamashita2019, Kazawa2020, Mimar2022, Ferrari2023, Tomassini2023, Borge-Holthoefer2012, Rosas-Casals2007}. 
Additionally, for several network models that include additional features (e.g., node labels or dynamics on networks), some researchers have compared the LCC size to one or more model-specific performance measures, and they have concluded that the LCC size is not a good proxy for performance measures that account for model-specific features \cite{Winterbach2011, LaRocca2015}. When a network model includes relevant additional features, these features should help guide researchers in their specification of performance measures. When one expects an additional feature (e.g., node labels, edge labels, or edge weights) to contribute to the outcome of a study of a system's robustness (see Section \ref{sec:choose_model_from_features}), it is desirable for a suitable performance measure to account for this feature. For example, when a supply system includes suppliers and consumers of electricity or other goods and one includes that information in a network model via node labels, the proportion of consumers that are connected to suppliers can be a more relevant measure of performance than the LCC size or reachability of a corresponding simple network. (See Section \ref{sec:model_specific_performance} for examples.) 

Performance measures that account for relevant additional model features are often more relevant for the study of system robustness than performance measures that are based only on a simple network's structure \cite{LaRocca2015, Dong2019}.

\subsubsection{Guidance from relevant types of walks and paths.}

It can also be helpful to consider the types of walks and paths that one expects to contribute to the functions of a system, especially when it is not possible to include additional features in a network model. For example, if one expects a system function to depend on the existence of a path between as many node pairs as possible and if one assumes that all paths are permissible, then the LCC size, the relative LCC size, and reachability can be relevant measures of performance. Alternatively, if one assumes that walks or paths need to be short to contribute to a system function, then the network diameter, the mean shortest-path length or efficiency can be relevant measures of performance. The mean-shortest path length and efficiency tend to be better performance measures than network diameter when a the network structure of a well-performing system needs to facilitate movement, transport, or communication along short paths for many (but not necessarily for all) nodes. Conversely, the diameter of a network is a better performance measure than the mean-shortest path length or efficiency if it is important for a system's function that every node is connected to every other node by a short path. 

If one expects that one can approximate the dynamics on a network that contribute to a system function reasonably well by a linear diffusion process or other linear spreading process, several spectral network properties (e.g., resistance distance, the spectral radius of $\bf A$, and natural connectivity) can be relevant measures of performance.

\subsubsection{Robustness of simple networks.}

Choosing a relevant performance measure is especially difficult if no additional features are available or if it is unclear which types of walks or paths contribute to a system function. In such cases, many researchers have either used the LCC size as a performance measure \cite{Cohen2000, Bollobas2004, Paul2004, Tanizawa2005, Wang2006, Sole2008, Herrmann2011, Zeng2012, Louzada2013, Rosas-Casals2007} or examined the robustness of several performance measures \cite{Albert2000, Hossain2013, Lordan2014, Ventresca2014}. Unfortunately, the results of such studies of robustness of simple networks can have limited relevance to the robustness of the systems that one models with these simple networks. Insights into a system's robustness tend to require an understanding of a system's function(s) and/or the dynamics that enable the system to fulfill a function. When no information about function or dynamics is available, it is difficult to obtain meaningful insights into a system's robustness.

\subsection{How should one choose a set of perturbations?}

In studies of the robustness of networks to structural perturbations, it is common to use a set of node removals or edge removals as the set of perturbations. We distinguish between the removal of individual nodes or edges, simultaneous node or edge removal, and sequential node or edge removal.

\subsubsection{Removal of a single node or a single edge.}

In the study of trophic cascades and cascading failures, researchers have investigated the impact of single-node removals and single-edge removals \cite{Dunne2002, Motter2002, Crucitti2004, Dunne2004}. One can use such an approach to study questions like the following one: How likely is the failure of a single node to cause a major disruption in a system? In systems in which the failure of a single node or single edge is unlikely to cause a major disruption, researchers have considered the proportion of nodes that, if simultaneously removed, one can expect to lead to a major disruption \cite{Cohen2000, Shargel2003, Paul2004, Tanizawa2005, Wang2006, Sole2008, Rosas-Casals2007}.

\subsubsection{Simultaneous and sequential removal of nodes and edges.}

In models that do not include cascades, many researchers have studied the impact of simultaneous node removal or simultaneous edge removal \cite{Tanizawa2005, Greco2012, Yang2015, Colladon2017, Galias2017, Rosas-Casals2007} and sequential node removal \cite{Holme2002, Herrmann2011, Schneider2011, Zeng2012, Zhou2014, Agreste2016, Hong2017} on a performance measure. 
If one is interested in modeling failures in a system, it is common to examine the impact of removing nodes uniformly at random \cite{Albert2000, Cohen2000, Crucitti2004a}. If a network model includes neither dynamics on a network nor dynamics of a network's structure, then simultaneous node removals and sequential removal of nodes that one chooses uniformly at random yield the same impact. When studying the impact of attacks on a network, it is common to consider node or edge removal according to a centrality-based ranking of nodes or edges \cite{Albert2000, Iyer2013, Lordan2014, Ventresca2014}. The distinction between the simultaneous and sequential removal of nodes or edges matters in such studies because the centrality-based rankings can change after a node or edge removal and researchers can decide to use continuously updated node or edge rankings to select perturbations \cite{Holme2002, Agreste2016}.

\subsubsection{Targeting strategies.}

To demonstrate that targeted node or edge removal can lead to larger impacts than removing nodes or edges uniformly at random, it usually suffices to compare targeting nodes by largest degree to removing nodes uniformly at random. To model the impact of an adversary's attack on a network, it is sensible to consider centrality measures that give a reasonable proxy of the adversary's attack strategy. 
It is also relevant to examine algorithms that seek to identify the sets of nodes or edges that, if removed, lead to the largest impacts. It is possible that no node or edge centrality measure can reliably identify such sets. Tools that researchers have developed to study collective influence \cite{Morone2015, Lue2016, Morone2016, Zhu2018, Pei2019} and network dismantling \cite{Braunstein2016, Zdeborova2016, Ren2019} are likely to find node sets and edge sets that, if removed, lead to larger impacts than sets that one can construct from traditional centrality-based rankings \cite{Braunstein2016, Ren2019}. 
Comparative studies of the impact of targeted attacks using different node and edge centrality measures may help identify centrality measures that are good heuristics for finding node sets or edge sets that, if removed, have a large impact on a system's performance.

\subsection{How should one choose a measure of robustness?}

After specifying a system or a model of a system, a measure of performance, and a set or sequence of perturbations, one still needs to decide how to measure the robustness (i.e., insensitivity to perturbations) or vulnerability (i.e., sensitivity to perturbations) of the measure of performance. A sensible choice of robustness measure can depend on (1) the level of desired detail for the results of a study of a system's robustness, (2) the frequency and severity of the considered perturbations, and (3) the implications of a partial or complete loss of performance.

\subsubsection{Scalar versus non-scalar measures of robustness.}

In a computational study of a robustness problem, it is common to choose a set of perturbations and compute the impact of each perturbation in that set. The resulting set of impacts reveals a lot of information about the robustness of a system. The study of non-scalar robustness measures (e.g., the joint distribution $P(I,p)$ of perturbations and impacts (see Section \ref{sec:robustness_dist}) or a distribution of impacts of single-node removals or a sequence of impacts of sequential node removals \cite{Albert2000, Dartnell2005, Newman2008, Pu2012, Hossain2013, Sha2013, Lordan2014, Nie2015, DeDomenico2017, Galias2017, Kaiser-Bunbury2010}) can yield detailed insights into a robustness problem. One inevitably loses information when computing a scalar robustness measure from a set of impacts. However, scalar robustness measures facilitate simple comparisons of the robustness values for different systems \cite{Ventresca2014}, different sets of perturbations \cite{Iyer2013}, and so on. For some scalar measures of robustness, it is possible to calculate their exact values or bounds on their values from theoretical models using mathematical tools (e.g., tools from percolation theory \cite{Kesten1982, Stauffer2018} or spectral graph theory \cite{VanMieghem2023}).

\subsubsection{Perturbations, impacts, and probabilities.}

The many approaches to characterizing a system's robustness or vulnerability to perturbations include (1) identifying a smallest perturbation (e.g., the smallest number of nodes or edges to remove) that leads to a prescribed loss of performance \cite{Cohen2000, Dunne2002, Shargel2003, Dunne2004, Paul2004, Tanizawa2005, VanMieghem2005, Wang2006, Sole2008, Wang2008, Cohen2010, Shang2012, Rosas-Casals2007}, (2) identifying the maximum or some other function of the impacts that are associated with the considered perturbations \cite{Latora2005, Herrmann2011, Mello2011, Mello2011a, Schneider2011, Zeng2012, Louzada2013, Zhou2014, Ahmadi2015, Hong2017}, and (3) calculating the probability that a system maintains a prescribed performance value \cite{Henneaux2015, Henneaux2015a, Panteli2015, Phan2018}.

The probability of achieving a prescribed performance goal is a potentially relevant measure of robustness when one can set a meaningful performance goal for a system. It can be very informative to use a probabilistic approach to quantify robustness or vulnerability, especially when perturbations to the system are frequent and usually have a low impact on its performance.

It is often relevant to identify a smallest perturbation that leads to a prescribed loss of performance  when it is possible to prevent perturbations to a system or to mitigate their effects at some cost. It is sometimes very important to determine the smallest perturbation that leads to an unacceptable loss of performance, especially when the impact of different perturbations are strongly heterogeneous and most perturbations have low impacts on a system's performance.
Such information can inform decisions regarding when and which perturbations to attempt to prevent or mitigate. When most perturbations are associated with low impacts on performance, knowing that these perturbations will not lead to an unacceptable performance loss can substantially decrease the costs of maintaining a system. 

Identifying the maximum impact, mean impact, or some other function of a set of impacts that are associated with a set of perturbations is a popular approach for characterizing a system's robustness or vulnerability. Studies of maximum impacts, mean impacts, and related measures are especially relevant when one cannot dismiss the possibility that a perturbation of one node or one edge can severely impact a system's performance. 
If a study of maximum or mean impacts of a set of perturbations on a system's performance leads to the conclusion that most or all perturbations have a small impact on the system's performance, 
then it may be reasonable to quantify the robustness of that system by the probability of meeting a prescribed performance goal or the smallest perturbations that cause a specified performance loss. 

\subsubsection{How should one choose a scalar measure of robustness?}

Robustness measures that are based on the mean value of a set of impacts are related to the expected impact of a perturbation on a system's performance. By contrast, robustness measures that are based on the maximum value of a set of impacts are related to a `worst-case scenario' of a system's performance under perturbation. The choice of mean impact or maximum impact to measure robustness thus reflects a researcher's focus on either an expected-case scenario or a worst-case scenario. Both scenarios can yield interesting insights into a system's robustness, and a system's robustness measures for an expected-case scenario and a worst-case scenario can differ substantially from each other. Obtaining a detailed understanding of a worst-case scenario and how to prevent it can be of great practical importance, especially when the implications of a loss of system performance are very severe.

\subsection{Outlook}

The search for `robust network structures' is a very active research area, but progress has been slow. A challenging aspect of studying network robustness is that a network's ability to perform a function often depends on a dynamical process on the network \cite{Dipple2001, Motter2002, Aldana2003, Demetrius2005, Gorban2007, Kwon2008, Simonsen2008, Tanaka2012, Goles2013, He2013}. Understanding the links between network robustness and network structure thus relies in part on understanding (1) the effects of heterogeneous roles of sets of nodes or edges of a network and (2) connections between dynamics on networks and network structure, which is also an active research area with many open problems \cite{Keeling2005, Newman2006, Pascual2006, Bascompte2010, Boccaletti2014, Porter2016, Nishikawa2017, Schwarze2020}.

We anticipate that the study of the robustness of network models with additional features (e.g., dynamics on networks, node labels, and multilayer structures) can help accurately characterize network structures that are beneficial for model-specific notions of robustness. A comparison of robust network structures for different notions of robustness has the potential to identify (1) robustness notions that lead to similar `robust network structures' and (2) robustness notions that lead to substantially different robust network structures from each other.

\section*{Acknowledgements}

ACS thanks the participants of the Santa Fe Institute's Complex Systems Summer School 2018 for highlighting the need for a comprehensive review of models of system robustness. This paper is based on a chapter of a PhD thesis \cite{Schwarze2019a} that ACS completed with the financial support of the UK Engineering and Physical Sciences Research Council (EPSRC) and the UK Medical Research Council (MRC) under grant number EP/L016044/1 and a Clarendon scholarship from the University of Oxford.
During the development of this paper, ACS was supported by funding to Bing Brunton's research group at the University of Washington, including the Air Force Office of Scientific Research and the Department of Defense (MURI award 1FA9550-19-1-0386), and funding to Peter J. Mucha's research group at Dartmouth College, including the Army Research Office (MURI award W911NF-18-1-0244) and startup funds provided by Dartmouth College. She additionally thanks Peter J. Mucha for helpful discussions and mentorship throughout the final stages of the manuscript preparation.  
She also thanks her cat for her unwavering support and her daughter for learning to sleep through the night early.





\end{document}